\documentclass[11pt]{article}
\pdfoutput=1
\usepackage{amsmath,bm,tensor}
\usepackage{jheppub}
\usepackage[utf8]{inputenc}
\usepackage{booktabs}
\usepackage{graphicx}
\usepackage{bbold}
\usepackage{subcaption}
\usepackage{mathtools}
\usepackage{slashed}

\usepackage{xkeyval}

\usepackage{tikz,fp}
\usetikzlibrary{external,fixedpointarithmetic,decorations.pathmorphing,positioning,calc,fadings,decorations.markings,decorations.pathreplacing,patterns,arrows}

\tikzset{
  label distance=-3pt,
  >=stealth,
  inner sep=1.5pt,
  boson/.style={
    decoration={snake, segment length=2mm, amplitude=0.5mm},
    decorate,
  },
  graviton/.style={
  	line width=0.1pt,
    decoration={snake, segment length=0.7mm, amplitude=0.2mm},
    decorate,
  },
  quark/.style={
    postaction={decorate},
    decoration={markings,mark=at position .5 with {\draw[->] (-0.01,0) -- (0.07,0);}}
  },
  aquark/.style={
    postaction={decorate},
    decoration={markings,mark=at position .5 with {\draw[->] (0.01,0) -- (-0.07,0);}}
  },
  gluonOld/.style={
    line width=0.7pt,
    decorate,
    decoration={coil,aspect=-0.5,amplitude=-2pt,segment length=2.2pt}
  },
  gluon/.style={
    line width=0.5pt,
    decorate,
    decoration={coil,aspect=-0.75,amplitude=-1.5pt,segment length=2pt}
  },
  agluon/.style={
    line width=0.5pt,
    decorate,
    decoration={coil,aspect=0.75,amplitude=1.5pt,segment length=2pt}
  },
  scalar/.style={
    dashed
  },
  dscalar/.style={
    dashed,
    postaction={decorate},
    decoration={markings,mark=at position 0.65 with {\draw[->] (-0.01,0) -- (0.07,0);}}
  },
  adscalar/.style={
    dashed,
    postaction={decorate},
    decoration={markings,mark=at position 0.65 with {\draw[->] (-0.01,0) -- (0.07,0);}}
  },
  line/.style={
  },
  doubleline/.style={
    double
  },
  middleArrow/.style={
    draw=white,
    decoration={markings,mark=at position 0.65 with{\arrow[scale=1,black]{>}}},
    decorate
  },
  dot/.style={
  	postaction={decorate},
    decoration={markings,mark=between positions 0.5 and 0.5 step 1 with {\draw[fill] (0,0) circle [radius=1.5pt];}}  
  },
  dots/.style={
  	postaction={decorate},
    decoration={markings,mark=between positions 0.33 and 0.66 step 0.33 with {\draw[fill] (0,0) circle [radius=1.5pt];}}  
  },
  dotss/.style={
  	postaction={decorate},
    decoration={markings,mark=between positions 0.25 and 0.75 step 0.25 with {\draw[fill] (0,0) circle [radius=1.5pt];}}  
  },
  blob/.style={
  	pattern=north west lines,
    draw=black
  },
  blobBlack/.style={
    draw=black,
    fill=black
  },
  blobWhite/.style={
    draw=black,
    fill=white
  }
}

\makeatletter
\def\gSetStdExtTypes#1{\presetkeys{gTypes}{eA=#1,eB=#1,eC=#1,eD=#1,eE=#1}{}}
\def\gSetStdIntTypes#1{\presetkeys{gTypes}{iA=#1,iB=#1,iC=#1,iD=#1,iE=#1,iF=#1,iG=#1,iH=#1,iI=#1,iJ=#1}{}}
\def\gSetStdTypes#1{\gSetStdExtTypes{#1}\gSetStdIntTypes{#1}}

\define@cmdkeys{general}{scale}
\define@cmdkeys{general}{rotate}
\define@cmdkeys{general}{font}
\define@cmdkeys{general}{yshift}
\presetkeys{general}{scale=1, rotate=0, font=\scriptsize, yshift=0pt}{}

\define@cmdkeys{gTypes}{eA,eB,eC,eD,eE,iA,iB,iC,iD,iE,iF,iG,iH,iI,iJ}
\gSetStdTypes{line}
\define@key{pre}{all}{\gSetStdTypes{#1}}
\define@key{pre}{allE}{\gSetStdExtTypes{#1}}
\define@key{pre}{allI}{\gSetStdIntTypes{#1}}

\define@cmdkeys{gLabels}{eLA,eLB,eLC,eLD,eLE,iLA,iLB,iLC,iLD,iLE,iLF,iLG,iLH,iLI,iLJ}
\presetkeys{gLabels}{eLA={},eLB={},eLC={},eLD={},eLE={},iLA={},iLB={},iLC={},iLD={},iLE={},iLF={},iLG={},iLH={},iLI={},iLJ={}}{}

\define@boolkeys{dots}{lDots,rDots}
\presetkeys{dots}{lDots=false, rDots=false}{}

\define@cmdkeys{blobs}{blobA,blobB,blobC}
\presetkeys{blobs}{blobA=blob,blobB=blob,blobC=blob}{}

\def\gTreeTri{\@ifnextchar[\@gTreeTri{\@gTreeTri[]}}
\def\@gTreeTri[#1]#2{%
  \setkeys*{pre}{#1}%
  \setrmkeys{general,gTypes,gLabels}%
  \begin{tikzpicture}
    [line width=1pt,
    baseline={([yshift=-0.5ex+\cmdKV@general@yshift]current bounding box.center)},
    scale=\cmdKV@general@scale,
    rotate=\cmdKV@general@rotate,
    font=\cmdKV@general@font]
    \draw[\cmdKV@gTypes@eA] (0.7,0.3) -- (0.4,0.3) node[label=(-180+\cmdKV@general@rotate):\cmdKV@gLabels@eLA] {};
    \draw[\cmdKV@gTypes@eB] (0.7,0.3) -- (0.9,0.6) node[label=(\cmdKV@general@rotate):\cmdKV@gLabels@eLB] {};
    \draw[\cmdKV@gTypes@eC] (0.7,0.3) -- (0.9,0) node[label=(\cmdKV@general@rotate):\cmdKV@gLabels@eLC] {};
    #2
  \end{tikzpicture}
  \gSetStdTypes{line}
}

\def\gTreeS{\@ifnextchar[\@gTreeS{\@gTreeS[]}}
\def\@gTreeS[#1]#2{%
  \setkeys*{pre}{#1}%
  \setrmkeys{general,gTypes,gLabels}%
  \begin{tikzpicture}
    [line width=1pt,
    baseline={([yshift=-0.5ex+\cmdKV@general@yshift]current bounding box.center)},
    scale=\cmdKV@general@scale,
    rotate=\cmdKV@general@rotate,
    font=\cmdKV@general@font]
    \draw[\cmdKV@gTypes@eA] (0.2,0.3) -- (0,0) node[label=(-180+\cmdKV@general@rotate):\cmdKV@gLabels@eLA] {};
    \draw[\cmdKV@gTypes@eB] (0.2,0.3) -- (0,0.6) node[label=(180+\cmdKV@general@rotate):\cmdKV@gLabels@eLB] {};
    \draw[\cmdKV@gTypes@eC] (0.7,0.3) -- (0.9,0.6) node[label=(\cmdKV@general@rotate):\cmdKV@gLabels@eLC] {};
    \draw[\cmdKV@gTypes@eD] (0.7,0.3) -- (0.9,0) node[label=(\cmdKV@general@rotate):\cmdKV@gLabels@eLD] {};    
    \draw[label distance=0,\cmdKV@gTypes@iA] (0.2,0.3) -- node[label=(90+\cmdKV@general@rotate):\cmdKV@gLabels@iLA] {} (0.7,0.3);
    #2
  \end{tikzpicture}
  \gSetStdTypes{line}
}

\def\gTreeT{\@ifnextchar[\@gTreeT{\@gTreeT[]}}
\def\@gTreeT[#1]#2{%
  \setkeys*{pre}{#1}%
  \setrmkeys{general,gTypes,gLabels}%
  \begin{tikzpicture}
    [line width=1pt,
    baseline={([yshift=-0.5ex+\cmdKV@general@yshift]current bounding box.center)},
    scale=\cmdKV@general@scale,
    rotate=\cmdKV@general@rotate,
    font=\cmdKV@general@font]
    \draw[\cmdKV@gTypes@eA] (0.45,0.1) -- (0,0) node[label=(-180+\cmdKV@general@rotate):\cmdKV@gLabels@eLA] {};
    \draw[\cmdKV@gTypes@eB] (0.45,0.5) -- (0,0.6) node[label=(180+\cmdKV@general@rotate):\cmdKV@gLabels@eLB] {};
    \draw[\cmdKV@gTypes@eC] (0.45,0.5) -- (0.9,0.6) node[label=(\cmdKV@general@rotate):\cmdKV@gLabels@eLC] {};
    \draw[\cmdKV@gTypes@eD] (0.45,0.1) -- (0.9,0) node[label=(\cmdKV@general@rotate):\cmdKV@gLabels@eLD] {};
    \draw[label distance=0,\cmdKV@gTypes@iA] (0.45,0.5) -- node[label=(\cmdKV@general@rotate):\cmdKV@gLabels@iLA] {} (0.45,0.1);
    #2
  \end{tikzpicture}
  \gSetStdTypes{line}
}

\def\gTreeU{\@ifnextchar[\@gTreeU{\@gTreeU[]}}
\def\@gTreeU[#1]#2{%
  \setkeys*{pre}{#1}%
  \setrmkeys{general,gTypes,gLabels}%
  \begin{tikzpicture}
    [line width=1pt,
    baseline={([yshift=-0.5ex+\cmdKV@general@yshift]current bounding box.center)},
    scale=\cmdKV@general@scale,
    rotate=\cmdKV@general@rotate,
    font=\cmdKV@general@font]
    \draw[\cmdKV@gTypes@eA] (0.45,0.5) -- (0,0) node[label=(-180+\cmdKV@general@rotate):\cmdKV@gLabels@eLA] {};
    \foreach \x in {5,...,14} \fill[opacity=0.12,white] (0.27,0.3) circle (\x/100);
    \draw[\cmdKV@gTypes@eB] (0.45,0.1) -- (0,0.6) node[label=(180+\cmdKV@general@rotate):\cmdKV@gLabels@eLB] {};
    \draw[\cmdKV@gTypes@eC] (0.45,0.5) -- (0.9,0.6) node[label=(\cmdKV@general@rotate):\cmdKV@gLabels@eLC] {};
    \draw[\cmdKV@gTypes@eD] (0.45,0.1) -- (0.9,0) node[label=(\cmdKV@general@rotate):\cmdKV@gLabels@eLD] {};
    \draw[label distance=0,\cmdKV@gTypes@iA] (0.45,0.5) -- node[label=(\cmdKV@general@rotate):\cmdKV@gLabels@iLA] {} (0.45,0.1);
    #2
  \end{tikzpicture}
  \gSetStdTypes{line}
}

\def\gTreeFourPt{\@ifnextchar[\@gTreeFourPt{\@gTreeFourPt[]}}
\def\@gTreeFourPt[#1]#2{%
  \setkeys*{pre}{#1}%
  \setrmkeys{general,gTypes,gLabels}%
  \begin{tikzpicture}
    [line width=1pt,
    baseline={([yshift=-0.5ex+\cmdKV@general@yshift]current bounding box.center)},
    scale=\cmdKV@general@scale,
    rotate=\cmdKV@general@rotate,
    font=\cmdKV@general@font]
    \draw[\cmdKV@gTypes@eA] (0,0) -- (-0.4,-0.4) node[label=(-180+\cmdKV@general@rotate):\cmdKV@gLabels@eLA] {};
    \draw[\cmdKV@gTypes@eB] (0,0) -- (-0.4,0.4) node[label=(180+\cmdKV@general@rotate):\cmdKV@gLabels@eLB] {};
    \draw[\cmdKV@gTypes@eC] (0,0) -- (0.4,0.4) node[label=(\cmdKV@general@rotate):\cmdKV@gLabels@eLC] {};
    \draw[\cmdKV@gTypes@eD] (0,0) -- (0.4,-0.4) node[label=(\cmdKV@general@rotate):\cmdKV@gLabels@eLD] {};
    #2
  \end{tikzpicture}
  \gSetStdTypes{line}
}

\def\gTadA{\@ifnextchar[\@gTadA{\@gTadA[]}}
\def\@gTadA[#1]#2{%
  \setkeys*{pre}{#1}%
  \setrmkeys{general,gTypes,gLabels}%
  \begin{tikzpicture}
    [line width=1pt,
    baseline={([yshift=-0.5ex+\cmdKV@general@yshift]current bounding box.center)},
    scale=\cmdKV@general@scale,
    rotate=\cmdKV@general@rotate,
    font=\cmdKV@general@font]
    \draw[\cmdKV@gTypes@eA] (-0.7,0.3) -- (-0.9,0) node[label=(180+\cmdKV@general@rotate):\cmdKV@gLabels@eLA] {};
    \draw[\cmdKV@gTypes@eB] (-0.7,0.3) -- (-0.9,0.6) node[label=(180+\cmdKV@general@rotate):\cmdKV@gLabels@eLB] {};
    \draw[\cmdKV@gTypes@eC] (0.15,0.45) -- (0.5,0.6) node[label=(\cmdKV@general@rotate):\cmdKV@gLabels@eLC] {};
    \draw[\cmdKV@gTypes@eD] (-0.3,0.3) -- (-0.2,0) node[label=(\cmdKV@general@rotate):\cmdKV@gLabels@eLD] {};
    \draw[label distance=0,\cmdKV@gTypes@iA] (-0.7,0.3) -- node[label=(90+\cmdKV@general@rotate):\cmdKV@gLabels@iLA] {} (-0.3,0.3);
    \draw[label distance=0,\cmdKV@gTypes@iB] (-0.3,0.3) -- node[label=(-80+\cmdKV@general@rotate):\cmdKV@gLabels@iLB] {} (0.15,0.45);
    \draw[label distance=0,\cmdKV@gTypes@iC] (0.15,0.45) -- node[label=(45+\cmdKV@general@rotate):\cmdKV@gLabels@iLC] {} (-0.18,0.69);
    \draw[label distance=-0.5,\cmdKV@gTypes@iD] (-0.18,0.69) .. controls ($(-0.18,0.69) + (-0.12,-0.12)$) and ($(-0.35,0.87) + (-0.12,-0.12)$) .. (-0.35,0.87) node[label=(180+\cmdKV@general@rotate):\cmdKV@gLabels@iLD] {} .. controls ($(-0.35,0.87) + (0.12,0.12)$) and ($(-0.18,0.69) + (0.12,0.12)$) .. (-0.18,0.69);
    #2
  \end{tikzpicture}
  \gSetStdTypes{line}
}

\def\gTadB{\@ifnextchar[\@gTadB{\@gTadB[]}}
\def\@gTadB[#1]#2{%
  \setkeys*{pre}{#1}%
  \setrmkeys{general,gTypes,gLabels}%
  \begin{tikzpicture}
    [line width=1pt,
    baseline={([yshift=-0.5ex+\cmdKV@general@yshift]current bounding box.center)},
    scale=\cmdKV@general@scale,
    rotate=\cmdKV@general@rotate,
    font=\cmdKV@general@font]
    \draw[\cmdKV@gTypes@eA] (0,0.3) -- (-0.2,0) node[label=(-180+\cmdKV@general@rotate):\cmdKV@gLabels@eLA] {};
    \draw[\cmdKV@gTypes@eB] (0,0.3) -- (-0.2,0.6) node[label=(180+\cmdKV@general@rotate):\cmdKV@gLabels@eLB] {};
    \draw[\cmdKV@gTypes@eC] (0.7,0.3) -- (0.9,0.6) node[label=(\cmdKV@general@rotate):\cmdKV@gLabels@eLC] {};
    \draw[\cmdKV@gTypes@eD] (0.7,0.3) -- (0.9,0) node[label=(\cmdKV@general@rotate):\cmdKV@gLabels@eLD] {};
    \draw[label distance=0,\cmdKV@gTypes@iA] (0,0.3) -- node[label=(-90+\cmdKV@general@rotate):\cmdKV@gLabels@iLA] {} (0.35,0.3);
    \draw[label distance=0,\cmdKV@gTypes@iB] (0.35,0.3) -- node[label=(\cmdKV@general@rotate):\cmdKV@gLabels@iLB] {} (0.35,0.6);
    \draw[label distance=-0.5,\cmdKV@gTypes@iC] (0.35,0.6) .. controls (0.18,0.6) and (0.18,0.85) .. (0.35,0.85) .. controls (0.52,0.85) and (0.52,0.6) .. node[label=(\cmdKV@general@rotate):\cmdKV@gLabels@iLC] {} (0.35,0.6);
    \draw[label distance=0,\cmdKV@gTypes@iD] (0.35,0.3) -- node[label=(-90+\cmdKV@general@rotate):\cmdKV@gLabels@iLD] {} (0.7,0.3);
    #2
  \end{tikzpicture}
  \gSetStdTypes{line}
}
\def\gBubA{\@ifnextchar[\@gBubA{\@gBubA[]}}
\def\@gBubA[#1]#2{%
  \setkeys*{pre}{#1}%
  \setrmkeys{general,gTypes,gLabels}%
  \begin{tikzpicture}
    [line width=1pt,
    baseline={([yshift=-0.5ex+\cmdKV@general@yshift]current bounding box.center)},
    scale=\cmdKV@general@scale,
    rotate=\cmdKV@general@rotate,
    font=\cmdKV@general@font]
    \draw[\cmdKV@gTypes@eA] (0.5,0.7) -- (0.2,0.7) node[label=(180+\cmdKV@general@rotate):\cmdKV@gLabels@eLA] {};
    \draw[\cmdKV@gTypes@eB] (0.9,0.7) -- (1.2,0.7) node[label=(\cmdKV@general@rotate):\cmdKV@gLabels@eLB] {};
    \draw[label distance=0,\cmdKV@gTypes@iA] (0.5,0.7) .. controls (0.5,0.95) and (0.9,0.95) .. node[label=(90+\cmdKV@general@rotate):\cmdKV@gLabels@iLA] {} (0.9,0.7);
    \draw[label distance=0,\cmdKV@gTypes@iB] (0.9,0.7) .. controls (0.9,0.45) and (0.5,0.45) .. node[label=(-90+\cmdKV@general@rotate):\cmdKV@gLabels@iLB] {} (0.5,0.7);
    #2
  \end{tikzpicture}
  \gSetStdTypes{line}
}
\def\gBubB{\@ifnextchar[\@gBubB{\@gBubB[]}}
\def\@gBubB[#1]#2{%
  \setkeys*{pre}{#1}%
  \setrmkeys{general,gTypes,gLabels}%
  \begin{tikzpicture}
    [line width=1pt,
    baseline={([yshift=-0.5ex+\cmdKV@general@yshift]current bounding box.center)},
    scale=\cmdKV@general@scale,
    rotate=\cmdKV@general@rotate,
    font=\cmdKV@general@font]
    \draw[\cmdKV@gTypes@eA] (0.2,0.7) -- (0,0.4) node[label=(-180+\cmdKV@general@rotate):\cmdKV@gLabels@eLA] {};
    \draw[\cmdKV@gTypes@eB] (0.2,0.7) -- (0,1) node[label=(180+\cmdKV@general@rotate):\cmdKV@gLabels@eLB] {};
    \draw[\cmdKV@gTypes@eC] (1.2,0.7) -- (1.4,1) node[label=(\cmdKV@general@rotate):\cmdKV@gLabels@eLC] {};
    \draw[\cmdKV@gTypes@eD] (1.2,0.7) -- (1.4,0.4) node[label=(\cmdKV@general@rotate):\cmdKV@gLabels@eLD] {};
    \draw[label distance=0,\cmdKV@gTypes@iA] (0.2,0.7) -- node[label=(90+\cmdKV@general@rotate):\cmdKV@gLabels@iLA] {} (0.5,0.7);
    \draw[label distance=0,\cmdKV@gTypes@iB] (0.5,0.7) .. controls (0.5,0.95) and (0.9,0.95) .. node[label=(90+\cmdKV@general@rotate):\cmdKV@gLabels@iLB] {} (0.9,0.7);
    \draw[label distance=0,\cmdKV@gTypes@iC] (0.9,0.7) .. controls (0.9,0.45) and (0.5,0.45) .. node[label=(-90+\cmdKV@general@rotate):\cmdKV@gLabels@iLC] {} (0.5,0.7);
    \draw[label distance=0,\cmdKV@gTypes@iD] (0.9,0.7) -- node[label=(90+\cmdKV@general@rotate):\cmdKV@gLabels@iLD] {} (1.2,0.7);
    #2
  \end{tikzpicture}
  \gSetStdTypes{line}
}

\def\gBubC{\@ifnextchar[\@gBubC{\@gBubC[]}}
\def\@gBubC[#1]#2{%
  \setkeys*{pre}{#1}%
  \setrmkeys{general,gTypes,gLabels}%
  \begin{tikzpicture}
    [line width=1pt,
    baseline={([yshift=-0.5ex+\cmdKV@general@yshift]current bounding box.center)},
    scale=\cmdKV@general@scale,
    rotate=\cmdKV@general@rotate,
    font=\cmdKV@general@font]
    \draw[\cmdKV@gTypes@eA] (-0.7,0.3) -- (-0.9,0) node[label=(180+\cmdKV@general@rotate):\cmdKV@gLabels@eLA] {};
    \draw[\cmdKV@gTypes@eB] (-0.7,0.3) -- (-0.9,0.6) node[label=(180+\cmdKV@general@rotate):\cmdKV@gLabels@eLB] {};
    \draw[\cmdKV@gTypes@eC] (0.3,0.5) -- (0.5,0.6) node[label=(\cmdKV@general@rotate):\cmdKV@gLabels@eLC] {};
    \draw[\cmdKV@gTypes@eD] (-0.3,0.3) -- (-0.2,0) node[label=(\cmdKV@general@rotate):\cmdKV@gLabels@eLD] {};
    \draw[label distance=0,\cmdKV@gTypes@iA] (-0.7,0.3) -- node[label=(90+\cmdKV@general@rotate):\cmdKV@gLabels@iLA] {} (-0.3,0.3);
    \draw[label distance=0,\cmdKV@gTypes@iB] (-0.3,0.3) -- node[label=(90+\cmdKV@general@rotate):\cmdKV@gLabels@iLB] {} (-0,0.4);
    \draw[label distance=0,\cmdKV@gTypes@iC] (0,0.4) .. controls (-0.03,0.6) and (0.22,0.65) .. node[label=(90+\cmdKV@general@rotate):\cmdKV@gLabels@iLC] {} (0.3,0.5);
    \draw[label distance=0,\cmdKV@gTypes@iD] (0.3,0.5) .. controls (0.38,0.3) and (0.08,0.21) .. node[label=(-90+\cmdKV@general@rotate):\cmdKV@gLabels@iLD] {} (0,0.4);
    #2
  \end{tikzpicture}
  \gSetStdTypes{line}
}
\def\gTriA{\@ifnextchar[\@gTriA{\@gTriA[]}}
\def\@gTriA[#1]#2{%
  \setkeys*{pre}{#1}%
  \setrmkeys{general,gTypes,gLabels}%
  \begin{tikzpicture}
    [line width=1pt,
    baseline={([yshift=-0.5ex+\cmdKV@general@yshift]current bounding box.center)},
    scale=\cmdKV@general@scale,
    rotate=\cmdKV@general@rotate,
    font=\cmdKV@general@font]
    \draw[\cmdKV@gTypes@eC] (0:0.3) -- (0:0.55) node[label=(\cmdKV@general@rotate):\cmdKV@gLabels@eLC] {};
    \draw[\cmdKV@gTypes@eA] (-120:0.3) -- (-120:0.55) node[label=(-180+\cmdKV@general@rotate):\cmdKV@gLabels@eLA] {};
    \draw[\cmdKV@gTypes@eB] (120:0.3) -- (120:0.55) node[label=(180+\cmdKV@general@rotate):\cmdKV@gLabels@eLB] {};    
    \draw[label distance=0,\cmdKV@gTypes@iA] (-120:0.3) -- node[label=(180+\cmdKV@general@rotate):\cmdKV@gLabels@iLA] {} (120:0.3);
    \draw[label distance=0,\cmdKV@gTypes@iB] (120:0.3) -- node[label=(60+\cmdKV@general@rotate):\cmdKV@gLabels@iLB] {} (0:0.3);
    \draw[label distance=0,\cmdKV@gTypes@iC] (0:0.3) -- node[label=(-60+\cmdKV@general@rotate):\cmdKV@gLabels@iLC] {} (-120:0.3);
    #2
  \end{tikzpicture}
  \gSetStdTypes{line}
}
\def\gTriB{\@ifnextchar[\@gTriB{\@gTriB[]}}
\def\@gTriB[#1]#2{%
  \setkeys*{pre}{#1}%
  \setrmkeys{general,gTypes,gLabels}%
  \begin{tikzpicture}
    [line width=1pt,
    baseline={([yshift=-0.5ex+\cmdKV@general@yshift]current bounding box.center)},
    scale=\cmdKV@general@scale,
    rotate=\cmdKV@general@rotate,
    font=\cmdKV@general@font]
    \draw[\cmdKV@gTypes@eA] (0.3,0.45) -- (0,0.3) node[label=(-180+\cmdKV@general@rotate):\cmdKV@gLabels@eLA] {};
    \draw[\cmdKV@gTypes@eB] (0.3,0.95) -- (0,1.1) node[label=(180+\cmdKV@general@rotate):\cmdKV@gLabels@eLB] {};
    \draw[\cmdKV@gTypes@eC] (1,0.7) -- (1.3,1.1) node[label=(\cmdKV@general@rotate):\cmdKV@gLabels@eLC] {};
    \draw[\cmdKV@gTypes@eD] (1,0.7) -- (1.3,0.3) node[label=(\cmdKV@general@rotate):\cmdKV@gLabels@eLD] {};
    \draw[label distance=0,\cmdKV@gTypes@iA] (0.3,0.45) -- node[label=(180+\cmdKV@general@rotate):\cmdKV@gLabels@iLA] {} (0.3,0.95);
    \draw[label distance=0,\cmdKV@gTypes@iB] (0.3,0.95) -- node[label=(70+\cmdKV@general@rotate):\cmdKV@gLabels@iLB] {} (0.7,0.7);
    \draw[label distance=0,\cmdKV@gTypes@iC] (0.7,0.7) -- node[label=(-70+\cmdKV@general@rotate):\cmdKV@gLabels@iLC] {} (0.3,0.45);
    \draw[label distance=0,\cmdKV@gTypes@iD] (1,0.7) -- node[label=(90+\cmdKV@general@rotate):\cmdKV@gLabels@iLD] {} (0.7,0.7);
    #2
  \end{tikzpicture}
  \gSetStdTypes{line}
}
\def\gBox{\@ifnextchar[\@gBox{\@gBox[]}}
\def\@gBox[#1]#2{%
  \setkeys*{pre}{#1}%
  \setrmkeys{general,gTypes,gLabels}%
  \begin{tikzpicture}
    [line width=1pt,
    baseline={([yshift=-0.5ex+\cmdKV@general@yshift]current bounding box.center)},
    scale=\cmdKV@general@scale,
    rotate=\cmdKV@general@rotate,
    font=\cmdKV@general@font]
    \draw[\cmdKV@gTypes@eA] (0.25,0.25) -- (0,0) node[label=(-180+\cmdKV@general@rotate):\cmdKV@gLabels@eLA] {};
    \draw[\cmdKV@gTypes@eB] (0.25,0.75) -- (0,1) node[label=(180+\cmdKV@general@rotate):\cmdKV@gLabels@eLB] {};
    \draw[\cmdKV@gTypes@eC] (0.75,0.75) -- (1,1) node[label=(\cmdKV@general@rotate):\cmdKV@gLabels@eLC] {};
    \draw[\cmdKV@gTypes@eD] (0.75,0.25) -- (1,0) node[label=(\cmdKV@general@rotate):\cmdKV@gLabels@eLD] {};
    \draw[label distance=0,\cmdKV@gTypes@iA] (0.25,0.25) -- node[label=(180+\cmdKV@general@rotate):\cmdKV@gLabels@iLA] {} (0.25,0.75);
    \draw[label distance=0,\cmdKV@gTypes@iB] (0.25,0.75) -- node[label=(90+\cmdKV@general@rotate):\cmdKV@gLabels@iLB] {} (0.75,0.75);
    \draw[label distance=0,\cmdKV@gTypes@iC] (0.75,0.75) -- node[label=(\cmdKV@general@rotate):\cmdKV@gLabels@iLC] {} (0.75,0.25);
    \draw[label distance=0,\cmdKV@gTypes@iD] (0.75,0.25) -- node[label=(-90+\cmdKV@general@rotate):\cmdKV@gLabels@iLD] {} (0.25,0.25);
    #2
  \end{tikzpicture}
  \gSetStdTypes{line}
}
\def\gPenta{\@ifnextchar[\@gPenta{\@gPenta[]}}
\def\@gPenta[#1]#2{%
  \setkeys*{pre}{#1}%
  \setrmkeys{general,gTypes,gLabels}%
  \begin{tikzpicture}
    [line width=1pt,
    baseline={([yshift=-0.5ex+\cmdKV@general@yshift]current bounding box.center)},
    scale=\cmdKV@general@scale,
    rotate=\cmdKV@general@rotate,
    font=\cmdKV@general@font]
    \draw[\cmdKV@gTypes@eA] (-144:0.35) -- (-144:0.75) node[label=(-144+\cmdKV@general@rotate):\cmdKV@gLabels@eLA] {};
    \draw[\cmdKV@gTypes@eB] (144:0.35) -- (144:0.75) node[label=(144+\cmdKV@general@rotate):\cmdKV@gLabels@eLB] {};
    \draw[\cmdKV@gTypes@eC] (72:0.35) -- (72:0.75) node[label=(72+\cmdKV@general@rotate):\cmdKV@gLabels@eLC] {};
    \draw[\cmdKV@gTypes@eD] (0:0.35) -- (0:0.75) node[label=(\cmdKV@general@rotate):\cmdKV@gLabels@eLD] {};
    \draw[\cmdKV@gTypes@eE] (-72:0.35) -- (-72:0.75) node[label=(-72+\cmdKV@general@rotate):\cmdKV@gLabels@eLE] {};
    \draw[label distance=0,\cmdKV@gTypes@iA] (-144:0.35) -- node[label=(180+\cmdKV@general@rotate):\cmdKV@gLabels@iLA] {} (144:0.35);
    \draw[label distance=0,\cmdKV@gTypes@iB] (144:0.35) -- node[label=(108+\cmdKV@general@rotate):\cmdKV@gLabels@iLB] {} (72:0.35);
    \draw[label distance=0,\cmdKV@gTypes@iC] (72:0.35) -- node[label=(36+\cmdKV@general@rotate):\cmdKV@gLabels@iLC] {} (0:0.35);
    \draw[label distance=0,\cmdKV@gTypes@iD] (0:0.35) -- node[label=(-36+\cmdKV@general@rotate):\cmdKV@gLabels@iLD] {} (-72:0.35);
    \draw[label distance=0,\cmdKV@gTypes@iE] (-72:0.35) -- node[label=(-108+\cmdKV@general@rotate):\cmdKV@gLabels@iLE] {} (-144:0.35);
    #2
  \end{tikzpicture}
  \gSetStdTypes{line}
}

\def\gBoxBox{\@ifnextchar[\@gBoxBox{\@gBoxBox[]}}
\def\@gBoxBox[#1]#2{%
  \setkeys*{pre}{#1}%
  \setrmkeys{general,gTypes,gLabels}%
  \begin{tikzpicture}
    [line width=1pt,
    baseline={([yshift=-0.5ex+\cmdKV@general@yshift]current bounding box.center)},
    scale=\cmdKV@general@scale,
    rotate=\cmdKV@general@rotate,
    font=\cmdKV@general@font]
    \draw[\cmdKV@gTypes@eA] (0.3,0.3) -- (0,0) node[label=(-180+\cmdKV@general@rotate):\cmdKV@gLabels@eLA] {};
    \draw[\cmdKV@gTypes@eB] (0.3,0.9) -- (0,1.2) node[label=(180+\cmdKV@general@rotate):\cmdKV@gLabels@eLB] {};
    \draw[\cmdKV@gTypes@eC] (1.5,0.9) -- (1.8,1.2) node[label=(\cmdKV@general@rotate):\cmdKV@gLabels@eLC] {};
    \draw[\cmdKV@gTypes@eD] (1.5,0.3) -- (1.8,0) node[label=(\cmdKV@general@rotate):\cmdKV@gLabels@eLD] {};
    \draw[label distance=0,\cmdKV@gTypes@iA] (0.3,0.3) -- node[label=(180+\cmdKV@general@rotate):\cmdKV@gLabels@iLA] {} (0.3,0.9);
    \draw[label distance=0,\cmdKV@gTypes@iB] (0.3,0.9) -- node[label=(90+\cmdKV@general@rotate):\cmdKV@gLabels@iLB] {} (0.9,0.9);
    \draw[label distance=0,\cmdKV@gTypes@iC] (0.9,0.9) -- node[label=(90+\cmdKV@general@rotate):\cmdKV@gLabels@iLC] {} (1.5,0.9);
    \draw[label distance=0,\cmdKV@gTypes@iD] (1.5,0.9) -- node[label=(\cmdKV@general@rotate):\cmdKV@gLabels@iLD] {} (1.5,0.3);
    \draw[label distance=0,\cmdKV@gTypes@iE] (1.5,0.3) -- node[label=(-90+\cmdKV@general@rotate):\cmdKV@gLabels@iLE] {} (0.9,0.3);
    \draw[label distance=0,\cmdKV@gTypes@iF] (0.9,0.3) -- node[label=(-90+\cmdKV@general@rotate):\cmdKV@gLabels@iLF] {} (0.3,0.3);
    \draw[label distance=0,\cmdKV@gTypes@iG] (0.9,0.9) -- node[label=(\cmdKV@general@rotate):\cmdKV@gLabels@iLG] {} (0.9,0.3);
    #2
  \end{tikzpicture}
  \gSetStdTypes{line}
}
\def\gBoxBoxNP{\@ifnextchar[\@gBoxBoxNP{\@gBoxBoxNP[]}}
\def\@gBoxBoxNP[#1]#2{%
  \setkeys*{pre}{#1}%
  \setrmkeys{general,gTypes,gLabels}%
  \begin{tikzpicture}
    [line width=1pt,
    baseline={([yshift=-0.5ex+\cmdKV@general@yshift]current bounding box.center)},
    scale=\cmdKV@general@scale,
    rotate=\cmdKV@general@rotate,
    font=\cmdKV@general@font]
    \draw[\cmdKV@gTypes@eA] (-1.9,0.3) -- (-2.2,0) node[label=(180+\cmdKV@general@rotate):\cmdKV@gLabels@eLA] {};
    \draw[\cmdKV@gTypes@eB] (-1.9,1.1) -- (-2.2,1.4) node[label=(180+\cmdKV@general@rotate):\cmdKV@gLabels@eLB] {};
    \draw[\cmdKV@gTypes@eC] (-0.3,0.7) -- (0,0.7) node[label=(\cmdKV@general@rotate):\cmdKV@gLabels@eLC] {};
    \draw[\cmdKV@gTypes@eD] (-1.1,0.7) -- (-1.4,0.7) node[label=(180+\cmdKV@general@rotate):\cmdKV@gLabels@eLD] {};
    \draw[label distance=0,\cmdKV@gTypes@iA] (-1.9,0.3) -- node[label=(180+\cmdKV@general@rotate):\cmdKV@gLabels@iLA] {} (-1.9,1.1);
    \draw[label distance=0,\cmdKV@gTypes@iB] (-1.9,1.1) -- node[label=(90+\cmdKV@general@rotate):\cmdKV@gLabels@iLB] {} (-0.7,1.1);
    \draw[label distance=0,\cmdKV@gTypes@iC] (-0.7,1.1) -- node[label=(45+\cmdKV@general@rotate):\cmdKV@gLabels@iLC] {} (-0.3,0.7);
    \draw[label distance=0,\cmdKV@gTypes@iD] (-0.3,0.7) -- node[label=(-45+\cmdKV@general@rotate):\cmdKV@gLabels@iLD] {} (-0.7,0.3);
    \draw[label distance=0,\cmdKV@gTypes@iE] (-0.7,0.3) -- node[label=(-90+\cmdKV@general@rotate):\cmdKV@gLabels@iLE] {} (-1.9,0.3);
    \draw[label distance=0,\cmdKV@gTypes@iF] (-0.7,1.1) -- node[label=(180+\cmdKV@general@rotate):\cmdKV@gLabels@iLF] {} (-1.1,0.7);
    \draw[label distance=0,\cmdKV@gTypes@iG] (-1.1,0.7) -- node[label=(-180+\cmdKV@general@rotate):\cmdKV@gLabels@iLG] {} (-0.7,0.3);
    #2
  \end{tikzpicture}
  \gSetStdTypes{line}
}

\def\gBoxBoxNPB{\@ifnextchar[\@gBoxBoxNPB{\@gBoxBoxNPB[]}}
\def\@gBoxBoxNPB[#1]#2{%
  \setkeys*{pre}{#1}%
  \setrmkeys{general,gTypes,gLabels}%
  \begin{tikzpicture}
    [line width=1pt,
    baseline={([yshift=-0.5ex+\cmdKV@general@yshift]current bounding box.center)},
    scale=\cmdKV@general@scale,
    rotate=\cmdKV@general@rotate,
    font=\cmdKV@general@font]
    \draw[\cmdKV@gTypes@eA] (0.3,0.3) -- (0,0) node[label=(-180+\cmdKV@general@rotate):\cmdKV@gLabels@eLA] {};
    \draw[\cmdKV@gTypes@eB] (0.3,0.9) -- (0,1.2) node[label=(180+\cmdKV@general@rotate):\cmdKV@gLabels@eLB] {};
    \draw[\cmdKV@gTypes@eC] (1.5,0.9) -- (1.8,1.2) node[label=(\cmdKV@general@rotate):\cmdKV@gLabels@eLC] {};
    \draw[\cmdKV@gTypes@eD] (1.5,0.3) -- (1.8,0) node[label=(\cmdKV@general@rotate):\cmdKV@gLabels@eLD] {};
    \draw[label distance=0,\cmdKV@gTypes@iA] (0.3,0.3) -- node[label=(180+\cmdKV@general@rotate):\cmdKV@gLabels@iLA] {} (0.3,0.9);
    \draw[label distance=0,\cmdKV@gTypes@iB] (0.3,0.9) -- node[label=(90+\cmdKV@general@rotate):\cmdKV@gLabels@iLB] {} (0.9,0.9);
    \draw[label distance=0,\cmdKV@gTypes@iC] (0.9,0.9) -- node[label=(90+\cmdKV@general@rotate):\cmdKV@gLabels@iLC] {} (1.5,0.3);
    \foreach \x in {5,...,14} \fill[opacity=0.12,white] (1.2,0.6) circle (\x/100);
    \draw[label distance=0,\cmdKV@gTypes@iD] (1.5,0.3) -- node[label=(\cmdKV@general@rotate):\cmdKV@gLabels@iLD] {} (1.5,0.9);
    \draw[label distance=0,\cmdKV@gTypes@iE] (1.5,0.9) -- node[label=(-90+\cmdKV@general@rotate):\cmdKV@gLabels@iLE] {} (0.9,0.3);
    \draw[label distance=0,\cmdKV@gTypes@iF] (0.9,0.3) -- node[label=(-90+\cmdKV@general@rotate):\cmdKV@gLabels@iLF] {} (0.3,0.3);
    \draw[label distance=-1,\cmdKV@gTypes@iG] (0.9,0.9) -- node[label=(180+\cmdKV@general@rotate):\cmdKV@gLabels@iLG] {} (0.9,0.3);
    #2
  \end{tikzpicture}
  \gSetStdTypes{line}
}

\def\gBoxBoxNPC{\@ifnextchar[\@gBoxBoxNPC{\@gBoxBoxNPC[]}}
\def\@gBoxBoxNPC[#1]#2{%
  \setkeys*{pre}{#1}%
  \setrmkeys{general,gTypes,gLabels}%
  \begin{tikzpicture}
    [line width=1pt,
    baseline={([yshift=-0.5ex+\cmdKV@general@yshift]current bounding box.center)},
    scale=\cmdKV@general@scale,
    rotate=\cmdKV@general@rotate,
    font=\cmdKV@general@font]
    \draw[\cmdKV@gTypes@eA] (0.3,0.3) -- (0,0) node[label=(-180+\cmdKV@general@rotate):\cmdKV@gLabels@eLA] {};
    \draw[\cmdKV@gTypes@eB] (0.3,0.9) -- (0,1.2) node[label=(180+\cmdKV@general@rotate):\cmdKV@gLabels@eLB] {};
    \draw[\cmdKV@gTypes@eC] (1.5,0.9) -- (1.8,1.2) node[label=(\cmdKV@general@rotate):\cmdKV@gLabels@eLC] {};
    \draw[\cmdKV@gTypes@eD] (1.5,0.3) -- (1.8,0) node[label=(\cmdKV@general@rotate):\cmdKV@gLabels@eLD] {};
    \draw[label distance=0,\cmdKV@gTypes@iA] (0.3,0.3) -- node[label=(180+\cmdKV@general@rotate):\cmdKV@gLabels@iLA] {} (0.3,0.9);
    \draw[label distance=0,\cmdKV@gTypes@iB] (0.3,0.9) -- node[label=(90+\cmdKV@general@rotate):\cmdKV@gLabels@iLB] {} (0.9,0.9);
    \draw[label distance=0,\cmdKV@gTypes@iC] (0.9,0.9) -- node[label=(90+\cmdKV@general@rotate):\cmdKV@gLabels@iLC] {} (1.5,0.9);
    \draw[label distance=0,\cmdKV@gTypes@iD] (1.5,0.9) -- node[label=(\cmdKV@general@rotate):\cmdKV@gLabels@iLD] {} (0.9,0.3);
    \foreach \x in {5,...,14} \fill[opacity=0.12,white] (1.2,0.6) circle (\x/100);
    \draw[label distance=0,\cmdKV@gTypes@iE] (1.5,0.3) -- node[label=(-90+\cmdKV@general@rotate):\cmdKV@gLabels@iLE] {} (0.9,0.3);
    \draw[label distance=0,\cmdKV@gTypes@iF] (0.9,0.3) -- node[label=(-90+\cmdKV@general@rotate):\cmdKV@gLabels@iLF] {} (0.3,0.3);
    \draw[label distance=-1,\cmdKV@gTypes@iG] (0.9,0.9) -- node[label=(180+\cmdKV@general@rotate):\cmdKV@gLabels@iLG] {} (1.5,0.3);
    #2
  \end{tikzpicture}
  \gSetStdTypes{line}
}

\def\gBoxTriNPA{\@ifnextchar[\@gBoxTriNPA{\@gBoxTriNPA[]}}
\def\@gBoxTriNPA[#1]#2{%
  \setkeys*{pre}{#1}%
  \setrmkeys{general,gTypes,gLabels}%
  \begin{tikzpicture}
    [line width=1pt,
    baseline={([yshift=-0.5ex+\cmdKV@general@yshift]current bounding box.center)},
    scale=\cmdKV@general@scale,
    rotate=\cmdKV@general@rotate,
    font=\cmdKV@general@font]
    \draw[\cmdKV@gTypes@eA] (0.3,0.6) -- (0,0) node[label=(-180+\cmdKV@general@rotate):\cmdKV@gLabels@eLA] {};
    \draw[\cmdKV@gTypes@eB] (0.3,0.6) -- (0,1.2) node[label=(180+\cmdKV@general@rotate):\cmdKV@gLabels@eLB] {};
    \draw[\cmdKV@gTypes@eC] (1.5,0.9) -- (1.8,1.2) node[label=(\cmdKV@general@rotate):\cmdKV@gLabels@eLC] {};
    \draw[\cmdKV@gTypes@eD] (1.5,0.3) -- (1.8,0) node[label=(\cmdKV@general@rotate):\cmdKV@gLabels@eLD] {};
    \draw[label distance=0,\cmdKV@gTypes@iA] (0.3,0.6) -- node[label=(90+\cmdKV@general@rotate):\cmdKV@gLabels@iLA] {} (0.9,0.9);
    \draw[label distance=0,\cmdKV@gTypes@iB] (0.9,0.9) -- node[label=(90+\cmdKV@general@rotate):\cmdKV@gLabels@iLB] {} (1.5,0.9);
    \draw[label distance=-1,\cmdKV@gTypes@iC] (0.9,0.9) -- node[label=(180+\cmdKV@general@rotate):\cmdKV@gLabels@iLC] {} (1.5,0.3);
    \foreach \x in {5,...,14} \fill[opacity=0.12,white] (1.2,0.6) circle (\x/100);
    \draw[label distance=0,\cmdKV@gTypes@iD] (1.5,0.3) -- node[label=(-90+\cmdKV@general@rotate):\cmdKV@gLabels@iLD] {} (0.9,0.3);
    \draw[label distance=0,\cmdKV@gTypes@iE] (0.9,0.3) -- node[label=(-90+\cmdKV@general@rotate):\cmdKV@gLabels@iLE] {} (0.3,0.6);
    \draw[label distance=-1,\cmdKV@gTypes@iF] (1.5,0.9) -- node[label=(180+\cmdKV@general@rotate):\cmdKV@gLabels@iLF] {} (0.9,0.3);
    #2
  \end{tikzpicture}
  \gSetStdTypes{line}
}

\def\gTriPenta{\@ifnextchar[\@gTriPenta{\@gTriPenta[]}}
\def\@gTriPenta[#1]#2{%
  \setkeys*{pre}{#1}%
  \setrmkeys{general,gTypes,gLabels}%
  \begin{tikzpicture}
    [line width=1pt,
    baseline={([yshift=-0.5ex+\cmdKV@general@yshift]current bounding box.center)},
    scale=\cmdKV@general@scale,
    rotate=\cmdKV@general@rotate,
    font=\cmdKV@general@font]
    \draw[\cmdKV@gTypes@eA] (-108:0.5) -- (-135:1) node[label=(145+\cmdKV@general@rotate):\cmdKV@gLabels@eLA] {};
    \draw[\cmdKV@gTypes@eB] (180:0.5) -- (180:1) node[label=(180+\cmdKV@general@rotate):\cmdKV@gLabels@eLB] {};
    \draw[\cmdKV@gTypes@eC] (108:0.5) -- (135:1) node[label=(-145+\cmdKV@general@rotate):\cmdKV@gLabels@eLC] {};
    \draw[\cmdKV@gTypes@eD] (0:1.2) -- (0:1.5) node[label=(\cmdKV@general@rotate):\cmdKV@gLabels@eLD] {};
    \draw[label distance=0,\cmdKV@gTypes@iA] (-108:0.5) -- node[label=(-135+\cmdKV@general@rotate):\cmdKV@gLabels@iLA] {} (-180:0.5);
    \draw[label distance=0,\cmdKV@gTypes@iB] (180:0.5) -- node[label=(135+\cmdKV@general@rotate):\cmdKV@gLabels@iLB] {} (108:0.5);
    \draw[label distance=0,\cmdKV@gTypes@iC] (108:0.5) -- node[label=(72+\cmdKV@general@rotate):\cmdKV@gLabels@iLC] {} (36:0.5);
    \draw[label distance=0,\cmdKV@gTypes@iD] (36:0.5) -- node[label=(72+\cmdKV@general@rotate):\cmdKV@gLabels@iLD] {} (0:1.2);
    \draw[label distance=2,\cmdKV@gTypes@iE] (0:1.2) -- node[label=(-90+\cmdKV@general@rotate):\cmdKV@gLabels@iLE] {} (-36:0.5);
    \draw[label distance=2,\cmdKV@gTypes@iF] (-36:0.5) -- node[label=(-90+\cmdKV@general@rotate):\cmdKV@gLabels@iLF] {} (-108:0.5);
    \draw[label distance=0,\cmdKV@gTypes@iG] (36:0.5) -- node[label=(180+\cmdKV@general@rotate):\cmdKV@gLabels@iLG] {} (-36:0.5);
    #2
  \end{tikzpicture}
  \gSetStdTypes{line}
}
\def\gTriTri{\@ifnextchar[\@gTriTri{\@gTriTri[]}}
\def\@gTriTri[#1]#2{%
  \setkeys*{pre}{#1}%
  \setrmkeys{general,gTypes,gLabels}%
  \begin{tikzpicture}
    [line width=1pt,
    baseline={([yshift=-0.5ex+\cmdKV@general@yshift]current bounding box.center)},
    scale=\cmdKV@general@scale,
    rotate=\cmdKV@general@rotate,
    font=\cmdKV@general@font]
    \draw[\cmdKV@gTypes@eA] (0.3,0.3) -- (0,0) node[label=(-180+\cmdKV@general@rotate):\cmdKV@gLabels@eLA] {};
    \draw[\cmdKV@gTypes@eB] (0.3,1.1) -- (0,1.4) node[label=(180+\cmdKV@general@rotate):\cmdKV@gLabels@eLB] {};
    \draw[\cmdKV@gTypes@eC] (1.9,1.1) -- (2.2,1.4) node[label=(\cmdKV@general@rotate):\cmdKV@gLabels@eLC] {};
    \draw[\cmdKV@gTypes@eD] (1.9,0.3) -- (2.2,0) node[label=(\cmdKV@general@rotate):\cmdKV@gLabels@eLD] {};
    \draw[label distance=0,\cmdKV@gTypes@iA] (0.3,0.3) -- node[label=(180+\cmdKV@general@rotate):\cmdKV@gLabels@iLA] {} (0.3,1.1);
    \draw[label distance=0,\cmdKV@gTypes@iB] (0.3,1.1) -- node[label=(45+\cmdKV@general@rotate):\cmdKV@gLabels@iLB] {} (0.9,0.7);
    \draw[label distance=0,\cmdKV@gTypes@iC] (0.9,0.7) -- node[label=(-45+\cmdKV@general@rotate):\cmdKV@gLabels@iLC] {} (0.3,0.3);
    \draw[label distance=0,\cmdKV@gTypes@iD] (0.9,0.7) -- node[label=(90+\cmdKV@general@rotate):\cmdKV@gLabels@iLD] {} (1.3,0.7);
    \draw[label distance=0,\cmdKV@gTypes@iE] (1.3,0.7) -- node[label=(135+\cmdKV@general@rotate):\cmdKV@gLabels@iLE] {} (1.9,1.1);
    \draw[label distance=0,\cmdKV@gTypes@iF] (1.9,1.1) -- node[label=(\cmdKV@general@rotate):\cmdKV@gLabels@iLF] {} (1.9,0.3);
    \draw[label distance=0,\cmdKV@gTypes@iG] (1.9,0.3) -- node[label=(-135+\cmdKV@general@rotate):\cmdKV@gLabels@iLG] {} (1.3,0.7);
    #2
  \end{tikzpicture}
  \gSetStdTypes{line}
}

\def\gBoxBubA{\@ifnextchar[\@gBoxBubA{\@gBoxBubA[]}}
\def\@gBoxBubA[#1]#2{%
  \setkeys*{pre}{#1}%
  \setrmkeys{general,gTypes,gLabels}%
  \begin{tikzpicture}
    [line width=1pt,
    baseline={([yshift=-0.5ex+\cmdKV@general@yshift]current bounding box.center)},
    scale=\cmdKV@general@scale,
    rotate=\cmdKV@general@rotate,
    font=\cmdKV@general@font]
    \draw[\cmdKV@gTypes@eA] (0.1,0.1) -- (-0.1,-0.1) node[label=(-180+\cmdKV@general@rotate):\cmdKV@gLabels@eLA] {};
    \draw[\cmdKV@gTypes@eB] (0.1,0.9) -- (-0.1,1.1) node[label=(180+\cmdKV@general@rotate):\cmdKV@gLabels@eLB] {};
    \draw[\cmdKV@gTypes@eC] (0.9,0.9) -- (1.1,1.1) node[label=(\cmdKV@general@rotate):\cmdKV@gLabels@eLC] {};
    \draw[\cmdKV@gTypes@eD] (0.9,0.1) -- (1.1,-0.1) node[label=(\cmdKV@general@rotate):\cmdKV@gLabels@eLD] {};
    \draw[label distance=0,\cmdKV@gTypes@iA] (0.1,0.1) -- node[label=(180+\cmdKV@general@rotate):\cmdKV@gLabels@iLA] {} (0.1,0.9);
    \draw[label distance=0,\cmdKV@gTypes@iB] (0.1,0.9) -- node[label=(90+\cmdKV@general@rotate):\cmdKV@gLabels@iLB] {} (0.32,0.9);
    \draw[label distance=0,\cmdKV@gTypes@iC] (0.32,0.9) .. controls (0.32,1.15) and (0.68,1.15) .. node[label=(90+\cmdKV@general@rotate):\cmdKV@gLabels@iLC] {} (0.68,0.9);
    \draw[label distance=0,\cmdKV@gTypes@iD] (0.68,0.9) .. controls (0.68,0.65) and (0.32,0.65) .. node[label=(-90+\cmdKV@general@rotate):\cmdKV@gLabels@iLD] {} (0.32,0.9);
    \draw[label distance=0,\cmdKV@gTypes@iE] (0.68,0.9) -- node[label=(90+\cmdKV@general@rotate):\cmdKV@gLabels@iLE] {} (0.9,0.9);
    \draw[label distance=0,\cmdKV@gTypes@iF] (0.9,0.9) -- node[label=(\cmdKV@general@rotate):\cmdKV@gLabels@iLF] {} (0.9,0.1);
    \draw[label distance=0,\cmdKV@gTypes@iG] (0.9,0.1) -- node[label=(-90+\cmdKV@general@rotate):\cmdKV@gLabels@iLG] {} (0.1,0.1);
    #2
  \end{tikzpicture}
  \gSetStdTypes{line}
}

\def\gBoxBubB{\@ifnextchar[\@gBoxBubB{\@gBoxBubB[]}}
\def\@gBoxBubB[#1]#2{%
  \setkeys*{pre}{#1}%
  \setrmkeys{general,gTypes,gLabels}%
  \begin{tikzpicture}
    [line width=1pt,
    baseline={([yshift=-0.5ex+\cmdKV@general@yshift]current bounding box.center)},
    scale=\cmdKV@general@scale,
    rotate=\cmdKV@general@rotate,
    font=\cmdKV@general@font]
    \draw[label distance=-1,\cmdKV@gTypes@eA] (0,-0.4) -- (0,-0.7) node[label=(180+\cmdKV@general@rotate):\cmdKV@gLabels@eLA] {};
    \draw[\cmdKV@gTypes@eB] (-0.4,0) -- (-0.7,0) node[label=(180+\cmdKV@general@rotate):\cmdKV@gLabels@eLB] {};
    \draw[label distance=-1,\cmdKV@gTypes@eC] (0,0.4) -- (0,0.7) node[label=(180+\cmdKV@general@rotate):\cmdKV@gLabels@eLC] {};
    \draw[\cmdKV@gTypes@eD] (1.1,0) -- (1.4,0) node[label=(\cmdKV@general@rotate):\cmdKV@gLabels@eLD] {};
    \draw[label distance=0,\cmdKV@gTypes@iA] (0,-0.4) -- node[label=(-135+\cmdKV@general@rotate):\cmdKV@gLabels@iLA] {} (-0.4,0);
    \draw[label distance=0,\cmdKV@gTypes@iB] (-0.4,0) -- node[label=(135+\cmdKV@general@rotate):\cmdKV@gLabels@iLB] {} (0,0.4);
    \draw[label distance=0,\cmdKV@gTypes@iC] (0,0.4) -- node[label=(45+\cmdKV@general@rotate):\cmdKV@gLabels@iLC] {} (0.4,0);
    \draw[label distance=0,\cmdKV@gTypes@iD] (0.4,0) -- node[label=(-45+\cmdKV@general@rotate):\cmdKV@gLabels@iLD] {} (0,-0.4);
    \draw[label distance=0,\cmdKV@gTypes@iE] (0.4,0) --  node[label=(90+\cmdKV@general@rotate):\cmdKV@gLabels@iLE] {} (0.7,0);
    \draw[label distance=0,\cmdKV@gTypes@iF] (0.7,0) .. controls (0.7,0.26) and (1.1,0.26) .. node[label=(90+\cmdKV@general@rotate):\cmdKV@gLabels@iLF] {} (1.1,0);
    \draw[label distance=0,\cmdKV@gTypes@iG] (1.1,0) .. controls (1.1,-0.26) and (0.7,-0.26) ..  node[label=(-90+\cmdKV@general@rotate):\cmdKV@gLabels@iLG] {} (0.7,0);
    #2
  \end{tikzpicture}
  \gSetStdTypes{line}
}

\def\gPentaTad{\@ifnextchar[\@gPentaTad{\@gPentaTad[]}}
\def\@gPentaTad[#1]#2{%
  \setkeys*{pre}{#1}%
  \setrmkeys{general,gTypes,gLabels}%
  \begin{tikzpicture}
    [line width=1pt,
    baseline={([yshift=-0.5ex+\cmdKV@general@yshift]current bounding box.center)},
    scale=\cmdKV@general@scale,
    rotate=\cmdKV@general@rotate,
    font=\cmdKV@general@font]
    \draw[label distance=-0.5,\cmdKV@gTypes@eA] (108:0.35) -- (108:0.75) node[label=(180+\cmdKV@general@rotate):\cmdKV@gLabels@eLA] {};
    \draw[\cmdKV@gTypes@eB] (36:0.35) -- (36:0.75) node[label=(36+\cmdKV@general@rotate):\cmdKV@gLabels@eLB] {};
    \draw[\cmdKV@gTypes@eC] (-36:0.35) -- (-36:0.75) node[label=(-36+\cmdKV@general@rotate):\cmdKV@gLabels@eLC] {};
    \draw[label distance=-0.5,\cmdKV@gTypes@eD] (-108:0.35) -- (-108:0.75) node[label=(-180+\cmdKV@general@rotate):\cmdKV@gLabels@eLD] {};
    \draw[label distance=0,\cmdKV@gTypes@iA] (108:0.35) -- node[label=(72+\cmdKV@general@rotate):\cmdKV@gLabels@iLA] {} (36:0.35);
    \draw[label distance=0,\cmdKV@gTypes@iB] (36:0.35) -- node[label=(\cmdKV@general@rotate):\cmdKV@gLabels@iLB] {} (-36:0.35);
    \draw[label distance=0,\cmdKV@gTypes@iC] (-36:0.35) -- node[label=(-72+\cmdKV@general@rotate):\cmdKV@gLabels@iLC] {} (-108:0.35);
    \draw[label distance=0,\cmdKV@gTypes@iD] (-108:0.35) -- node[label=(-144+\cmdKV@general@rotate):\cmdKV@gLabels@iLD] {} (-180:0.35);
    \draw[label distance=0,\cmdKV@gTypes@iE] (180:0.35) -- node[label=(144+\cmdKV@general@rotate):\cmdKV@gLabels@iLE] {} (108:0.35);
    \draw[label distance=0,\cmdKV@gTypes@iF] (180:0.35) -- node[label=(90+\cmdKV@general@rotate):\cmdKV@gLabels@iLF] {} (180:0.75);
    \draw[label distance=-0.5,\cmdKV@gTypes@iG] (180:0.75) .. controls ($(180:0.75)+(0,-0.18)$) and ($(180:1)+(0,-0.18)$) .. (180:1) node[label=(180+\cmdKV@general@rotate):\cmdKV@gLabels@iLG] {} .. controls ($(180:1)+(0,0.18)$) and ($(180:0.75)+(0,0.18)$) .. (180:0.75);
    #2
  \end{tikzpicture}
  \gSetStdTypes{line}
}

\def\gBoxTadA{\@ifnextchar[\@gBoxTadA{\@gBoxTadA[]}}
\def\@gBoxTadA[#1]#2{%
  \setkeys*{pre}{#1}%
  \setrmkeys{general,gTypes,gLabels}%
  \begin{tikzpicture}
    [line width=1pt,
    baseline={([yshift=-0.5ex+\cmdKV@general@yshift]current bounding box.center)},
    scale=\cmdKV@general@scale,
    rotate=\cmdKV@general@rotate,
    font=\cmdKV@general@font]
    \draw[label distance=-0.5,\cmdKV@gTypes@eA] (0,0.4) -- (0,0.7) node[label=(\cmdKV@general@rotate):\cmdKV@gLabels@eLA] {};
    \draw[\cmdKV@gTypes@eB] (0.4,0) -- (0.7,0) node[label=(\cmdKV@general@rotate):\cmdKV@gLabels@eLB] {};
    \draw[label distance=-0.5,\cmdKV@gTypes@eC] (0,-0.4) -- (0,-0.7) node[label=(\cmdKV@general@rotate):\cmdKV@gLabels@eLC] {};
    \draw[\cmdKV@gTypes@eD] (-0.7,0) -- (-0.7,-0.3) node[label=(-90+\cmdKV@general@rotate):\cmdKV@gLabels@eLD] {};
    \draw[label distance=0,\cmdKV@gTypes@iA] (0,0.4) -- node[label=(45+\cmdKV@general@rotate):\cmdKV@gLabels@iLA] {} (0.4,0);
    \draw[label distance=0,\cmdKV@gTypes@iB] (0.4,0) -- node[label=(-45+\cmdKV@general@rotate):\cmdKV@gLabels@iLB] {} (0,-0.4);
    \draw[label distance=0,\cmdKV@gTypes@iC] (0,-0.4) -- node[label=(-135+\cmdKV@general@rotate):\cmdKV@gLabels@iLC] {} (-0.4,0);
    \draw[label distance=0,\cmdKV@gTypes@iD] (-0.4,0) -- node[label=(135+\cmdKV@general@rotate):\cmdKV@gLabels@iLD] {} (0,0.4);
    \draw[label distance=0,\cmdKV@gTypes@iE] (-0.4,0) --  node[label=(90+\cmdKV@general@rotate):\cmdKV@gLabels@iLE] {} (-0.7,0);
    \draw[label distance=0,\cmdKV@gTypes@iF] (-0.7,0) -- node[label=(90+\cmdKV@general@rotate):\cmdKV@gLabels@iLF] {} (-1.1,0);
    \draw[label distance=-0.5,\cmdKV@gTypes@iG] (-1.1,0) .. controls (-1.1,-0.18) and (-1.35,-0.18) .. (-1.35,0) node[label=(180+\cmdKV@general@rotate):\cmdKV@gLabels@iLG] {} .. controls (-1.35,0.18) and (-1.1,0.18) .. (-1.1,0);
    #2
  \end{tikzpicture}
  \gSetStdTypes{line}
}

\def\gSunsetA{\@ifnextchar[\@gSunsetA{\@gSunsetA[]}}
\def\@gSunsetA[#1]#2{%
  \setkeys*{pre}{#1}%
  \setrmkeys{general,gTypes,gLabels}%
  \begin{tikzpicture}
    [line width=1pt,
    baseline={([yshift=-0.5ex+\cmdKV@general@yshift]current bounding box.center)},
    scale=\cmdKV@general@scale,
    rotate=\cmdKV@general@rotate,
    font=\cmdKV@general@font]
    \draw[\cmdKV@gTypes@eA] (0.3,0.6) -- (0,0) node[label=(-180+\cmdKV@general@rotate):\cmdKV@gLabels@eLA] {};
    \draw[\cmdKV@gTypes@eB] (0.3,0.6) -- (0,1.2) node[label=(180+\cmdKV@general@rotate):\cmdKV@gLabels@eLB] {};
    \draw[\cmdKV@gTypes@eC] (1.5,0.6) -- (1.8,1.2) node[label=(\cmdKV@general@rotate):\cmdKV@gLabels@eLC] {};
    \draw[\cmdKV@gTypes@eD] (1.5,0.6) -- (1.8,0) node[label=(\cmdKV@general@rotate):\cmdKV@gLabels@eLD] {};
    \draw[label distance=0,\cmdKV@gTypes@iA] (0.3,0.6) to[bend left=90] node[label=(90+\cmdKV@general@rotate):\cmdKV@gLabels@iLA] {} (1.5,0.6);
    \draw[label distance=0,\cmdKV@gTypes@iB] (1.5,0.6) to[bend left=90] node[label=(-90+\cmdKV@general@rotate):\cmdKV@gLabels@iLB] {} (0.3,0.6);
    \draw[label distance=-2,\cmdKV@gTypes@iC] (0.3,0.6) -- node[label=(90+\cmdKV@general@rotate):\cmdKV@gLabels@iLC] {} (1.5,0.6);
    #2
  \end{tikzpicture}
  \gSetStdTypes{line}
}

\def\gTriBubA{\@ifnextchar[\@gTriBubA{\@gTriBubA[]}}
\def\@gTriBubA[#1]#2{%
  \setkeys*{pre}{#1}%
  \setrmkeys{general,gTypes,gLabels}%
  \begin{tikzpicture}
    [line width=1pt,
    baseline={([yshift=-0.5ex+\cmdKV@general@yshift]current bounding box.center)},
    scale=\cmdKV@general@scale,
    rotate=\cmdKV@general@rotate,
    font=\cmdKV@general@font]
    \draw[\cmdKV@gTypes@eA] (0.4,0.3) -- (0,0) node[label=(-180+\cmdKV@general@rotate):\cmdKV@gLabels@eLA] {};
    \draw[\cmdKV@gTypes@eB] (0.4,0.9) -- (0,1.2) node[label=(180+\cmdKV@general@rotate):\cmdKV@gLabels@eLB] {};
    \draw[\cmdKV@gTypes@eC] (1.4,0.6) -- (1.8,1.2) node[label=(\cmdKV@general@rotate):\cmdKV@gLabels@eLC] {};
    \draw[\cmdKV@gTypes@eD] (1.4,0.6) -- (1.8,0) node[label=(\cmdKV@general@rotate):\cmdKV@gLabels@eLD] {};
    \draw[label distance=0,\cmdKV@gTypes@iA] (0.4,0.3) to[bend left=40] node[label=(180+\cmdKV@general@rotate):\cmdKV@gLabels@iLA] {} (0.4,0.9);
    \draw[label distance=0,\cmdKV@gTypes@iB] (0.4,0.9) -- node[label=(90+\cmdKV@general@rotate):\cmdKV@gLabels@iLB] {} (1.4,0.6);
    \draw[label distance=0,\cmdKV@gTypes@iC] (1.4,0.6) -- node[label=(-90+\cmdKV@general@rotate):\cmdKV@gLabels@iLC] {} (0.4,0.3);
    \draw[label distance=-1,\cmdKV@gTypes@iD] (0.4,0.9) to[bend left=40] node[label=(\cmdKV@general@rotate):\cmdKV@gLabels@iLD] {} (0.4,0.3);
    #2
  \end{tikzpicture}
  \gSetStdTypes{line}
}

\def\gBubBubA{\@ifnextchar[\@gBubBubA{\@gBubBubA[]}}
\def\@gBubBubA[#1]#2{%
  \setkeys*{pre}{#1}%
  \setrmkeys{general,gTypes,gLabels}%
  \begin{tikzpicture}
    [line width=1pt,
    baseline={([yshift=-0.5ex+\cmdKV@general@yshift]current bounding box.center)},
    scale=\cmdKV@general@scale,
    rotate=\cmdKV@general@rotate,
    font=\cmdKV@general@font]
    \draw[\cmdKV@gTypes@eA] (0.3,0.6) -- (0,0) node[label=(-180+\cmdKV@general@rotate):\cmdKV@gLabels@eLA] {};
    \draw[\cmdKV@gTypes@eB] (0.3,0.6) -- (0,1.2) node[label=(180+\cmdKV@general@rotate):\cmdKV@gLabels@eLB] {};
    \draw[\cmdKV@gTypes@eC] (1.5,0.6) -- (1.8,1.2) node[label=(\cmdKV@general@rotate):\cmdKV@gLabels@eLC] {};
    \draw[\cmdKV@gTypes@eD] (1.5,0.6) -- (1.8,0) node[label=(\cmdKV@general@rotate):\cmdKV@gLabels@eLD] {};
    \draw[label distance=0,\cmdKV@gTypes@iA] (0.3,0.6) to[bend left=90] node[label=(90+\cmdKV@general@rotate):\cmdKV@gLabels@iLA] {} (0.9,0.6);
    \draw[label distance=0,\cmdKV@gTypes@iB] (0.9,0.6) to[bend left=90] node[label=(90+\cmdKV@general@rotate):\cmdKV@gLabels@iLB] {} (1.5,0.6);
    \draw[label distance=0,\cmdKV@gTypes@iC] (1.5,0.6) to[bend left=90] node[label=(-90+\cmdKV@general@rotate):\cmdKV@gLabels@iLC] {} (0.9,0.6);
    \draw[label distance=0,\cmdKV@gTypes@iD] (0.9,0.6) to[bend left=90] node[label=(-90+\cmdKV@general@rotate):\cmdKV@gLabels@iLD] {} (0.3,0.6);
    #2
  \end{tikzpicture}
  \gSetStdTypes{line}
}

\def\gBoxBubC{\@ifnextchar[\@gBoxBubC{\@gBoxBubC[]}}
\def\@gBoxBubC[#1]#2{%
  \setkeys*{pre}{#1}%
  \setrmkeys{general,gTypes,gLabels}%
  \begin{tikzpicture}
    [line width=1pt,
    baseline={([yshift=-0.5ex+\cmdKV@general@yshift]current bounding box.center)},
    scale=\cmdKV@general@scale,
    rotate=\cmdKV@general@rotate,
    font=\cmdKV@general@font]
    \draw[\cmdKV@gTypes@eA] (0.1,0.1) -- (-0.1,-0.1) node[label=(-180+\cmdKV@general@rotate):\cmdKV@gLabels@eLA] {};
    \draw[\cmdKV@gTypes@eB] (0.1,0.9) -- (-0.1,1.1) node[label=(180+\cmdKV@general@rotate):\cmdKV@gLabels@eLB] {};
    \draw[\cmdKV@gTypes@eC] (0.9,0.9) -- (1.1,1.1) node[label=(\cmdKV@general@rotate):\cmdKV@gLabels@eLC] {};
    \draw[\cmdKV@gTypes@eD] (0.9,0.1) -- (1.1,-0.1) node[label=(\cmdKV@general@rotate):\cmdKV@gLabels@eLD] {};
    \draw[label distance=0,\cmdKV@gTypes@iA] (0.1,0.1) -- node[label=(180+\cmdKV@general@rotate):\cmdKV@gLabels@iLA] {} (0.1,0.9);
    \draw[label distance=0,\cmdKV@gTypes@iB] (0.1,0.9) to[bend left] node[label=(90+\cmdKV@general@rotate):\cmdKV@gLabels@iLB] {} (0.9,0.9);
    \draw[label distance=0,\cmdKV@gTypes@iC] (0.9,0.9) to[bend left] node[label=(-90+\cmdKV@general@rotate):\cmdKV@gLabels@iLC] {} (0.1,0.9);
    \draw[label distance=0,\cmdKV@gTypes@iD] (0.9,0.9) -- node[label=(\cmdKV@general@rotate):\cmdKV@gLabels@iLD] {} (0.9,0.1);
    \draw[label distance=0,\cmdKV@gTypes@iE] (0.9,0.1) -- node[label=(-90+\cmdKV@general@rotate):\cmdKV@gLabels@iLE] {} (0.1,0.1);
    #2
  \end{tikzpicture}
  \gSetStdTypes{line}
}

\def\gBoxZ{\@ifnextchar[\@gBoxZ{\@gBoxZ[]}}
\def\@gBoxZ[#1]#2{%
  \setkeys*{pre}{#1}%
  \setrmkeys{general,gTypes,gLabels}%
  \begin{tikzpicture}
    [line width=1pt,
    baseline={([yshift=-0.5ex+\cmdKV@general@yshift]current bounding box.center)},
    scale=\cmdKV@general@scale,
    rotate=\cmdKV@general@rotate,
    font=\cmdKV@general@font]
    \draw[\cmdKV@gTypes@eA] (0.15,0.15) -- (-0.1,-0.1) node[label=(-180+\cmdKV@general@rotate):\cmdKV@gLabels@eLA] {};
    \draw[\cmdKV@gTypes@eB] (0.15,0.85) -- (-0.1,1.1) node[label=(180+\cmdKV@general@rotate):\cmdKV@gLabels@eLB] {};
    \draw[\cmdKV@gTypes@eC] (0.85,0.85) -- (1.1,1.1) node[label=(\cmdKV@general@rotate):\cmdKV@gLabels@eLC] {};
    \draw[\cmdKV@gTypes@eD] (0.85,0.15) -- (1.1,-0.1) node[label=(\cmdKV@general@rotate):\cmdKV@gLabels@eLD] {};
    \draw[label distance=0,\cmdKV@gTypes@iA] (0.15,0.15) -- node[label=(180+\cmdKV@general@rotate):\cmdKV@gLabels@iLA] {} (0.15,0.85);
    \draw[label distance=0,\cmdKV@gTypes@iB] (0.15,0.85) -- node[label=(90+\cmdKV@general@rotate):\cmdKV@gLabels@iLB] {} (0.85,0.85);
    \draw[label distance=0,\cmdKV@gTypes@iC] (0.85,0.85) -- node[label=(\cmdKV@general@rotate):\cmdKV@gLabels@iLC] {} (0.85,0.15);
    \draw[label distance=0,\cmdKV@gTypes@iD] (0.85,0.15) -- node[label=(-90+\cmdKV@general@rotate):\cmdKV@gLabels@iLD] {} (0.15,0.15);
    \draw[label distance=-4,\cmdKV@gTypes@iE] (0.15,0.15) -- node[label=(135+\cmdKV@general@rotate):\cmdKV@gLabels@iLE] {} (0.85,0.85);
    #2
  \end{tikzpicture}
  \gSetStdTypes{line}
}

\def\gBoxBoxBoxA{\@ifnextchar[\@gBoxBoxBoxA{\@gBoxBoxBoxA[]}}
\def\@gBoxBoxBoxA[#1]#2{%
  \setkeys*{pre}{#1}%
  \setrmkeys{general,gTypes,gLabels}%
  \begin{tikzpicture}
    [line width=1pt,
    baseline={([yshift=-0.5ex+\cmdKV@general@yshift]current bounding box.center)},
    scale=\cmdKV@general@scale,
    rotate=\cmdKV@general@rotate,
    font=\cmdKV@general@font]
    \draw[\cmdKV@gTypes@eA] (-0.3,0.3) -- (-0.6,0) node[label=(-180+\cmdKV@general@rotate):\cmdKV@gLabels@eLA] {};
    \draw[\cmdKV@gTypes@eB] (-0.3,0.9) -- (-0.6,1.2) node[label=(180+\cmdKV@general@rotate):\cmdKV@gLabels@eLB] {};
    \draw[\cmdKV@gTypes@eC] (1.5,0.9) -- (1.8,1.2) node[label=(\cmdKV@general@rotate):\cmdKV@gLabels@eLC] {};
    \draw[\cmdKV@gTypes@eD] (1.5,0.3) -- (1.8,0) node[label=(\cmdKV@general@rotate):\cmdKV@gLabels@eLD] {};
    \draw[label distance=0,\cmdKV@gTypes@iA] (-0.3,0.3) -- node[label=(180+\cmdKV@general@rotate):\cmdKV@gLabels@iLA] {} (-0.3,0.9);
    \draw[label distance=0,\cmdKV@gTypes@iB] (-0.3,0.9) -- node[label=(90+\cmdKV@general@rotate):\cmdKV@gLabels@iLB] {} (0.3,0.9);
    \draw[label distance=0,\cmdKV@gTypes@iC] (0.3,0.9) -- node[label=(90+\cmdKV@general@rotate):\cmdKV@gLabels@iLC] {} (0.9,0.9);
    \draw[label distance=0,\cmdKV@gTypes@iD] (0.9,0.9) -- node[label=(90+\cmdKV@general@rotate):\cmdKV@gLabels@iLD] {} (1.5,0.9);
    \draw[label distance=0,\cmdKV@gTypes@iE] (1.5,0.9) -- node[label=(\cmdKV@general@rotate):\cmdKV@gLabels@iLE] {} (1.5,0.3);
    \draw[label distance=0,\cmdKV@gTypes@iF] (1.5,0.3) -- node[label=(-90+\cmdKV@general@rotate):\cmdKV@gLabels@iLF] {} (0.9,0.3);
    \draw[label distance=0,\cmdKV@gTypes@iG] (0.9,0.3) -- node[label=(-90+\cmdKV@general@rotate):\cmdKV@gLabels@iLG] {} (0.3,0.3);
    \draw[label distance=0,\cmdKV@gTypes@iH] (0.3,0.3) -- node[label=(-90+\cmdKV@general@rotate):\cmdKV@gLabels@iLH] {} (-0.3,0.3);
    \draw[label distance=0,\cmdKV@gTypes@iI] (0.3,0.9) -- node[label=(\cmdKV@general@rotate):\cmdKV@gLabels@iLI] {} (0.3,0.3);
    \draw[label distance=0,\cmdKV@gTypes@iJ] (0.9,0.9) -- node[label=(\cmdKV@general@rotate):\cmdKV@gLabels@iLJ] {} (0.9,0.3);
    #2
  \end{tikzpicture}
  \gSetStdTypes{line}
}

\def\gBoxBoxBoxB{\@ifnextchar[\@gBoxBoxBoxB{\@gBoxBoxBoxB[]}}
\def\@gBoxBoxBoxB[#1]#2{%
  \setkeys*{pre}{#1}%
  \setrmkeys{general,gTypes,gLabels}%
  \begin{tikzpicture}
    [line width=1pt,
    baseline={([yshift=-0.5ex+\cmdKV@general@yshift]current bounding box.center)},
    scale=\cmdKV@general@scale,
    rotate=\cmdKV@general@rotate,
    font=\cmdKV@general@font]
    \draw[\cmdKV@gTypes@eA] (0.3,0.3) -- (0,0) node[label=(-180+\cmdKV@general@rotate):\cmdKV@gLabels@eLA] {};
    \draw[\cmdKV@gTypes@eB] (0.3,1.5) -- (0,1.8) node[label=(180+\cmdKV@general@rotate):\cmdKV@gLabels@eLB] {};
    \draw[\cmdKV@gTypes@eC] (1.5,1.5) -- (1.8,1.8) node[label=(\cmdKV@general@rotate):\cmdKV@gLabels@eLC] {};
    \draw[\cmdKV@gTypes@eD] (1.5,0.3) -- (1.8,0) node[label=(\cmdKV@general@rotate):\cmdKV@gLabels@eLD] {};
    \draw[label distance=0,\cmdKV@gTypes@iA] (0.3,0.3) -- node[label=(180+\cmdKV@general@rotate):\cmdKV@gLabels@iLA] {} (0.3,0.9);
    \draw[label distance=0,\cmdKV@gTypes@iB] (0.3,0.9) -- node[label=(180+\cmdKV@general@rotate):\cmdKV@gLabels@iLB] {} (0.3,1.5);
    \draw[label distance=0,\cmdKV@gTypes@iC] (0.3,1.5) -- node[label=(90+\cmdKV@general@rotate):\cmdKV@gLabels@iLC] {} (0.9,1.5);
    \draw[label distance=0,\cmdKV@gTypes@iD] (0.9,1.5) -- node[label=(90+\cmdKV@general@rotate):\cmdKV@gLabels@iLD] {} (1.5,1.5);
    \draw[label distance=0,\cmdKV@gTypes@iE] (1.5,1.5) -- node[label=(\cmdKV@general@rotate):\cmdKV@gLabels@iLE] {} (1.5,0.3);
    \draw[label distance=0,\cmdKV@gTypes@iF] (1.5,0.3) -- node[label=(-90+\cmdKV@general@rotate):\cmdKV@gLabels@iLF] {} (0.9,0.3);
    \draw[label distance=0,\cmdKV@gTypes@iG] (0.9,0.3) -- node[label=(-90+\cmdKV@general@rotate):\cmdKV@gLabels@iLG] {} (0.3,0.3);
    \draw[label distance=0,\cmdKV@gTypes@iH] (0.9,1.5) -- node[label=(\cmdKV@general@rotate):\cmdKV@gLabels@iLH] {} (0.9,0.9);
    \draw[label distance=0,\cmdKV@gTypes@iI] (0.9,0.9) -- node[label=(\cmdKV@general@rotate):\cmdKV@gLabels@iLI] {} (0.9,0.3);
    \draw[label distance=0,\cmdKV@gTypes@iJ] (0.9,0.9) -- node[label=(90+\cmdKV@general@rotate):\cmdKV@gLabels@iLJ] {} (0.3,0.9);
    #2
  \end{tikzpicture}
  \gSetStdTypes{line}
}

\def\gEFTTree{\@ifnextchar[\@gEFTTree{\@gEFTTree[]}}
\def\@gEFTTree[#1]#2{%
  \setkeys*{pre}{#1}%
  \setrmkeys{general,gTypes,gLabels}%
  \begin{tikzpicture}
    [line width=1pt,
    baseline={([yshift=-0.5ex+\cmdKV@general@yshift]current bounding box.center)},
    scale=\cmdKV@general@scale,
    rotate=\cmdKV@general@rotate,
    font=\cmdKV@general@font]
    \draw[\cmdKV@gTypes@iA] (0,0.5) node[label=(90+\cmdKV@general@rotate):\cmdKV@gLabels@eLA] {} -- node[label=(\cmdKV@general@rotate):\cmdKV@gLabels@iLA] {} (0,-0.5) node[label=(-90+\cmdKV@general@rotate):\cmdKV@gLabels@eLB] {};
    \draw[fill] (0,0.5) circle (0.02);
    \draw[fill] (0,-0.5) circle (0.02);
    #2
  \end{tikzpicture}
  \gSetStdTypes{line}
}

\def\gEFTV{\@ifnextchar[\@gEFTV{\@gEFTV[]}}
\def\@gEFTV[#1]#2{%
  \setkeys*{pre}{#1}%
  \setrmkeys{general,gTypes,gLabels}%
  \begin{tikzpicture}
    [line width=1pt,
    baseline={([yshift=-0.5ex+\cmdKV@general@yshift]current bounding box.center)},
    scale=\cmdKV@general@scale,
    rotate=\cmdKV@general@rotate,
    font=\cmdKV@general@font]
    \draw[\cmdKV@gTypes@iA] (0,0.5) node[label=(90+\cmdKV@general@rotate):\cmdKV@gLabels@eLA] {} -- node[label=(200+\cmdKV@general@rotate):\cmdKV@gLabels@iLA] {} (0.3,-0.5) node[label=(-90+\cmdKV@general@rotate):\cmdKV@gLabels@eLC] {};
    \draw[\cmdKV@gTypes@iB] (0.6,0.5) node[label=(90+\cmdKV@general@rotate):\cmdKV@gLabels@eLB] {} -- node[label=(-20+\cmdKV@general@rotate):\cmdKV@gLabels@iLB] {} (0.3,-0.5);
    \draw[fill] (0,0.5) circle (0.02);
    \draw[fill] (0.3,-0.5) circle (0.02);
    \draw[fill] (0.6,0.5) circle (0.02);
    #2
  \end{tikzpicture}
  \gSetStdTypes{line}
}

\def\gEFTY{\@ifnextchar[\@gEFTY{\@gEFTY[]}}
\def\@gEFTY[#1]#2{%
  \setkeys*{pre}{#1}%
  \setrmkeys{general,gTypes,gLabels}%
  \begin{tikzpicture}
    [line width=1pt,
    baseline={([yshift=-0.5ex+\cmdKV@general@yshift]current bounding box.center)},
    scale=\cmdKV@general@scale,
    rotate=\cmdKV@general@rotate,
    font=\cmdKV@general@font]
    \draw[\cmdKV@gTypes@iA] (0,0.5) node[label=(90+\cmdKV@general@rotate):\cmdKV@gLabels@eLA] {} -- node[label=(225+\cmdKV@general@rotate):\cmdKV@gLabels@iLA] {} (0.3,0);
    \draw[\cmdKV@gTypes@iB] (0.6,0.5) node[label=(90+\cmdKV@general@rotate):\cmdKV@gLabels@eLB] {} -- node[label=(-45+\cmdKV@general@rotate):\cmdKV@gLabels@iLB] {} (0.3,0);
    \draw[\cmdKV@gTypes@iC] (0.3,0) -- node[label=(\cmdKV@general@rotate):\cmdKV@gLabels@iLC] {} (0.3,-0.5) node[label=(-90+\cmdKV@general@rotate):\cmdKV@gLabels@eLC] {};
    \draw[fill] (0,0.5) circle (0.02);
    \draw[fill] (0.6,0.5) circle (0.02);
    \draw[fill] (0.3,-0.5) circle (0.02);
    #2
  \end{tikzpicture}
  \gSetStdTypes{line}
}

\def\gEFTPsi{\@ifnextchar[\@gEFTPsi{\@gEFTPsi[]}}
\def\@gEFTPsi[#1]#2{%
  \setkeys*{pre}{#1}%
  \setrmkeys{general,gTypes,gLabels}%
  \begin{tikzpicture}
    [line width=1pt,
    baseline={([yshift=-0.5ex+\cmdKV@general@yshift]current bounding box.center)},
    scale=\cmdKV@general@scale,
    rotate=\cmdKV@general@rotate,
    font=\cmdKV@general@font]
    \draw[\cmdKV@gTypes@iA] (0,0.5) node[label=(90+\cmdKV@general@rotate):\cmdKV@gLabels@eLA] {} -- node[label=(225+\cmdKV@general@rotate):\cmdKV@gLabels@iLA] {} (0.5,0);
    \draw[\cmdKV@gTypes@iB] (0.5,0.5) node[label=(90+\cmdKV@general@rotate):\cmdKV@gLabels@eLB] {} -- node[label=(\cmdKV@general@rotate):\cmdKV@gLabels@iLB] {} (0.5,0);
    \draw[\cmdKV@gTypes@iC] (1,0.5) node[label=(90+\cmdKV@general@rotate):\cmdKV@gLabels@eLC] {} -- node[label=(-45+\cmdKV@general@rotate):\cmdKV@gLabels@iLC] {} (0.5,0);
    \draw[\cmdKV@gTypes@iD] (0.5,0) -- node[label=(\cmdKV@general@rotate):\cmdKV@gLabels@iLD] {} (0.5,-0.5) node[label=(-90+\cmdKV@general@rotate):\cmdKV@gLabels@eLD] {};
    \draw[fill] (0,0.5) circle (0.02);
    \draw[fill] (0.5,0.5) circle (0.02);
    \draw[fill] (1,0.5) circle (0.02);
    \draw[fill] (0.5,-0.5) circle (0.02);
    #2
  \end{tikzpicture}
  \gSetStdTypes{line}
}

\def\gEFTVinY{\@ifnextchar[\@gEFTVinY{\@gEFTVinY[]}}
\def\@gEFTVinY[#1]#2{%
  \setkeys*{pre}{#1}%
  \setrmkeys{general,gTypes,gLabels}%
  \begin{tikzpicture}
    [line width=1pt,
    baseline={([yshift=-0.5ex+\cmdKV@general@yshift]current bounding box.center)},
    scale=\cmdKV@general@scale,
    rotate=\cmdKV@general@rotate,
    font=\cmdKV@general@font]
    \draw[\cmdKV@gTypes@iA] (0,0.5) node[label=(90+\cmdKV@general@rotate):\cmdKV@gLabels@eLA] {} -- node[label=(225+\cmdKV@general@rotate):\cmdKV@gLabels@iLA] {} (0.25,0.25);
    \draw[\cmdKV@gTypes@iE] (0.25,0.25) -- node[label=(225+\cmdKV@general@rotate):\cmdKV@gLabels@iLE] {} (0.5,0);
    \draw[\cmdKV@gTypes@iB] (0.5,0.5) node[label=(90+\cmdKV@general@rotate):\cmdKV@gLabels@eLB] {} -- node[label=(\cmdKV@general@rotate):\cmdKV@gLabels@iLB] {} (0.25,0.25);
    \draw[\cmdKV@gTypes@iC] (1,0.5) node[label=(90+\cmdKV@general@rotate):\cmdKV@gLabels@eLC] {} -- node[label=(-45+\cmdKV@general@rotate):\cmdKV@gLabels@iLC] {} (0.5,0);
    \draw[\cmdKV@gTypes@iD] (0.5,0) -- node[label=(\cmdKV@general@rotate):\cmdKV@gLabels@iLD] {} (0.5,-0.5) node[label=(-90+\cmdKV@general@rotate):\cmdKV@gLabels@eLD] {};
    \draw[fill] (0,0.5) circle (0.02);
    \draw[fill] (0.5,0.5) circle (0.02);
    \draw[fill] (1,0.5) circle (0.02);
    \draw[fill] (0.5,-0.5) circle (0.02);
    #2
  \end{tikzpicture}
  \gSetStdTypes{line}
}

\def\gEFTYinvY{\@ifnextchar[\@gEFTYinvY{\@gEFTYinvY[]}}
\def\@gEFTYinvY[#1]#2{%
  \setkeys*{pre}{#1}%
  \setrmkeys{general,gTypes,gLabels}%
  \begin{tikzpicture}
    [line width=1pt,
    baseline={([yshift=-0.5ex+\cmdKV@general@yshift]current bounding box.center)},
    scale=\cmdKV@general@scale,
    rotate=\cmdKV@general@rotate,
    font=\cmdKV@general@font]
    \draw[\cmdKV@gTypes@iA] (0,0.5) node[label=(90+\cmdKV@general@rotate):\cmdKV@gLabels@eLA] {} -- node[label=(225+\cmdKV@general@rotate):\cmdKV@gLabels@iLA] {} (0.3,0.166);
    \draw[\cmdKV@gTypes@iB] (0.6,0.5) node[label=(90+\cmdKV@general@rotate):\cmdKV@gLabels@eLB] {} -- node[label=(-45+\cmdKV@general@rotate):\cmdKV@gLabels@iLB] {} (0.3,0.166);
    \draw[\cmdKV@gTypes@iC] (0.3,-0.166) -- node[label=(135+\cmdKV@general@rotate):\cmdKV@gLabels@iLC] {} (0,-0.5) node[label=(-90+\cmdKV@general@rotate):\cmdKV@gLabels@eLC] {}; 
    \draw[\cmdKV@gTypes@iD] (0.3,-0.166) -- node[label=(45+\cmdKV@general@rotate):\cmdKV@gLabels@iLD] {} (0.6,-0.5) node[label=(-90+\cmdKV@general@rotate):\cmdKV@gLabels@eLD] {};
    \draw[\cmdKV@gTypes@iE] (0.3,0.166) -- node[label=(\cmdKV@general@rotate):\cmdKV@gLabels@iLE] {} (0.3,-0.166);
    \draw[fill] (0,0.5) circle (0.02);
    \draw[fill] (0.6,0.5) circle (0.02);
    \draw[fill] (0,-0.5) circle (0.02);
    \draw[fill] (0.6,-0.5) circle (0.02);
    #2
  \end{tikzpicture}
  \gSetStdTypes{line}
}

\def\gEFTX{\@ifnextchar[\@gEFTX{\@gEFTX[]}}
\def\@gEFTX[#1]#2{%
  \setkeys*{pre}{#1}%
  \setrmkeys{general,gTypes,gLabels}%
  \begin{tikzpicture}
    [line width=1pt,
    baseline={([yshift=-0.5ex+\cmdKV@general@yshift]current bounding box.center)},
    scale=\cmdKV@general@scale,
    rotate=\cmdKV@general@rotate,
    font=\cmdKV@general@font]
    \draw[\cmdKV@gTypes@iA] (0,0.5) node[label=(90+\cmdKV@general@rotate):\cmdKV@gLabels@eLA] {} -- node[label=(225+\cmdKV@general@rotate):\cmdKV@gLabels@iLA] {} (0.3,0);
    \draw[\cmdKV@gTypes@iB] (0.6,0.5) node[label=(90+\cmdKV@general@rotate):\cmdKV@gLabels@eLB] {} -- node[label=(-45+\cmdKV@general@rotate):\cmdKV@gLabels@iLB] {} (0.3,0);
    \draw[\cmdKV@gTypes@iC] (0.3,0) -- node[label=(135+\cmdKV@general@rotate):\cmdKV@gLabels@iLC] {} (0,-0.5) node[label=(-90+\cmdKV@general@rotate):\cmdKV@gLabels@eLC] {}; 
    \draw[\cmdKV@gTypes@iD] (0.3,0) -- node[label=(45+\cmdKV@general@rotate):\cmdKV@gLabels@iLD] {} (0.6,-0.5) node[label=(-90+\cmdKV@general@rotate):\cmdKV@gLabels@eLD] {};
    \draw[fill] (0,0.5) circle (0.02);
    \draw[fill] (0.6,0.5) circle (0.02);
    \draw[fill] (0,-0.5) circle (0.02);
    \draw[fill] (0.6,-0.5) circle (0.02);
    #2
  \end{tikzpicture}
  \gSetStdTypes{line}
}

\def\gEFTH{\@ifnextchar[\@gEFTH{\@gEFTH[]}}
\def\@gEFTH[#1]#2{%
  \setkeys*{pre}{#1}%
  \setrmkeys{general,gTypes,gLabels}%
  \begin{tikzpicture}
    [line width=1pt,
    baseline={([yshift=-0.5ex+\cmdKV@general@yshift]current bounding box.center)},
    scale=\cmdKV@general@scale,
    rotate=\cmdKV@general@rotate,
    font=\cmdKV@general@font]
    \draw[\cmdKV@gTypes@iA] (0,0.5) node[label=(90+\cmdKV@general@rotate):\cmdKV@gLabels@eLA] {} -- node[label=(180+\cmdKV@general@rotate):\cmdKV@gLabels@iLA] {} (0,0);
    \draw[\cmdKV@gTypes@iB] (0.6,0.5) node[label=(90+\cmdKV@general@rotate):\cmdKV@gLabels@eLB] {} -- node[label=(\cmdKV@general@rotate):\cmdKV@gLabels@iLB] {} (0.6,0);
    \draw[\cmdKV@gTypes@iC] (0.0,0) -- node[label=(180+\cmdKV@general@rotate):\cmdKV@gLabels@iLC] {} (0,-0.5) node[label=(-90+\cmdKV@general@rotate):\cmdKV@gLabels@eLC] {}; 
    \draw[\cmdKV@gTypes@iD] (0.6,0) -- node[label=(\cmdKV@general@rotate):\cmdKV@gLabels@iLD] {} (0.6,-0.5) node[label=(-90+\cmdKV@general@rotate):\cmdKV@gLabels@eLD] {};
    \draw[\cmdKV@gTypes@iE] (0,0) -- node[label=(90+\cmdKV@general@rotate):\cmdKV@gLabels@iLE] {} (0.6,0);
    \draw[fill] (0,0.5) circle (0.02);
    \draw[fill] (0.6,0.5) circle (0.02);
    \draw[fill] (0,-0.5) circle (0.02);
    \draw[fill] (0.6,-0.5) circle (0.02);
    #2
  \end{tikzpicture}
  \gSetStdTypes{line}
}

\def\cTree{\@ifnextchar[\@cTree{\@cTree[]}}
\def\@cTree[#1]#2{%
  \setkeys*{pre}{#1}%
  \setrmkeys{general,gTypes,gLabels,blobs}%
  \begin{tikzpicture}
    [line width=1pt,
    baseline={([yshift=-0.5ex+\cmdKV@general@yshift]current bounding box.center)},
    scale=\cmdKV@general@scale,
    rotate=\cmdKV@general@rotate,
    font=\cmdKV@general@font]
    \fill[\cmdKV@blobs@blobA] (0,0) ellipse (0.25 and 0.25);
    \draw[\cmdKV@gTypes@eA] ($ (0,0) + (-135:0.25) $) -- ($ (0,0) + (-135:0.6) $) node[label=(-180+\cmdKV@general@rotate):\cmdKV@gLabels@eLA] {};
    \draw[\cmdKV@gTypes@eB] ($ (0,0) + (135:0.25) $) -- ($ (0,0) + (135:0.6) $) node[label=(180+\cmdKV@general@rotate):\cmdKV@gLabels@eLB] {};
    \draw[\cmdKV@gTypes@eC] ($ (0,0) + (45:0.25) $) -- ($ (0,0) + (45:0.6) $) node[label=(\cmdKV@general@rotate):\cmdKV@gLabels@eLC] {};
    \draw[\cmdKV@gTypes@eD] ($ (0,0) + (-45:0.25) $) -- ($ (0,0) + (-45:0.6) $) node[label=(\cmdKV@general@rotate):\cmdKV@gLabels@eLD] {};
    \node[label=(90+\cmdKV@general@rotate):\cmdKV@gLabels@iLA] at (0,0.3) {};
    \node[label=(\cmdKV@general@rotate):\cmdKV@gLabels@iLB] at (0.3,0) {};
    #2
  \end{tikzpicture}
  \gSetStdTypes{line}
}

\def\cBoxA{\@ifnextchar[\@cBoxA{\@cBoxA[]}}
\def\@cBoxA[#1]#2{%
  \setkeys*{pre}{#1}%
  \setrmkeys{general,gTypes,gLabels,dots,blobs}%
  \begin{tikzpicture}
    [line width=1pt,
    baseline={([yshift=-0.5ex+\cmdKV@general@yshift]current bounding box.center)},
    scale=\cmdKV@general@scale,
    rotate=\cmdKV@general@rotate,
    font=\cmdKV@general@font]
    \fill[\cmdKV@blobs@blobA] (-0.4,0) ellipse (0.2 and 0.4);
    \fill[\cmdKV@blobs@blobB] (0.4,0) ellipse (0.2 and 0.4);
    \draw[\cmdKV@gTypes@eA] ($ (-0.4,0) + (-135:0.26) $) -- ($ (-0.4,0) + (-135:0.7) $) node[label=(-180+\cmdKV@general@rotate):\cmdKV@gLabels@eLA] {};
    \draw[\cmdKV@gTypes@eB] ($ (-0.4,0) + (135:0.26) $) -- ($ (-0.4,0) + (135:0.7) $) node[label=(180+\cmdKV@general@rotate):\cmdKV@gLabels@eLB] {};
    \draw[\cmdKV@gTypes@eC] ($ (0.4,0) + (45:0.26) $) -- ($ (0.4,0) + (45:0.7) $) node[label=(\cmdKV@general@rotate):\cmdKV@gLabels@eLC] {};
    \draw[\cmdKV@gTypes@eD] ($ (0.4,0) + (-45:0.26) $) -- ($ (0.4,0) + (-45:0.7) $) node[label=(\cmdKV@general@rotate):\cmdKV@gLabels@eLD] {};
    \draw[label distance=1,\cmdKV@gTypes@iA] ($ (-0.4,0) + (45:0.26) $) -- node[label=(90+\cmdKV@general@rotate):\cmdKV@gLabels@iLA] {}($ (0.4,0) + (135:0.26) $);
    \draw[label distance=1,\cmdKV@gTypes@iB] ($ (0.4,0) + (-135:0.26) $) -- node[label=(-90+\cmdKV@general@rotate):\cmdKV@gLabels@iLB] {}($ (-0.4,0) + (-45:0.26) $);
    \node[label=(180+\cmdKV@general@rotate):\cmdKV@gLabels@iLC] at (-0.6,0) {};
    \node[label=(\cmdKV@general@rotate):\cmdKV@gLabels@iLD] at (0.6,0) {};
    \ifKV@dots@lDots
      \draw[dotted] ($(-0.4,0) + (-150:0.7)$) to[bend left] ($(-0.4,0) + (150:0.7)$);%
    \fi
    \ifKV@dots@rDots
      \draw[dotted] ($(0.4,0) + (30:0.7)$) to[bend left] ($(0.4,0) + (-30:0.7)$);%
    \fi
    #2
  \end{tikzpicture}
  \gSetStdTypes{line}
}

\def\cBoxAA{\@ifnextchar[\@cBoxAA{\@cBoxAA[]}}
\def\@cBoxAA[#1]#2{%
  \setkeys*{pre}{#1}%
  \setrmkeys{general,gTypes,gLabels,dots,blobs}%
  \begin{tikzpicture}
    [line width=1pt,
    baseline={([yshift=-0.5ex+\cmdKV@general@yshift]current bounding box.center)},
    scale=\cmdKV@general@scale,
    rotate=\cmdKV@general@rotate,
    font=\cmdKV@general@font]
    \fill[\cmdKV@blobs@blobA] (-0.4,0) ellipse (0.2 and 0.4);
    \fill[\cmdKV@blobs@blobB] (0.4,0) ellipse (0.2 and 0.4);
    \draw[\cmdKV@gTypes@eA] ($ (-0.4,0) + (-135:0.26) $) -- ($ (-0.4,0) + (-135:0.7) $) node[label=(-180+\cmdKV@general@rotate):\cmdKV@gLabels@eLA] {};
    \draw[\cmdKV@gTypes@eB] ($ (-0.4,0) + (135:0.26) $) -- ($ (-0.4,0) + (135:0.7) $) node[label=(180+\cmdKV@general@rotate):\cmdKV@gLabels@eLB] {};
    \draw[\cmdKV@gTypes@eC] ($ (0.4,0) + (45:0.26) $) -- ($ (0.4,0) + (45:0.7) $) node[label=(\cmdKV@general@rotate):\cmdKV@gLabels@eLC] {};
    \draw[\cmdKV@gTypes@eD] ($ (0.4,0) + (-45:0.26) $) -- ($ (0.4,0) + (-45:0.7) $) node[label=(\cmdKV@general@rotate):\cmdKV@gLabels@eLD] {};
    \draw[label distance=1,\cmdKV@gTypes@iA] ($ (-0.4,0) + (60:0.32) $) to[bend left] node[label=(90+\cmdKV@general@rotate):\cmdKV@gLabels@iLA] {}($ (0.4,0) + (120:0.32) $);
    \draw[label distance=1,\cmdKV@gTypes@iB] ($ (0.4,0) + (-120:0.32) $) to[bend left]  node[label=(-90+\cmdKV@general@rotate):\cmdKV@gLabels@iLB] {}($ (-0.4,0) + (-60:0.32) $);
    \draw[dotted] (0,0.25) -- (0,-0.25);
    \node[label=(180+\cmdKV@general@rotate):\cmdKV@gLabels@iLC] at (-0.6,0) {};
    \node[label=(\cmdKV@general@rotate):\cmdKV@gLabels@iLD] at (0.6,0) {};
    \ifKV@dots@lDots
      \draw[dotted] ($(-0.4,0) + (-150:0.7)$) to[bend left] ($(-0.4,0) + (150:0.7)$);%
    \fi
    \ifKV@dots@rDots
      \draw[dotted] ($(0.4,0) + (30:0.7)$) to[bend left] ($(0.4,0) + (-30:0.7)$);%
    \fi
    #2
  \end{tikzpicture}
  \gSetStdTypes{line}
}

\def\cBoxB{\@ifnextchar[\@cBoxB{\@cBoxB[]}}
\def\@cBoxB[#1]#2{%
  \setkeys*{pre}{#1}%
  \setrmkeys{general,gTypes,gLabels,blobs}%
  \begin{tikzpicture}
    [line width=1pt,
    baseline={([yshift=-0.5ex+\cmdKV@general@yshift]current bounding box.center)},
    scale=\cmdKV@general@scale,
    rotate=\cmdKV@general@rotate,
    font=\cmdKV@general@font]
    \fill[\cmdKV@blobs@blobA] (0,-0.3) ellipse (0.4 and 0.2);
    \fill[\cmdKV@blobs@blobB] (0,0.3) ellipse (0.4 and 0.2);
    \draw[\cmdKV@gTypes@eA] ($ (0,-0.3) + (180:0.4) $) -- ($ (0,-0.3) + (-170:0.7) $) node[label=(180+\cmdKV@general@rotate):\cmdKV@gLabels@eLA] {};
    \draw[\cmdKV@gTypes@eB] ($ (0,0.3) + (180:0.4) $) -- ($ (0,0.3) + (170:0.7) $) node[label=(180+\cmdKV@general@rotate):\cmdKV@gLabels@eLB] {};
    \draw[\cmdKV@gTypes@eC] ($ (0,0.3) + (0:0.4) $) -- ($ (0,0.3) + (10:0.7) $) node[label=(\cmdKV@general@rotate):\cmdKV@gLabels@eLC] {};
    \draw[\cmdKV@gTypes@eD] ($ (0,-0.3) + (0:0.4) $) -- ($ (0,-0.3) + (-10:0.7) $) node[label=(\cmdKV@general@rotate):\cmdKV@gLabels@eLD] {};
    \draw[label distance=1,\cmdKV@gTypes@iA] ($ (0,-0.3) + (135:0.26) $) -- node[label=(-180+\cmdKV@general@rotate):\cmdKV@gLabels@iLA] {}($ (0,0.3) + (-135:0.26) $);
    \draw[label distance=1,\cmdKV@gTypes@iB] ($ (0,0.3) + (-45:0.26) $) -- node[label=(\cmdKV@general@rotate):\cmdKV@gLabels@iLB] {}($ (0,-0.3) + (45:0.26) $);
    \node[label=(90+\cmdKV@general@rotate):\cmdKV@gLabels@iLC] at (0,0.5) {};
    \node[label=(-90+\cmdKV@general@rotate):\cmdKV@gLabels@iLD] at (0,-0.5) {};
    #2
  \end{tikzpicture}
  \gSetStdTypes{line}
}

\def\cBoxBoxA{\@ifnextchar[\@cBoxBoxA{\@cBoxBoxA[]}}
\def\@cBoxBoxA[#1]#2{%
  \setkeys*{pre}{#1}%
  \setrmkeys{general,gTypes,gLabels,blobs}%
  \begin{tikzpicture}
    [line width=1pt,
    baseline={([yshift=-0.5ex+\cmdKV@general@yshift]current bounding box.center)},
    scale=\cmdKV@general@scale,
    rotate=\cmdKV@general@rotate,
    font=\cmdKV@general@font]
    \fill[\cmdKV@blobs@blobA] (-0.4,0) ellipse (0.2 and 0.4);
    \fill[\cmdKV@blobs@blobB] (0.4,0) ellipse (0.2 and 0.4);
    \fill[\cmdKV@blobs@blobC] (1.2,0) ellipse (0.2 and 0.4);
    \draw[\cmdKV@gTypes@eA] ($ (-0.4,0) + (-135:0.26) $) -- ($ (-0.4,0) + (-135:0.7) $) node[label=(-180+\cmdKV@general@rotate):\cmdKV@gLabels@eLA] {};
    \draw[\cmdKV@gTypes@eB] ($ (-0.4,0) + (135:0.26) $) -- ($ (-0.4,0) + (135:0.7) $) node[label=(180+\cmdKV@general@rotate):\cmdKV@gLabels@eLB] {};
    \draw[\cmdKV@gTypes@eC] ($ (1.2,0) + (45:0.26) $) -- ($ (1.2,0) + (45:0.7) $) node[label=(\cmdKV@general@rotate):\cmdKV@gLabels@eLC] {};
    \draw[\cmdKV@gTypes@eD] ($ (1.2,0) + (-45:0.26) $) -- ($ (1.2,0) + (-45:0.7) $) node[label=(\cmdKV@general@rotate):\cmdKV@gLabels@eLD] {};
    \draw[label distance=1,\cmdKV@gTypes@iA] ($ (-0.4,0) + (45:0.26) $) -- node[label=(90+\cmdKV@general@rotate):\cmdKV@gLabels@iLA] {}($ (0.4,0) + (135:0.26) $);
    \draw[label distance=1,\cmdKV@gTypes@iB] ($ (0.4,0) + (45:0.26) $) -- node[label=(90+\cmdKV@general@rotate):\cmdKV@gLabels@iLB] {}($ (1.2,0) + (135:0.26) $);
    \draw[label distance=1,\cmdKV@gTypes@iC] ($ (1.2,0) + (-135:0.26) $) -- node[label=(-90+\cmdKV@general@rotate):\cmdKV@gLabels@iLC] {}($ (0.4,0) + (-45:0.26) $);
    \draw[label distance=1,\cmdKV@gTypes@iD] ($ (0.4,0) + (-135:0.26) $) -- node[label=(-90+\cmdKV@general@rotate):\cmdKV@gLabels@iLD] {}($ (-0.4,0) + (-45:0.26) $);
    \node[label=(180+\cmdKV@general@rotate):\cmdKV@gLabels@iLE] at (-0.6,0) {};
    \node[label=(\cmdKV@general@rotate):\cmdKV@gLabels@iLF] at (0.6,0) {};
    \node[label=(\cmdKV@general@rotate):\cmdKV@gLabels@iLG] at (1.4,0) {};
    #2
  \end{tikzpicture}
  \gSetStdTypes{line}
}

\def\cBoxBoxB{\@ifnextchar[\@cBoxBoxB{\@cBoxBoxB[]}}
\def\@cBoxBoxB[#1]#2{%
  \setkeys*{pre}{#1}%
  \setrmkeys{general,gTypes,gLabels}%
  \begin{tikzpicture}
    [line width=1pt,
    baseline={([yshift=-0.5ex+\cmdKV@general@yshift]current bounding box.center)},
    scale=\cmdKV@general@scale,
    rotate=\cmdKV@general@rotate,
    font=\cmdKV@general@font]
    \fill[\cmdKV@blobs@blobA] (0,0.3) ellipse (0.4 and 0.2);
    \fill[\cmdKV@blobs@blobB] (0,-0.3) ellipse (0.4 and 0.2);
    \fill[\cmdKV@blobs@blobC] (1,0) ellipse (0.2 and 0.4);
    \draw[\cmdKV@gTypes@eA] ($ (0,-0.3) + (180:0.4) $) -- ($ (0,-0.3) + (-170:0.7) $) node[label=(180+\cmdKV@general@rotate):\cmdKV@gLabels@eLA] {};
    \draw[\cmdKV@gTypes@eB] ($ (0,0.3) + (180:0.4) $) -- ($ (0,0.3) + (170:0.7) $) node[label=(180+\cmdKV@general@rotate):\cmdKV@gLabels@eLB] {};
    \draw[\cmdKV@gTypes@eC] ($ (1,0) + (45:0.26) $) -- ($ (1,0) + (45:0.7) $) node[label=(\cmdKV@general@rotate):\cmdKV@gLabels@eLC] {};
    \draw[\cmdKV@gTypes@eD] ($ (1,0) + (-45:0.26) $) -- ($ (1,0) + (-45:0.7) $) node[label=(\cmdKV@general@rotate):\cmdKV@gLabels@eLD] {};
    \draw[label distance=1,\cmdKV@gTypes@iA] ($ (0,-0.3) + (135:0.26) $) -- node[label=(-180+\cmdKV@general@rotate):\cmdKV@gLabels@iLA] {}($ (0,0.3) + (-135:0.26) $);
    \draw[label distance=1,\cmdKV@gTypes@iB] ($ (0,0.3) + (0:0.4) $) -- node[label=(90+\cmdKV@general@rotate):\cmdKV@gLabels@iLB] {} ($ (1,0) + (135:0.26) $);
    \draw[label distance=1,\cmdKV@gTypes@iC] ($ (1,0) + (-135:0.26) $) -- node[label=(-90+\cmdKV@general@rotate):\cmdKV@gLabels@iLC] {} ($ (0,-0.3) + (0:0.4) $);
    \draw[label distance=1,\cmdKV@gTypes@iD] ($ (0,0.3) + (-45:0.26) $) -- node[label=(\cmdKV@general@rotate):\cmdKV@gLabels@iLD] {}($ (0,-0.3) + (45:0.26) $);
    \node[label=(90+\cmdKV@general@rotate):\cmdKV@gLabels@iLE] at (0,0.5) {};
    \node[label=(\cmdKV@general@rotate):\cmdKV@gLabels@iLF] at (1.2,0) {};
    \node[label=(-90+\cmdKV@general@rotate):\cmdKV@gLabels@iLG] at (0,-0.5) {};
    #2
  \end{tikzpicture}
  \gSetStdTypes{line}
}

\def\cNPBox{\@ifnextchar[\@cNPBox{\@cNPBox[]}}
\def\@cNPBox[#1]#2{%
  \setkeys*{pre}{#1}%
  \setrmkeys{general,gTypes,gLabels,blob}%
  \begin{tikzpicture}
    [line width=1pt,
    baseline={([yshift=-0.5ex+\cmdKV@general@yshift]current bounding box.center)},
    scale=\cmdKV@general@scale,
    rotate=\cmdKV@general@rotate,
    font=\cmdKV@general@font]
    \fill[\cmdKV@blobs@blobA] (-0.6,0) ellipse (0.2 and 0.4);
    \fill[\cmdKV@blobs@blobB] (0.4,0) ellipse (0.2 and 0.4);
    \draw[\cmdKV@gTypes@eA] ($ (-0.6,0) + (-180:0.2) $) -- ($ (-0.6,0) + (-180:0.6) $) node[label=(-180+\cmdKV@general@rotate):\cmdKV@gLabels@eLA] {};
    \draw[\cmdKV@gTypes@eB] ($ (-0.6,0) + (0:0.2) $) -- (-0.18,0) node[label=(\cmdKV@general@rotate):\cmdKV@gLabels@eLB] {};
    \draw[\cmdKV@gTypes@eC] ($ (0.4,0) + (45:0.26) $) -- ($ (0.4,0) + (45:0.7) $) node[label=(\cmdKV@general@rotate):\cmdKV@gLabels@eLC] {};
    \draw[\cmdKV@gTypes@eD] ($ (0.4,0) + (-45:0.26) $) -- ($ (0.4,0) + (-45:0.7) $) node[label=(\cmdKV@general@rotate):\cmdKV@gLabels@eLD] {};
    \draw[label distance=1,\cmdKV@gTypes@iA] ($ (-0.6,0) + (60:0.3) $) -- node[label=(90+\cmdKV@general@rotate):\cmdKV@gLabels@iLA] {}($ (0.4,0) + (135:0.26) $);
    \draw[label distance=1,\cmdKV@gTypes@iB] ($ (0.4,0) + (-135:0.26) $) -- node[label=(-90+\cmdKV@general@rotate):\cmdKV@gLabels@iLB] {}($ (-0.6,0) + (-60:0.3) $);
    \node[label=(\cmdKV@general@rotate):\cmdKV@gLabels@iLD] at (0.6,0) {};
    #2
  \end{tikzpicture}
  \gSetStdTypes{line}
}

\def\cFourPtRec{\@ifnextchar[\@cFourPtRec{\@cFourPtRec[]}}
\def\@cFourPtRec[#1]#2{%
  \setkeys*{pre}{#1}%
  \setrmkeys{general,gTypes,gLabels,blobs}%
  \begin{tikzpicture}
    [line width=1pt,
    baseline={([yshift=-0.5ex+\cmdKV@general@yshift]current bounding box.center)},
    scale=\cmdKV@general@scale,
    rotate=\cmdKV@general@rotate,
    font=\cmdKV@general@font]
    \fill[\cmdKV@blobs@blobA] (-0.5,0) ellipse (0.2 and 0.3);
    \fill[\cmdKV@blobs@blobB] (0.5,0) ellipse (0.3 and 0.4);
    \filldraw[fill=white,draw=black] (0.55,0.15) circle (0.1);
    \filldraw[fill=white,draw=black] (0.45,-0.1) circle (0.1);
    \draw[\cmdKV@gTypes@eA] ($(-0.5,0) + (135:0.23)$) -- ($(-0.5,0) + (135:0.5)$) node[label=(180+\cmdKV@general@rotate):\cmdKV@gLabels@eLA] {};
    \draw[\cmdKV@gTypes@eB] ($(-0.5,0) + (-135:0.23)$) -- ($(-0.5,0) + (-135:0.5)$) node[label=(180+\cmdKV@general@rotate):\cmdKV@gLabels@eLB] {};
    \draw[\cmdKV@gTypes@eC] ($(0.5,0) + (45:0.35)$) -- ($(0.5,0) + (45:0.7)$) node[label=(\cmdKV@general@rotate):\cmdKV@gLabels@eLC] {};
    \draw[\cmdKV@gTypes@eD] ($(0.5,0) + (-45:0.35)$) -- ($(0.5,0) + (-45:0.7)$) node[label=(\cmdKV@general@rotate):\cmdKV@gLabels@eLD] {};
    \draw[label distance=1,\cmdKV@gTypes@iA] ($(-0.5,0) + (30:0.23)$) -- node[label=(90+\cmdKV@general@rotate):\cmdKV@gLabels@iLA] {} ($(0.5,0) + (157:0.3)$);
    \draw[label distance=1,\cmdKV@gTypes@iB] ($(0.5,0) + (-157:0.3)$) -- node[label=(-90+\cmdKV@general@rotate):\cmdKV@gLabels@iLB] {} ($(-0.5,0) + (-30:0.23)$);
    #2
 \end{tikzpicture}
  \gSetStdTypes{line}
}

\makeatother

\numberwithin{equation}{section}

\newcommand\vecbf[1]{{\boldsymbol #1}}

\newcommand{\nn}{\nonumber}

\newcommand\beq{\begin{equation}}
\newcommand\eeq{\end{equation}}
\newcommand\beal{\begin{aligned}}
\newcommand\eeal{\end{aligned}}
\newcommand\bea{\begin{eqnarray}}
\newcommand\eea{\end{eqnarray}}

\newcommand\dd{{\mathrm d}}
\usepackage{xcolor}

\newcommand{\bu}{{\boldsymbol u}}
\newcommand{\bb}{{\boldsymbol b}}
\newcommand{\bk}{{\boldsymbol k}}

\newcommand{\bp}{{\boldsymbol p}}

\newcommand{\bL}{{\boldsymbol L}}
\newcommand{\bJ}{{\boldsymbol J}}

\newcommand{\bx}{{\boldsymbol x}}

\newcommand{\bS}{{\boldsymbol S}}
\newcommand{\ba}{{\boldsymbol a}}
\newcommand{\bSi}{{\boldsymbol \Xi}}

\newcommand{\cD}{\mathcal{D}}
\newcommand\cE{\mathcal{E}}

\newcommand\Mp{M_{\rm Pl}}

\newcommand\cR{\mathcal{R}}

\makeatletter
\newcommand{\Biggg}{\bBigg@{3.5}}
\makeatother

\begin{document}
\preprint{\texttt{DESY\,21-020}}
\title{\center Spin Effects in the Effective Field Theory Approach to \\ [0.2cm] Post-Minkowskian Conservative Dynamics  }

\author{\large \,\, Zhengwen Liu,\,}
\author{\large Rafael A. Porto}
\author{\large and Zixin Yang}
\affiliation{Deutsches Elektronen-Synchrotron DESY, Notkestrasse 85, 22607 Hamburg, Germany}
\emailAdd{zhengwen.liu@desy.de} \emailAdd{rafael.porto@desy.de} \emailAdd{zixin.yang@desy.de}

\abstract{Building upon the worldline effective field theory (EFT) formalism for spinning bodies developed for the Post-Newtonian regime, we generalize the EFT approach to Post-Minkowskian (PM) dynamics to include rotational degrees of freedom in a manifestly covariant framework. We introduce a systematic procedure to compute the total change in momentum and spin in the gravitational scattering of compact objects. For the special case of spins aligned with the orbital angular momentum, we show how to construct the radial action for elliptic-like orbits using the Boundary-to-Bound correspondence. As~a paradigmatic example, we solve the scattering problem to next-to-leading PM order with~linear and bilinear spin effects and arbitrary initial conditions, incorporating for the first time finite-size corrections. We~obtain the aligned-spin radial action from the resulting scattering data, and derive the periastron advance and binding energy for circular orbits. We also provide the (square of the) center-of-mass momentum to ${\cal O}(G^2)$, which may be used to reconstruct a Hamiltonian. Our results are in perfect agreement with the existent literature, while at the same time extend the knowledge of the PM dynamics of compact binaries at quadratic order in spins.} 
\maketitle
\newpage

\section{Introduction} \label{sec:intro}

The power of gravitational wave (GW) science \cite{LIGO} is predicated on the precise reconstruction of the GW signal as a function of the parameters of the sources, notably binary compact objects~\cite{buosathya,tune,music}. As one may anticipate, effects due to spin play a key role in the problem,~e.g.~\cite{salvo}, particularly due to the expectation that  binary black holes in the observable universe may be rapidly rotating,~e.g.~\cite{Zackay:2019tzo}. Spinning black holes have also attracted interest in recent years due their ability to harvest clouds of putative ultralight particles~\cite{axiverse}, which can be fleshed out either through mass/spin distributions, in GW stochastic backgrounds \cite{qcd1,cardoso,salvo2} or, more promising, the precise reconstruction of the GW signal emitted from binary systems \cite{gcollider1,gcollider2}. Rotating~black holes have puzzled relativists for decades,  taking almost 50 years after the discovery of Schwarzschild's solution to arrive at the Kerr metric \cite{Kerr}. Not surprisingly, the situation does not improve in the two-body problem. Consequently, prior to the development of the effective field theory (EFT) approach \cite{nrgr,nrgrs} (see \cite{walterLH,Foffa:2013qca,review} for detailed reviews), incorporating spin effects in the gravitational dynamics of binary systems was a daunting task.\footnote{Needless to say, numerical simulations for spinning black holes in the strongly coupled regime are also significantly more involved than non-rotating counterparts, e.g.~\cite{Hinder:2018fsy}.} This was the case even~for conservative contributions in the perturbative Post-Newtonian (PN) regime of small-velocity/weak-gravity, with only spin-orbit results known at the time to next-to-leading order (NLO) \cite{Faye1}. Presently, spin effects in the PN conservative dynamics of binary compact systems are known up to N$^{2}$LO \cite{Faye1,prl,Porto:2007px,nrgrss,nrgrs2,jan1,jan2,nrgrso,Levi:2015uxa,Levi:2016ofk}, with partial results also at higher orders, e.g.~\cite{Levi:2020uwu,Levi:2020kvb}.\footnote{On the other hand,  spin-dependent radiation effects are only known to NLO and up to quadratic order in the spins \cite{Faye2,rads1,amps,bohe,natalia1,natalia2,zixin,Pardo}. The radiated power without spin was (re-)obtained in the EFT framework of \cite{nrgr,andirad} to N$^{2}$LO in \cite{radnrgr}. Absorption effects can also be studied within an EFT worldline theory, see e.g.~\cite{dis1,dis2,dis3}.} Moreover, spin-independent conservative contributions --- both from potential and radiation-reaction effects --- are known up to N$^{5}$LO  \cite{Blanchet:2003gy,nrgr3pn,Foffa:2012rn,tail,nrgrG5,apparent,Damour:2014jta,Marchand:2017pir,nrgr4pn1,nrgr4pn2,5pn1,5pn2,tail2,Blumlein:2020pyo}, with partial results known at higher orders using various methodologies,~e.g.~\cite{blum,blum2,bini2,tail3}. 
Most of these results in the PN regime, notably spin effects, were obtained for the first time following variants of the EFT approach developed in \cite{nrgr,nrgrs}. The goal of this paper is therefore to repurpose the worldline theory for spinning bodies, originally introduced for the PN expansion, to calculate spin effects in the Post-Minkowskian (PM) regime using the EFT approach and boundary-to-bound (B2B) correspondence recently developed in \cite{paper1,paper2,pmeft,3pmeft,tidaleft}. \vskip 4pt

One of the main advantages of an EFT framework for rotating bodies --- presently widely adopted \cite{review,blanchet} --- is the introduction of a point-particle effective action to describe compact objects in gravitational backgrounds \cite{nrgrs}, in contrast to applying Mathisson-Papapetrou-Dixon (MPD) equations-of-motion (EoM) independently for momentum and spin \cite{Mathisson,Papa,Dixon}. In~addition, finite-size effects can be readily incorporated as a series of corrections beyond minimal coupling constrained solely by diffeomorphism invariance \cite{nrgr,nrgrs}, without the need of an {\it ansatz} for the stress energy tensor. The EFT framework therefore reduces the number of free parameters in previous (more traditional) approaches, e.g.~\cite{jan2}. There are a few other subtleties when dealing with the spin dynamics of point-like objects. For instance, the gauge redundancy in describing rotational degrees of freedom in relativistic theories means that {\it Spin Supplementarity Conditions} (SCCs) are often invoked \cite{hanson}. Moreover, rather than a {\it position} the spin angular momentum behaves as a {\it conjugated} variable. This naturally leads us to an effective theory written in terms of a Routhian \cite{yee,Porto:2007px,nrgrs2,nrgrss,review}; that is, half as a Lagrangian (for the position/velocity) and half as a Hamiltonian (for the spin variables). The EFT machinery then sets in, systematically ``integrating out"  in the saddle point approximation the {\it potential} and {\it radiation} modes of the gravitational field, via a series of Feynman diagrams. As it is customary, divergences of the point-particle approximation are then naturally handled via regularization/renormalization.\vskip 4pt

Up until recently, efforts to solve the conservative binary dynamics of compact objects have focused on the direct calculation of the Hamiltonian \cite{Damour:2014jta} or Lagrangian \cite{nrgr4pn1,nrgr4pn2,Marchand:2017pir,5pn1,5pn2} as an intermedia step towards building waveforms. This was no different for spin effects~\cite{blanchet,review}. However, building upon novel ideas from scattering amplitudes \cite{elvang,reviewdc,Henn:2014qga}, in the last years we experienced an explosion of work using the classical limit of amplitudes, either to compute the impulse or to extract an effective Hamiltonian which can then be used to study generic orbits, e.g.~\cite{ira1,cheung,zvi1,zvi2,donal,donalvines, withchad,Holstein:2008sx,Bjerrum-Bohr:2013bxa,Vaidya:2014kza,Guevara:2017csg,Chung:2018kqs,Guevara:2018wpp,Guevara:2019fsj,bohr,cristof1,simon,Arkani-Hamed:2019ymq,Bjerrum-Bohr:2019kec,Chung:2019duq,Bautista:2019tdr,Bautista:2019evw,KoemansCollado:2019ggb,Johansson:2019dnu,Aoude:2020onz,Cristofoli:2020uzm,Chung:2020rrz,zvispin,Bern:2020gjj,DiVecchia:2019myk,Antonelli:2019ytb,Brandhuber:2019qpg,Cheung:2020gyp,Parra,soloncheung,AccettulliHuber:2020oou,Bern:2020uwk,Cheung:2020gbf,spinsheet,4pmzvi}. These developments, which notably belong to the realm of the PM expansion, have produced the state-of-the-art for the conservative dynamics of non-spinning bodies in the PM regime at N$^{2}$LO order \cite{cheung,zvi1,zvi2},  and very recently also partial results (with potential modes) at N$^{3}$LO \cite{4pmzvi}. On the other hand, for spinning bodies, spin-orbit and spin$_{1}$-spin$_{2}$ contributions are presently known to NLO in the PM regime~\cite{justin1,justin2,donalvines,Guevara:2018wpp,Guevara:2019fsj,Chung:2020rrz,zvispin}. (Radiation effects in the PM expansion have also been recently approached in e.g.~\cite{Parra2,Gabriele,janmogul,janmogul2,Mougiakakos:2021ckm}.)\vskip 4pt While the derivation of a (classical) Hamiltonian from a (quantum) scattering amplitude as an intermedia step is a perfectly viable option (dating back to the work of Iwasaki \cite{iwasaki}), one of the main paradigms in modern approaches is to avoid the introduction of gauge-dependent objects \cite{elvang,reviewdc}. For instance, this  was adopted in \cite{donal,donalvines} to solve for the (classical) scattering problem, although without providing yet the necessary link to bound states. This then became the main motivation for the B2B correspondence --- to remain entirely within the {\it on-shell} philosophy. The B2B dictionary was introduced in~\cite{paper1,paper2}, mapping scattering data to observables for elliptic-like orbits via a radial action and the analytic continuation in binding energy and angular momentum. This allowed us to directly relate gravitational observables without ever invoking a Hamiltonian. In its first incarnation \cite{paper1}, the B2B map relied on the connection (dubbed `impetus formula') between the center-of-mass momentum and the (infrared-finite) scattering amplitude in the classical limit, which was used to compute the radial action.\footnote{See \cite{4pmzvi} for further developments in the amplitude-action link motivated by the B2B map \cite{paper1,paper2}.} In~the second version, suggested~in~\cite{paper1} and elaborated in \cite{paper2}, the dictionary was entirely constructed from the knowledge of the scattering angle instead. Moreover, in \cite{paper2} we showed how spin effects are incorporated in the B2B correspondence through the connection between the periastron advance and scattering angle, albeit for configurations where spins are aligned with the orbital angular momentum. Once the B2B correspondence is written in terms of the deflection angle, bypassing the need to go through the classical limit of a scattering amplitude, the remaining task is to systematically compute the former entirely within the classical domain. Following the pioneering work in~\cite{nrgr}, an EFT formalism was developed~in~\cite{pmeft} to solve for the impulse and scattering angle in the PM regime via Feynman diagrams, originally without spin effects. Shortly after the EFT approach was introduced, and benefiting from the simplifications of the classical framework together with powerful tools for computing `loop' integrals via differential equations \cite{Henn:2014qga,Parra}, the EFT formalism~in~\cite{pmeft} rapidly achieved the state-of-the-art at 3PM \cite{3pmeft}, subsequently yielding also new results for tidal effects beyond leading order \cite{tidaleft}. In this paper, building upon the EFT in \cite{nrgrs,nrgrss,nrgrs2,nrgrso}, we continue the development of the EFT approach in the PM regime by incorporating spin effects in the scattering problem, insofar for the conservative sector. We then implement the B2B dictionary to derive observables for bound orbits with aligned spins. Mirroring the simplifications already reflected in \cite{pmeft,3pmeft,tidaleft}, the inclusion of spin in classical scattering and consequently in elliptic-like motion become remarkably simpler than computing the Hamiltonian. As a result, we will readily achieve the state-of-the-art for spin-orbit and spin$_{1}$-spin$_{2}$ effects in the PM regime, and present  spin$_{1}$-spin$_{1}$ contributions to NLO, including finite-size effects, for the first~time.\vskip 4pt

This paper is organized as follows. In~\S\ref{sec:intros} we review the worldline EFT for spinning compact objects developed in \cite{nrgrs,nrgrss,nrgrs2,nrgrso}, and subsequently adapt it to the PM expansion along the lines of \cite{pmeft}. In~\S\ref{sec:b2b} we discuss the construction of the bound radial action via the B2B map with (aligned-)spin effects, as well as the matching between the coefficients of the deflection angle and the square of the CoM momentum (impetus). In \S\ref{2pms} we apply the EFT approach within the covariant SSC to compute the total momentum impulse and spin kick to 2PM and quadratic order in the spins, with generic initial orientations. We also discuss the map to canonical variables and derive the (aligned-spin) deflection angle. Finally, in \S\ref{2pmo} we derive the bound radial action through the B2B correspondence and compute observables to NLO in the PM regime and quadratic order in the spins, including finite-size effects beyond leading order. We also display the coefficients of the impetus to 2PM. We~conclude in \S\ref{disc} with a few remarks on future directions. Aspects of the calculations are relegated to appendices. We provide also an ancillary file in the {\tt arXiv} submission with more detailed results.\vskip 4pt

{\it Conventions}: We use $\eta_{\mu\nu} = {\rm diag}(+,-,-,-)$ for the Minkowski metric. The product of four-vectors is denoted as $k \cdot x = \eta_{\mu\nu} k^\mu x^\nu$, and $\bk \cdot \bx = \delta^{ij} \bk^i \bx^j$ for the Euclidean case, with boldface letters representing three-vectors. We use the convention $\epsilon_{0123}=1$ from the \texttt{xCoba} package. We~work in dimensional regularization in $D= d-2\epsilon$ dimensions, with $d$ either $4,3$ or~$2$. We use the notation $\int_k \equiv \int d^Dk/(2\pi)^D$, as well as $\hat\delta(x) \equiv 2\pi \delta(x)$. We use $\Mp^{-1} \equiv \sqrt{32\pi G}$ for the Planck mass, in $\hbar=c=1$ units, with $G$ Newton's constant.  \newpage
\section{Spinning bodies in the PM EFT approach} \label{sec:intros}

We start by briefly reviewing the worldline effective theory approach for rotating compact bodies \cite{nrgrs,nrgrss,nrgrs2}. Afterwards we show how to solve for the momentum and spin impulse in gravitational scattering to all orders in $G$. For more details in the EFT framework see \cite{review}.

\subsection{Worldline effective theory} \label{sec:pp}

As it is well-known, e.g.~\cite{hanson}, additional constrained variables are needed in order to introduce a local (off-shell) effective action describing a spinning body in a relativistic framework. Following the analogy with angular momentum, we use a spin tensor, $S^{\alpha\beta}$, that is subject to a SSC (technically a second class constraint) \cite{nrgrs}. In order to preserve covariance (without background fields) it is customary to resort to a covariant one,
\beq
S^{\alpha\beta} p_\beta =0\,,\label{ssc1}
\eeq
with $p_\mu$ the particle's momentum. 
The preservation of the SSC upon evolution  implies \cite{review} 
\beq
p^\alpha = \frac{1}{\sqrt{v^2}}\left( m v^\alpha + \frac{1}{2m} R_{\beta\rho\mu\nu} S^{\alpha\beta}S^{\mu\nu} v^\rho + \cdots\right)\,,
\eeq
with $v^\mu \equiv \frac{dx^\mu}{d\sigma}$ the particle's velocity and $\sigma$ and affine parameter. The ellipses account for higher orders in spin and curvature. To bilinear order in the spin,  the SSC in \eqref{ssc1} becomes
\beq
S^{\alpha\beta} v_\beta =0 + {\cal O}(S^3) \,.\label{ssc2}
\eeq
The mass, $m \equiv m(S^2)$, can be read-off from the on-shell condition, $p^2=m^2$, which also serves as a constraint enforcing reparameterization invariance~\cite{hanson}.\vskip 4pt

To introduce a worldline action, from which the equations of motion can be derived, it is convenient to use a tetrad field, $e^I_\mu$, which co-rotates with the (compact) body \cite{nrgrs}. Using a locally-flat frame, $e^a_\mu$ (with $g^{\mu\nu} e^a_\mu e^b_\nu = \eta^{ab}$), the co-rotating tetrad can be parameterized in terms of an element of the Lorentz algebra, $\Lambda_a^I$, via $e^I_\mu = \Lambda^I_a e^a_\mu$. Using these fields (and time derivatives) as degrees of freedom we can introduce an action such that we obtain the MPD equations of motions, with the spin tensor emerging as a momentum variable conjugate to the angular velocity of the co-rotating field~\cite{nrgrs}. Because of this, rather than a Lagrangian (or a Hamiltonian), it turns out to be useful to use a Routhian instead to describe rotating bodies, with the spin promoted to a lead-actor in the effective theory.  Furthermore, it is also convenient to write the worldline theory using the spin variables projected onto the locally-flat frame,
\beq
S^{ab} \equiv S^{\mu\nu} e^a_\mu e^b_\nu\,,
\eeq
such that the $S^{ab}$ matrices obey the $SO(1,3)$  algebra, 
\beq
\{S^{ab}, S^{cd}\} = \eta^{ac}S^{bd}+ \eta^{bd}S^{ac}- \eta^{ad}S^{bc}-\eta^{bc}S^{ad}\,.\label{algebra}
\eeq
The SSC is then easily incorporated through Lagrange multipliers which are fixed by the preservation upon evolution. These extra parameters then yield an additional (curvature-dependent) term in the worldline Routhian (see \eqref{routhian} below) \cite{review}. As in the non-spinning case, the worldline theory can also readily include spin-dependent finite-size effects through diffeomorphism invariant contributions beyond minimal coupling \cite{nrgr,nrgrs}. For instance, the self-induced quadrupole moment of a rotating body is described by the  coupling, first introduced in \cite{nrgrs,nrgrss,nrgrs2},
\beq
\frac{C_{ES^2}}{2m} \int \, \frac{E_{\mu\nu}}{\sqrt{v^2}} \, e^\mu_a e^\nu_b S^{ac}{S_c}^b \, d \sigma \,,
\eeq
where $E_{\mu\nu}$ is the electric compoment of the Weyl tensor. The {\it Wilson coefficient}, $C_{ES^2}$, parameterizes our ignorance about the internal degrees of freedom of the compact object, either a black hole, neutron star, or any other exotic possibility. For example, for a Kerr black hole we have $C^{\rm Kerr}_{ ES^2}=1$ \cite{nrgrs2}, but (much) larger values may be obtained in other scenarios, for instance with clouds of ultralight particles surrounding black holes \cite{gcollider1,gcollider2}.\vskip 4pt

Before we move on, there is yet another important simplification that occurs when studying scattering processes. As it was discussed in \cite{pmeft,3pmeft,tidaleft}, without spin, we can introduce an einbein, $e$, and a Polyakov-type action linear in the metric field. We can then choose the gauge $e=1$, which coincides with the proper-time for incoming and outgoing states. It is straightforward to extend the same reasoning to the case of spinning bodies, resulting in a point-particle wordline action that can be written as 
\beq
S_{\rm pp} \equiv  \int_{-\infty}^{+\infty} d\tau\, {\cal R}\,,\label{seff}
\eeq
with $\tau$ the proper-time (at $\pm \infty$). The Routhian, $\cal R$, in the covariant SSC then takes the form\beq
\begin{aligned}
\quad\quad{\cal R} = - \frac{1}{2} &  \bigg( m\, g_{\mu\nu} v^\mu v^\nu +  \omega_\mu^{ab}S_{ab} v^{\mu} \\
&+ \frac{1}{m} R_{\beta\rho\mu\nu}e_a^\alpha e^\beta_b e^\mu_c e^\nu_d  S^{ab}S^{cd} v^\rho v_\alpha - \frac{C_{ES^2}}{m} E_{\mu\nu} e^\mu_a e^\nu_b S^{ac}{S_{c}}^b  +\cdots  \bigg)\,, \label{routhian}
\end{aligned}
\eeq
to linear order in curvature and quadratic order in the spins, with $\omega^{ab}_\mu$ the Ricci rotation coefficients. The last two (curvature-dependent) terms account for the conservation of the SSC as well as finite-size effects to quadratic order in the spins, respectively. The EoM are obtained via \cite{review}
\beq
{\delta  \over \delta x^\mu} S_{\rm pp}= 0 \,, \quad \frac{d}{d\tau} S^{ab} =  
\{S^{ab},\,{\cal R}\}\,.\label{eom}
\eeq
Notice that after expanding in the weak field limit, 
\beq
g_{\mu\nu} = \eta_{\mu\nu} + h_{\mu\nu}/\Mp\,\,,
\eeq
the mass coupling remains linear in the metric \cite{pmeft}, but that is not the case for the other terms, which instead yield non-linear gravitational interactions both at linear and bilinear order in the spin. In what follows we show how to use this formalism to compute the total momentum and spin impulses in gravitational encounters.

\subsection{Momentum \& Spin impulses} \label{sec:impulse}

The computation follows similar steps as described in \cite{pmeft}. We start by `integrating out' the gravitational field in the potential region in a saddle-point approximation ($A=1,2$)
\beq
e^{i S_{\rm eff}[x_A,S_A^{ab}] } = \int \cD h_{\mu\nu} \, e^{i S_{\rm EH}[h] + i S_{\rm GF}[h] + i \int d\tau\, {\cal R}[x_A,S^{ab}_A,h]}\,,\label{eff1}
\eeq
 where $S_{\rm EH}$ and $S_{\rm GF}$ are the Einstein-Hilbert action and gauge-fixing terms, respectively. As explained in \cite{pmeft}, we adapt $S_{\rm GF}$ (as well as total time-derivatives) to simplify the resulting Feynman rules. The effective Routhian/action then becomes a (local-in-time)\footnote{We are ignoring here the non-local contributions due to radiation-reaction (tail) effects \cite{tail,apparent,nrgr4pn2}, which include also spin-dependent effects at higher PM orders.} function of the position and spin of the two-body systems,
\beq
S_{\rm eff} = \sum_n \int d\tau_1 {\cal R}_{n} [x_1(\tau_1),S_1(\tau_1);x_1(\tau_2),S_2(\tau_1) ] \,.
\eeq 
The ${\cal R}_n$'s are the ${\cal O}(G^n)$ contribution to the worldline Routhian after evaluating the Feynman integrals for generic configurations. From here we can then use \eqref{eom} to obtain the EoM, which we can solve iteratively in powers of $G$, both for the position variables \cite{pmeft},
\beq
\label{pmexp1}
\begin{aligned}
x^\mu_A(\tau_A) &= b^\mu_A + u^\mu_a \tau_a + \sum_n \delta^{(n)} x^\mu_A (\tau_A)\,,\\
v^\nu_A(\tau_A) &= u^\nu_A  + \sum_n \delta^{(n)} v^\nu_A(\tau_A) \,,
\end{aligned}
\eeq
as well as the spin in a locally-flat frame,
\beq
\label{pmexp2}
\begin{aligned}
S^{ab}_A(\tau_A) &= {\cal S}_A^{ab} + \sum_n \delta^{(n)} S^{ab}_A(\tau_A) \,.\\
\end{aligned}
\eeq
The initial values, $\{ b_A^\mu, u_A^\mu ,{\cal S}_A^{ab}\}$, are related to the impact parameter, $b \equiv b_1-b_2$, incoming velocity and spin, respectively. Since the perturbation vanishes at infinity, we have $e^a_\mu \to \delta^a_\mu$. Hence, the locally-flat frame and Lorentzian one (where the initial spins and velocities are defined) coincide. This observation allows us to enforce the SSC in \eqref{ssc2} via the constraint
\beq
{\cal S}_{\mu\nu} u^\nu = 0\,,\label{ssc3}
\eeq
on the initial data, which is then preserved by the evolution equations in \eqref{eom}. The total momentum impulse is obtained as in \cite{pmeft}, but with a Routhian rather than a Lagrangian,
\beq
\Delta p^\mu_A = - \eta^{\mu\nu} \sum_n \int_{-\infty}^{+\infty} \dd\tau_A {\partial {\cR}_n  \over\partial x^\nu_A} \,.\label{spacetime}
\eeq 
Similarly to the non-spinning case, the {\it iterations} of the EoM on lower order contributions to the effective action play an important role \cite{pmeft}, and we have the  same type of decomposition
\beq
\Delta^{(n)} p^\mu_A = \sum_{k \leq n} \Delta^{(n)}_{\cR_{k}} p^\mu_A,\, \label{dp}
\eeq
at $n$PM order, with
\beq
\Delta^{(n)}_{\cR_k}\, p^\mu_A \equiv  -\eta^{\mu\nu}  \int_{-\infty}^{+\infty}  \dd\tau_A \left({\partial   \over\partial x^\nu_A} \cR_{k} \Bigg[b_{A(B)} + u_{A(B)} \tau_{A(B)}+ \sum_{r=0}^{n-k} \delta^{(r)} x_{A(B)}; \, {\cal S}^{ab}_{A(B)}+ \sum_{r=0}^{n-k} \delta^{(r)} S_{A(B)}^{ab}\Bigg]\right)_{{\cal O}(G^n)}\label{ogn}\,.
\eeq
Likewise, the total change of spin follows from
\beq
\Delta S_A^{ab} = \sum_n  \int_{-\infty}^{+\infty} d\tau_A \left\{ S_A^{ab},{\cal R}_n\right\}\,, 
\eeq
which must be evaluated iteratively on solutions to the EoM, yielding the same structure, i.e. $\Delta^{(n)}_{\cR_k}\, S^{ab}_A$, as in \eqref{ogn}. \vskip 4pt

It is somewhat convenient to re-write the final covariant expressions, obtained after using the EoM though \eqref{eom},  in terms of the initial Pauli-Lubanski vector, 
\beq
 {\cal S}^\mu_A = m_A a_A^\mu \equiv \frac{1}{2} {\epsilon^\mu}_{\nu\alpha\beta} {\cal S}_A^{\alpha\beta} u_A^{\nu}\,,\label{sa}
\eeq
where we take advantage of the fact that both the incoming velocity and spin tensor live in the same (inertial) frame. This observation drastically simplifies the handling of the SSC in the scattering problem. By considering only incoming/outgoing states in Minkowski space, the complexity due to the mismatch between the locally-flat and `PN frame' disappears. Furthermore, for the case of spins aligned with the angular momentum, it is easy to see that the motion remains in a plane, and we can compute the standard deflection angle, e.g.~\cite{pmeft},
\beq
2\sin\left(\frac{\chi}{2}\right) = \frac{\sqrt{-\Delta p_1^2}}{p_\infty}\,,\label{eq:apm}
\eeq
where the momentum at infinity, $p_\infty$, is given by
\beq
p_\infty = \mu \frac{\sqrt{\gamma^2-1}}{\Gamma} = \mu\, \hat p_\infty \,,\label{pinf}
\eeq
and 
\bea
\gamma &\equiv&  u_1\cdot u_2 \label{gamma}\,,\\
 \Gamma &\equiv& E/M = \sqrt{1+2\nu(\gamma-1)}\label{Egam}\,,
\eea
with $E$ the total energy in the CoM frame. Throughout the remaining of this paper we use the notation  $M=m_1+m_2$ for the total mass, $\mu = m_1m_2/M$ for the reduced mass, and $\nu \equiv \mu/M$ for the symmetric mass ratio. We also introduce the (reduced) binding energy, $\cE$, such that  \beq  E = M(1+\nu\cE)\,. \eeq 
 
\section{Aligned-spin Boundary-to-Bound correspondence} \label{sec:b2b}

In principle, the B2B dictionary with generic spins would require a map for non-planar motion. However, a major simplification arises for aligned-spin configurations, which we have shown in \cite{paper2}  is amenable to the same correspondence between the periastron advanced, $\Delta\Phi$, and scattering angle, $\chi$, 
\beq
\frac{\Delta\Phi(J,\cE)}{2\pi}=\frac{\chi(J,\cE)+\chi(-J,\cE)}{2\pi}\,, \qquad \cE<0 \,,
\label{eq:chiphiS}
\eeq
albeit with the {\it canonical} total angular momentum,~$\bJ \equiv \bL + \bS_1 + \bS_2$, as opposite to the orbital angular momentum, $\bL$, which enters in the non-spinning case. We review in what follows how to use \eqref{eq:chiphiS} to reconstruct the bound radial action from scattering data for aligned spins. For convenience, we will write various results in terms of the spin vector in \eqref{sa}, which for aligned spins obeys $a_A^\mu u_{A\mu}=a_A^\mu b_\mu =0$, and introduce the scalar variables $a_A \equiv \ba_A \cdot \bL$. Moreover, we often use the spin parameters  $a_\pm =  a_1 \pm  a_2$ for the two-body state, as well as the re-scaled variables $\tilde a_\pm \equiv a_\pm/(GM)$, $\ell \equiv L/GM\mu$  for the spin and orbital angular momentum.

\subsection{Bound radial action I: Angle}

As it was shown in \cite{paper2},  for the case of non-spinning bodies the relationship in \eqref{eq:chiphiS} allows us to construct the (reduced) radial action for the bound problem, $i_r(\cE,\ell)$,  in terms of the analytic continuation to negative binding energy of the PM coefficient of the scattering angle,
\beq
\frac{\chi}{2} = \sum_n\chi_b^{(n)}(\cE) \left(\frac{GM}{b}\right)^n  = \sum_n \frac{\chi^{(n)}_\ell(\cE) }{\ell^n} \,,\label{pmangle}
\eeq
yielding $\big(\text{with} \,\,{\rm sg}(\hat p_\infty) \equiv \hat p_\infty/\sqrt{-\hat p_\infty^2}\big)$
\beq
i_r(\cE,\ell) =  {\rm sg}(\hat p_\infty
)\chi^{(1)}_\ell(\cE) - \ell \left(1 + \frac{2}{\pi} \sum_{n=1}  \frac{\chi^{(2n)}_\ell({\cE})}{(1-2n)\ell^{2n}}\right)\quad\quad  \rm{(without~spin)\,.} \label{eq:ir}
\eeq

The expression in \eqref{eq:ir} does not translate directly to the spinning case. For starters, 
the expansion in \eqref{pmangle} gets modified into a two-scale expansion with spin effects, such that in addition to the standard factors of $GM/b$ we also have an expansion in $a_\pm/b$. This can be circumvented by the introduction of the dimensionless variables $\tilde a_\pm = a_\pm/(GM)$, which allows us to conveniently keep the same type of expansion as in \eqref{pmangle} (with $\chi^{(n)}_\ell (\cE,\tilde a_\pm)$ coefficients) at the expenses of a minor mismatch in the $G$ power-counting. Hence, using the fact that the relationship in \eqref{eq:chiphiS} involving both the orbital and spin angular momentum still applies, we can once again integrate with respect to $L$ and perform the same manipulations as in \cite{paper2} to obtain, after some trivial re-arrangement,
\beq
i_r(\cE,\ell,\tilde a_\pm) =  {\rm sg}(\hat p_\infty
)\chi^{(1)}_\ell(\cE) + \ell \left(-1 + \frac{2}{\pi} \sum^{\infty}_{n=1} \left(    \frac{\chi^{(2n+1)}_{\ell,{\rm odd}}({\cE},\tilde a_\pm)}{2n\, \ell^{2n+1}}+\frac{\chi^{(2n)}_{\ell,{\rm even}}({\cE},\tilde a_\pm)}{(2n-1)\ell^{2n}}\right)\right)\,. \label{eq:irs1}
\eeq
The $\chi^{(k)}_{\ell,{\rm odd(even)}}({\cE},\tilde a_\pm)$ are the odd (and even) contributions in the $\tilde a_\pm$ spin variables, analytically continued to negative binding energies. The expression in \eqref{eq:irs1} plays a similar role as \eqref{eq:ir}, with the addition of the odd contributions accounting for spin-orbit corrections. The even terms including not only spin-independent factors, but also effects quadratic in spin. Higher orders in spin follow the same pattern.\vskip 4pt

There is still an important caveat in the B2B dictionary for spinning bodies. The solution to the scattering problem produces results in an expansion in $GM/b$, with $b$ the {\it covariant} impact parameter, as in \eqref{pmangle}. However, for rotating bodies the latter is not directly related to $\ell$, the canonical orbital angular momentum. Instead we have \cite{justin1,justin2} 
\beq
\ell = \hat p_\infty \frac{b}{GM} +  \frac{\Gamma-1}{2\nu }\left(\tilde a_+ - \frac{\delta}{\Gamma} \tilde a_-\right)\,,\label{eqLb}
\eeq
where  $\delta \equiv \sqrt{1-4\nu}\, (m_1-m_2)/|m_1-m_2|.$ This introduces an additional expansion in $\tilde a_\pm/\ell$ once the scattering angle in \eqref{pmangle} is written in covariant form, mixing the power-counting. For example, it leads to spin-dependent contributions stemming off of the spin-independent deflection angle in impact-parameter space \cite{paper2}.

\subsection{Bound radial action II: Impetus}

As it was demonstrated in \cite{paper1,paper2}, the B2B dictionary relies on the connection between the orbital elements for hyperbolic- and elliptic-like motion \cite{paper1,paper2}. The orbital elements are obtained from the roots of the radial momentum, which can be solved as a function of the binding energy using a gauge where the (canonical) impetus takes the quasi-isotropic form
\beq
P_r^2 =  p^2_\infty \Bigg( 1 + \sum_{i=1}^{\infty} f_i(\cE, \ell \tilde a_\pm,\tilde a^2_\pm, \cdots) \frac{(GM)^i}{r^i}\Bigg)-\frac{L^2}{r^2}\,.\label{eq:Pr2}
\eeq
The expression in \eqref{eq:Pr2} also allows us to re-write the radial action in terms of the $f_i$'s, using the same algebraic relationships uncovered in \cite{paper1,paper2}. For instance, for the case of non-spinning bodies, the coefficients in the PM expansion of \eqref{pmangle} are related  to the CoM momentum in \eqref{eq:Pr2}, via \cite{paper1}
  \beq
\chi_\ell^{(n)} (\cE)= \frac{\sqrt{\pi}}{2} \hat \Gamma\left(\frac{n+1}{2}\right)\sum_{\sigma\in\mathcal{P}(n)}\frac{\hat p_\infty^n }{\Gamma\left(1+\frac{n}{2} -\Sigma^k\right)}\prod_{k} \frac{f_{\sigma_{k}}^{\sigma^{k}}(\cE)}{\sigma^{k}!}\quad\quad \rm{(without~spin)}\label{eq:fi2}\,,
\eeq
which follows from Firsov's solution to the scattering problem \cite{firsov}. (See \cite{paper1} for details on the combinatorial manipulations involved in \eqref{eq:fi2}.)  Using the expression in \eqref{eq:ir}, the relation in \eqref{eq:fi2} then leads to an alternative representation for the radial action ---  so far for the case of non-spinning bodies. However, as demonstrated in \cite{paper2} (see its Appendix A), the resulting form in terms of the $f_i$'s coincides with the PM expansion of the radial action that follows from the direct integration of the radial momentum, i.e. 
\beq i_r(\ell,\cE,\tilde a_\pm)= \frac{1}{2\pi GM\mu} \oint P_r(\ell,\cE,\tilde a_\pm)\, dr\,,\quad\quad ({\rm bound})\eeq 
using Sommerfeld's contour in the complex plane (originally performed to all orders in~\cite{paper1}). Hence, after noticing that the spin and angular momentum are simple spectators in all manipulations involving integration over the radial coordinate, it is straightforward to conclude that the general solution for the radial action in terms of the coefficients of the CoM momentum (in isotropic gauge) carries over unscathed onto the spinning case, obtaining \beq
\beal
\label{eq:irs2}
  i_r (\cE,\ell,\tilde a_\pm) = \frac{\hat p_\infty^2}{ \sqrt{-\hat p_\infty^2}} \frac{f_1}{2}
  + &\frac{\ell}{2\sqrt{\pi}} \sum_{n=0}^\infty  \left(\frac{\hat p_\infty}{\ell}\right)^{2n}
  \Gamma\left(\frac{2n-1}{2}\right) \quad\quad {\rm (with~spin)} \\ &\times \sum_{\sigma\in\mathcal{P}(2n)}\frac{1}{\Gamma\left(1+ n-\Sigma^{k}\right)}\prod_{k} \frac{f_{\sigma_{k}}^{\sigma^{k}}(\cE,\ell, \tilde a_\pm)}{\sigma^{k}!} \,,
  \eeal
\eeq
in terms of the coefficients in \eqref{eq:Pr2}. For instance, we have~\cite{paper1,paper2} \beq
\beal
\label{eq:irs6}
i_r (\cE,\ell,\tilde a_\pm) = -\ell &+ \frac{\hat p_\infty^2}{\sqrt{-\hat p_\infty^2}} \frac{f_1}{2} + \frac{\hat p_\infty^2}{2\ell} f_2 +  \frac{\hat p_\infty^4}{4\ell^3}\left(\frac{f_2^2}{2} +  f_1 f_3 +  f_4\right) \\
&+ \frac{\hat p_\infty^6}{16\ell^5} \Big( f_2^3 + 6(f_1f_3+f_4) f_2 + 3(f_4 f_1^2+2 f_5 f_1+f_3^2+f_6)\Big)  \\
&+ \frac{5\hat p_\infty^8}{128 \ell^7}\big( \cdots + 6 f_1^2 f_3^2  + \cdots \big) + \cdots \,,
\eeal
\eeq
where we kept only the one piece in the final term which will be needed later on. (The reader should keep in mind that the $f_i$'s themselves may also depend on $\ell$.)\vskip 4pt 
Notice that, similarly to the non-spinning case, the $f_k$'s contribute also at  $n$PM order (for $n\geq k$). This will allow us to perform a consistent PN-truncation, as discussed in \cite{paper1,paper2,pmeft,3pmeft,tidaleft}. We will return to this point in \S\ref{2pmo}.\vskip 4pt

\subsection{Impetus from angle}\label{secim}

It is useful to relate the PM coefficients in \eqref{eq:Pr2} to the deflection angle in \eqref{pmangle}, also with spin effects. This will allow us to perform the analytic continuation of the CoM momentum to negative binding energies, and also find a Hamiltonian if so desired. The main observation is the same we used to arrive at the equivalent representation for the radial action in \eqref{eq:irs2}. That is, spin and angular momentum are going for the ride when the radial action is constructed via the integral of the radial momentum, regardless of whether we consider bound or unbound orbits. Hence, the result of the integral
\beq i_r(\cE,\ell,\tilde a_\pm)= \frac{1}{2\pi GM\mu} \int_{-\infty}^\infty P_r(\cE,\ell,\tilde a_\pm)\, dr\quad\quad ({\rm unbound})\,,\eeq  
remains also the same, with the $f_i$'s in \eqref{eq:Pr2} including spin-dependent parts. Moreover, since the scattering angle obeys 
\beq
-\frac {\partial }{\partial \ell} i_r(\cE,\tilde a_\pm) = \frac{1}{2} +\frac{\chi(\ell,\cE, \tilde a_\pm)}{2\pi}\quad\quad ({\rm unbound})\,,\label{eq:chir}
\eeq
we can solve for the unbound radial action, which can then be written as
\beq
\beal
\label{eq:irs3}
  i_r (\ell,\cE,\tilde a_\pm) &=  -\frac{\ell}{2} - \chi^{(1)}_\ell(\cE) \frac{\log\ell}{\pi} - \frac{\ell}{\pi}\left\{ \sum_{n \geq 2}  \frac{\chi_\ell^{(n)}[f_i]}{(1-n)\ell^n} \right\}_{f_i \to f_i(\cE,\ell,\tilde a_\pm)} \quad\quad ({\rm unbound})\,,
  \eeal
\eeq
with the functional form of $\chi^{(n)}_\ell [f_i]$ given exactly by the expression in~\eqref{eq:fi2}, to all PM orders.\vskip 4pt Let us stress two related important points regarding  \eqref{eq:irs3}. First of all, there could be a constant of integration (depending only on the binding energy) as in the bound case \cite{paper1}. Moreover, the $n=1$ term ($\propto \log\ell$) is a bit subtle when considering the analytic continuation. The constant of integration may be fixed by using the expression in \eqref{eq:irs2} and~imposing 
\beq
i_r^{({\rm bound})}(\cE<0,\ell,\tilde a_\pm) = i_r^{({\rm unbound})}(\cE<0,\ell,\tilde a_\pm) - i_r^{({\rm unbound})}(\cE<0,-\ell,-\tilde a_\pm)\,, 
\eeq
which follows directly from the B2B relation in \eqref{eq:chiphiS}.\footnote{\label{foot}This analytic continuation is behind the relationship between the total radiated energy for unbound   orbits and the energy emitted over a period, discussed in \cite{bini3}. This relationship follows immediately from the B2B map applied to the (local part of the) conservative tail effect \cite{tail,tail2,tail3}.} However, this requires a choice for the branch of the logarithm, when performing the analytical continuation to negative orbital angular momentum. We find the choice $\log(\ell)/\pi - \log(-\ell)/\pi \to \mp i$, in combination with $\hat p_\infty \to \pm i \hat p_\infty$ for the analytic continuation in the binding energy, leads to
\beq
-\chi^{(1)}_\ell(\cE)\left(\frac{\log\ell}{\pi} -  \frac{\log(-\ell)}{\pi}\right) \to {\rm sg}(\hat p_\infty) \chi^{(1)}_\ell(\cE)\,,
\eeq
uniquely fixing the unbound radial action.\footnote{The choice must be uniformly adopted to be consistent with the analytic continuation in the (covariant) impact parameter, which was used in \cite{paper2} to connect the orbital elements.}\vskip 4pt 

For convenience, since we work here to quadratic order in the spins, in what follows we decompose the coefficients of the CoM momentum as
\beq
f_i(\cE,\ell, \tilde a_\pm)  = f^{0}_i(\cE) + \ell \sum_{A= \pm} \tilde a_A f_i^{A}(\cE) + \sum_{\{A,B\}=\pm} \tilde a_A \tilde a_B f^{AB}_i(\cE) + \cdots\,,\label{eq:Pr2n}
\eeq
with $f^{0}_i(\cE)$ the spin-independent part, and $\{f^{A}_i(\cE),f^{AB}_i(\cE)\}$  (dimensionless) functions of the binding energy (and masses). Likewise for the coefficients in \eqref{pmangle},
\beq
\chi^{(n)}_\ell(\cE,\tilde a_\pm) = \chi^{(n)}_{0}(\cE) +  \sum_{A= \pm} \tilde a_A  \chi^{(n)}_{A}(\cE) + \sum_{\{A,B\}=\pm} \tilde a_A \tilde a_B \, \chi^{(n)}_{AB}(\cE)+ \cdots\,\label{eq:angAB}
\eeq
(we suppress the $\ell$-subscript on the RHS for notational convenience). Hence, applying \eqref{eq:chir} to \eqref{eq:irs3}, while keeping track of all the $\ell$'s inside the $f_i$'s in \eqref{eq:Pr2n}, we can derive the scattering angle in terms of the CoM momentum including spin effects to all PM orders. As we will see momentarily,  terms linear and quadratic  in the spin first show up at $n=2$ and $n=3$ in \eqref{eq:angAB}, respectively. This is intuitively simple to understand, and it follows directly from the expansion in impact-parameter of the deflection angle yielding extra factors of $a_\pm/b$ once spin is included.\footnote{Notice the leading linear and quadratic terms, scaling as $(GM/b) (a_\pm/b)$ and $(GM/b) (a_\pm/b)^2$, have the wrong parity through the B2B map and therefore do not contribute to \eqref{eq:irs1}.} As a consequence,  \beq f_{1,2}^{A}(\cE)=f_{1,2}^{AB}(\cE)=0\,.\eeq
For the remaining coefficients, we find
\bea
 \chi^{(2)}_{A}&=& \frac{\hat p_\infty^3}{2}  f_3^{A}\label{2A} \,,\\
  \chi^{(3)}_{A} &=&  \frac{\pi \hat p_\infty^4}{4} \left(f_1^{0} f_3^{A}+ f_4^{A} \right) \,,\\
   \chi^{(3)}_{AB}  &=& \hat p_\infty^3  f_3^{(AB)} \,,\\
  \chi^{(4)}_{AB}  &=& \frac{3\pi \hat p_\infty^4}{8} \left( f_1^{(0)} f_3^{AB}+ f_4^{AB}+\frac{3 \hat p_\infty^2}{4} f_3^{A} f_3^{B} \right)\label{4AB}\,,
\eea
where (recall $\tilde a_\pm = a_\pm/(GM)$) we kept only terms which contribute to ${\cal O}(G^2)$. Incidentally, notice these values are consistent with the equivalence between the two representations of the radial action, in \eqref{eq:irs1} and \eqref{eq:irs2}-\eqref{eq:irs6}. It is  straightforward to invert these equations to obtain the value of the CoM impetus. See \S\ref{2pmo} for more details.

\section{Scattering to 2PM: linear and bilinear (generic) spin effects}  \label{2pms}

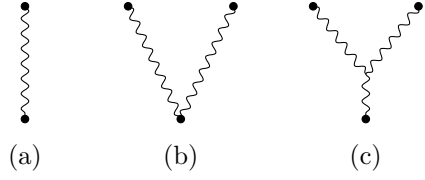
\begin{figure} 
  \centering
  \begin{subfigure}[b]{0.1\textwidth}
    \centering
    \begin{tikzpicture}
      [scale=1]
      \draw[boson] (0,0) -- (0,-1.5);
      \filldraw (0,0) circle (0.05);
      \filldraw (0,-1.5) circle (0.05);
    \end{tikzpicture}
    \caption{}
          \label{fig1a}
  \end{subfigure}
  \begin{subfigure}[b]{0.15\textwidth}
    \centering
    \begin{tikzpicture}
      [scale=1]
      \draw[boson] (-0.7,0) -- (0,-1.5);
      \draw[boson] (0.7,0) -- (0,-1.5);
      \filldraw (-0.7,0) circle (0.05);
      \filldraw (0.7,0) circle (0.05);
      \filldraw (0,-1.5) circle (0.05);
    \end{tikzpicture}
    \caption{}
          \label{fig1b}
  \end{subfigure}
  \begin{subfigure}[b]{0.15\textwidth}
    \centering
    \begin{tikzpicture}
      [scale=1]
      \draw[boson] (-0.7,0) -- (0,-0.9);
      \draw[boson] (0.7,0) -- (0,-0.9);
      \draw[boson] (0,-0.9) -- (0,-1.5);
      \filldraw (-0.7,0) circle (0.05);
      \filldraw (0.7,0) circle (0.05);
      \filldraw (0,-1.5) circle (0.05);
    \end{tikzpicture}
    \caption{}
      \label{fig1c}
  \end{subfigure}
  \caption{Feynman topologies needed to 2PM order. (See text.)}
\label{fig1}
\end{figure}

In this section we apply the EFT formalism for spinning bodies to compute the total momentum and spin impulses to 2PM and quadratic order in the spins. The needed topologies are shown in Fig.~\ref{fig1} to 2PM order. The vertices at the worldline may include mass and spin couplings, both linear (from the Ricci-rotation coefficients) and bilinear (from the SSC and finite-size terms) in the spins. Because the coupling to the mass is linear in the metric perturbation in our (Polyakov-type) gauge \cite{pmeft}, the diagram in Fig.~\ref{fig1b} only contributes spin-dependent effects.  The `tree-level' diagram in Fig.~\ref{fig1a} contributes both at 1PM and 2PM order, the latter through the iteration of the EoM described in \S\ref{sec:impulse}. Notice we need both spin-dependent iterations on the spin-independent tree-level and vice versa. The `one-loop' diagrams shown in Fig~\ref{fig1b} and \ref{fig1c} is evaluated on the unperturbed solutions. For the sake of comparison, in this section we quote the variation of the spin four-vector, defined as
\beq
  S^\mu_A \equiv \frac{1}{2 m_A} {\epsilon^\mu}_{\nu\alpha\beta}  S_A^{\alpha\beta} p_A^{\nu}\,,\label{san}
 \eeq
which coincides with the value in \eqref{sa} at early times. We obtain the spin impulse in terms of the total change in the spin tensor, using the spin algebra in \eqref{algebra} on the Routhian/action, in combination with the momentum impulse.  As a non-trivial check, the results below can be shown to be consistent with the preservation of the SSC, $S_\mu p^\mu=0$, the on-shell condition, $p^2=m^2$, and the constancy of the magnitude of the spin, $S_\mu S^\mu = a^2$, to 2PM order.\vskip 4pt

We illustrate the basic ideas and quote the results in what follows, with supplemental material in appendix~\ref{data}, and a few comments on the integration procedure in appendix~\ref{bubble}.   Throughout this section we use the notation $|b| \equiv \sqrt{-b^\mu b_\mu }$ and $\hat b^\mu  \equiv b^\mu /|b|$. Moreover, we also use $\kappa_\pm=C_{ES^2}^{(1)}\pm C_{ES^2}^{(2)}$, and the tensorial structure \cite{donalvines}
\begin{equation}
	\begin{aligned}
\Pi^{\mu}{ }_{\nu} &\equiv\epsilon^{\mu \rho \alpha \beta} \epsilon_{\nu \rho \gamma \delta} \frac{u_{1 \alpha} u_{2 \beta} u_{1}^{\gamma} u_{2}^{\delta}}{\gamma^{2}-1}\,,\\
T^{\mu \nu \rho}& \equiv \tensor{\hat{b}}{^{\rho }} \tensor{\Pi}{^{\mu }^{\nu }}+\tensor{\hat{b}}{^{\nu }} \tensor{\Pi}{^{\rho }^{\mu }}+\tensor{\hat{b}}{^{\mu }} \tensor{\Pi}{^{\rho }^{\nu }} \,,\\
u_{2(1) \perp}^{\mu} &\equiv u_{2(1)}^{\mu} - \gamma u_{1(2)}^{\mu}\,.
	\end{aligned}
\end{equation}

\subsection{Momentum impulse}

\subsubsection{Leading order}

The derivation of the tree-level Routhian/action is straightforward. Following the same steps as in \cite{pmeft} and evaluating on the unperturbed solutions in \eqref{pmexp1} and \eqref{pmexp2}, we obtain
\begin{equation}
\begin{aligned}
	\Delta^{(1)}_{a_{1}} p_{1}^{\mu} &=\frac{\nu G M^2}{|b|^2} \frac{4 \gamma }{ \sqrt{\gamma ^2-1}} \tensor{\epsilon}{_{\alpha }_{\rho }_{\beta }_{\sigma }}a_{1}^{\rho } u_1^{\beta} u_{2}^{\sigma } \left(\tensor{\Pi}{^{\mu }^{\alpha }}+2 \tensor{\hat{b}}{^{\mu }} \tensor{\hat{b}}{^{\alpha }}\right) - (1\leftrightarrow 2)\,,
	\label{eq:p_1_a1}
\end{aligned}
\end{equation}
for the spin-orbit contributions, whereas at quadratic order \begin{equation}
\begin{aligned}
	\Delta^{(1)}_{a_{1}a_{2}} p_{1}^{\mu} = \frac{\nu G M^2}{|b|^3}\frac{4 \left(2 \gamma ^2-1\right) }{\sqrt{\gamma ^2-1}}\tensor{a}{_1_{\alpha }} \tensor{a}{_2_{\beta }} \left(\tensor{T}{^{\alpha }^{\beta }^{\mu }}+4  \tensor{\hat{b}}{^{\alpha }}\tensor{\hat{b}}{^{\beta }} \tensor{\hat{b}}{^{\mu }} \right) - (1\leftrightarrow 2)\,,
	\label{eq:p_1_S1S2}
\end{aligned}
\end{equation}
and
\begin{equation}
\begin{aligned}
	\Delta^{(1)}_{a^2_{1}} p_{1}^{\mu} =  \frac{\nu G M^2}{|b|^3}\frac{2C_{ES^2}^{(1)}\left(2 \gamma ^2-1\right) }{\sqrt{\gamma ^2-1}} \tensor{a}{_1_{\alpha }} \tensor{a}{_1_{\beta }} \left(\tensor{T}{^{\alpha }^{\beta }^{\mu }}+4 \tensor{\hat{b}}{^{\alpha }} \tensor{\hat{b}}{^{\beta }} \tensor{\hat{b}}{^{\mu }}\right) - (1\leftrightarrow 2)\,,
	\label{eq:p_1_a1a1}
\end{aligned}
\end{equation}
the latter including the insertion of the finite-size term in \eqref{routhian}. In all of these expressions we have (anti-)symmetrized the result (under which $b^\mu \to -b^\mu$ and $\delta \to -\delta$). It is straightforward to show that all of these 1PM values coincide with the results reported in \cite{justin1,justin2,donalvines} for the case of Kerr black holes (with $C_{ES^2}=1$).

\subsubsection{Next-to-leading order}
For the NLO results we must evaluate the one-loop diagrams in Figs.~\ref{fig1b} \& \ref{fig1c} on the unperturbed solution, and use the trajectories (shown in Appendix~\ref{data}) to compute the iteration with the tree-level diagram in Fig~\ref{fig1a}. The results are:
\begin{equation}
\begin{aligned}
	\Delta^{(2)}_{a} p^{\mu}_{1} = &\frac{\nu G^2 M^3}{|b|^3} \left[D_{1}\tensor{\epsilon}{_{\alpha }_{\rho }_{\beta }_{\sigma }}a_{1}^{\rho } u_1^{\beta} u_{2}^{\sigma } \left(\tensor{\Pi}{^{\mu }^{\alpha }}+3 \tensor{\hat{b}}{^{\alpha }} \tensor{\hat{b}}{^{\mu }}\right) \right.\\
	&+ D_{2}\left.  \tensor{\epsilon}{^{\mu }^{\alpha }^{\rho }^{\beta }}\tensor{a}{_1_{\rho }}\tensor{u}{_1_{\beta }} \tensor{\hat{b}}{_{\alpha }} + \left(a_{1}^{\rho }  u_1^{\beta} u_{2}^{\sigma } \tensor{\hat{b}}{^{\alpha }}\tensor{\epsilon}{_{\alpha }_{\rho }_{\beta }_{\sigma }}\right) \left(D_{3}  u_{1}^{\mu }  + D_{4} u_{2}^{\mu} \right) \right]- (1\leftrightarrow 2)\,,
	\label{eq:p_2_S1}
\end{aligned}
\end{equation}
\begin{equation}
\begin{aligned}	
	\Delta^{(2)}_{a^2} p^{\mu}_{1} = &\frac{\nu G^2 M^3}{|b|^4}  \left[ D_{5}\tensor{a}{_1_{\alpha }} \tensor{a}{_1_{\beta }} \left(\tensor{T}{^{\alpha }^{\beta }^{\mu }}+5 \tensor{\hat{b}}{^{\alpha }} \tensor{\hat{b}}{^{\beta }}\tensor{\hat{b}}{^{\mu }} \right) + D_{6} \tensor{a}{_1_{\alpha }}(\tensor{a}{_1}\cdot \tensor{u}{_2})\left(\tensor{\Pi}{^{\alpha }^{\mu }}+4 \tensor{\hat{b}}{^{\alpha }} \tensor{\hat{b}}{^{\mu }}\right) \right.\\
	&+   \tensor{a}{_1_{\alpha }}  \tensor{a}{_1_{\beta }} \left(D_{7} u_{1}^{\mu } - D_{8} u_{2}^{\mu} \right)\left(\tensor{\Pi}{^{\alpha }^{\beta }}+4 \tensor{\hat{b}}{^{\alpha }} \tensor{\hat{b}}{^{\beta }}\right)+  \tensor{\hat{b}}{^{\mu }} \left( D_{9} a_1^2 + D_{10} (\tensor{a}{_1}\cdot \tensor{u}{_2})^2 \right)  \\
	&\left.+ 2D_{6} a_{1}^{\mu } (\tensor{a}{_1}\cdot \tensor{u}{_2})  + (\tensor{a}{_1}\cdot \tensor{u}{_2})^2 \left(D_{11} u_{1}^{\mu } + D_{12}u_{2}^{\mu}\right) - a_1^2 \left(D_{13} u_{1}^{\mu } - D_{14}u_{2}^{\mu}\right)\right] \\
	&- (1\leftrightarrow 2)\,,
	\label{eq:p_2_a1a1}
\end{aligned}	
\end{equation}

\begin{equation}
\begin{aligned}
	\Delta^{(2)}_{a_1a_2} p^{\mu}_{1} =& \frac{\nu G^2 M^3}{|b|^4}  \left[ \frac{1}{2}D_{15}\tensor{a}{_1_{\alpha }} \tensor{a}{_2_{\beta }}\left(\tensor{T}{^{\alpha }^{\beta }^{\mu }}+5 \tensor{\hat{b}}{^{\alpha }} \tensor{\hat{b}}{^{\beta }} \tensor{\hat{b}}{^{\mu }}\right) + D_{16}  \tensor{a}{_1_{\alpha }} (\tensor{a}{_2}\cdot \tensor{u}{_1})  \left(\tensor{\Pi}{^{\alpha }^{\mu }}+4 \tensor{\hat{b}}{^{\alpha }} \tensor{\hat{b}}{^{\mu }}\right) \right. \\
	& + D_{17}\tensor{a}{_1_{\alpha }} \tensor{a}{_2_{\beta }}  u_{1}^{\mu }\left(\tensor{\Pi}{^{\alpha }^{\beta }}+4 \tensor{\hat{b}}{^{\alpha }} \tensor{\hat{b}}{^{\beta }}\right) + \frac{1}{2}\tensor{\hat{b}}{^{\mu }}\left( D_{15}(\tensor{a}{_1}\cdot \tensor{a}{_2}) +D_{18} (\tensor{a}{_1}\cdot \tensor{u}{_2}) (\tensor{a}{_2}\cdot \tensor{u}{_1}) \right)  \\
	&\left. + 2D_{16} a_{1}^{\mu } (\tensor{a}{_2}\cdot \tensor{u}{_1}) + D_{19} (\tensor{a}{_1}\cdot \tensor{u}{_2}) (\tensor{a}{_2}\cdot \tensor{u}{_1})  u_{1}^{\mu} + D_{20}(\tensor{a}{_1}\cdot \tensor{a}{_2})  u_{1}^{\mu}  \right] - (1\leftrightarrow 2)\,.
\end{aligned}	
\end{equation}
The $D_i$ coefficients are displayed in Appendix \ref{data}. 

\subsection{Spin kick}

\subsubsection{Leading order}

We now move to the computation of the spin dynamics. As we discussed earlier, we quote the result in terms of the spin vector in \eqref{san}. We find,
\begin{equation}
	\begin{aligned}
		\Delta^{(1)}_{a} S_{1}^{\mu} = -\frac{\nu G  M^2}{|b|} \frac{2}{ \sqrt{\gamma ^2-1}}\left((\hat{b}\cdot \tensor{a}{_1}) (u_{1}^{\mu}-2 \gamma  u_2^{\mu})+2 \gamma  \tensor{\hat{b}}{^{\mu }} (\tensor{a}{_1}\cdot \tensor{u}{_2})\right)
	\end{aligned}
\end{equation}
at linear order in the spins, while at quadratic order we arrive at
\begin{equation}
	\begin{aligned}
		\Delta^{(1)}_{a_1a_2} S_{1}^{\mu} = \frac{\nu G  M^2}{|b|^2}\frac{2 }{ \sqrt{\gamma ^2-1}} \tensor{\epsilon}{^{\mu }_{\beta }_{\sigma }_{\rho }} &\left(\tensor{\Pi}{^{\alpha }^{\beta }}+2 \tensor{\hat{b}}{^{\alpha }} \tensor{\hat{b}}{^{\beta }}\right) \Big(  u_{1}^{\rho } u_{2}^{\sigma } \left(a_{2\alpha}(\tensor{a}{_1}\cdot \tensor{u}{_2})- \gamma \tensor{a}{_1_{\alpha }}(\tensor{a}{_2}\cdot \tensor{u}{_1})\right)  \\
		&+\gamma ^2 a_{1}^{\rho } a_{2\alpha} u_{1}^{\sigma } -\tensor{a}{_1_{\alpha }}a_{2}^{\rho } \left((\gamma ^2 - 1) u_{1}^{\sigma }+2 \gamma  u_{2\perp}^{\sigma }\right)\Big) 
	\end{aligned}
\end{equation}
\begin{equation}
	\begin{aligned}
		\Delta^{(1)}_{a^2} S_{1}^{\mu} = -\frac{\nu G  M^2}{|b|^2}\frac{2 }{ \sqrt{\gamma ^2-1}} &\left(\left(2 \gamma ^2-1\right) C_{ES^2}^{(1)}   u_{1}^{\rho } + 2 \gamma   u_{2\perp}^{\rho} \right) \tensor{\epsilon}{^{\mu }_{\beta }_{\sigma }_{\rho }}\tensor{a}{_1_{\alpha }} a_{1}^{\sigma } \left(\tensor{\Pi}{^{\alpha }^{\beta }}+2 \tensor{\hat{b}}{^{\alpha }} \tensor{\hat{b}}{^{\beta }}\right) \,,
	\end{aligned}
\end{equation}
These results are, once again, in agreement with the 1PM variation obtained in \cite{donalvines} for the case of Kerr black holes.

\subsubsection{Next-to-leading order}

The total change of spin at NLO is significantly more cumbersome. While, based on various arguments, we suspect an underlying structure that extends the compact expressions at 1PM order \cite{justin1}, we have not been able to uncover it so far. Yet, we believe these (manifestly covariant) expressions are perhaps the best hope to unravel a deeper (spacetime) structure. See \ref{disc} for more on this point. The results are:
\begin{equation}
	\begin{aligned}
		\Delta^{(2)}_{a} S_{1}^{\mu} = \frac{\nu G^2  M^3}{|b|^2} &\left[ D_{21}\tensor{a}{_1_{\alpha }} (\tensor{\Pi}{^{\mu }^{\alpha }}+2 \tensor{\hat{b}}{^{\alpha }} \tensor{\hat{b}}{^{\mu }}) -D_{21}a_{1}^{\mu } + D_{1}\left((\hat{b}\cdot \tensor{a}{_1}) u_{2\perp}^{\mu }-\tensor{\hat{b}}{^{\mu }} (\tensor{a}{_1}\cdot \tensor{u}{_2})\right)  \right.\\
		&\left.+ \left(D_{22}u_{1}^{\mu} + D_{23}\tensor{\hat{b}}{^{\mu }}\right) (\hat{b}\cdot \tensor{a}{_1})  + \left(D_{24} u_{1}^{\mu } + D_{25} u_{2\perp}^{\mu}\right) (\tensor{a}{_1}\cdot \tensor{u}{_2}) \right]
	\end{aligned}
\end{equation}

\begin{equation}
	\begin{aligned}
		\Delta^{(2)}_{a_1a_2} S_{1}^{\mu} = &\frac{\nu G^2  M^3}{|b|^2} \tensor{\epsilon}{^{\mu }_{\nu }_{\alpha }_{\beta }}\left[-D_{28}   u_1^{\nu}u_{2}^{\beta } \tensor{a}{_1_{\sigma }}\tensor{a}{_2_{\rho }}\left(\tensor{T}{^{\alpha }^{\rho }^{\sigma }}+4 \tensor{\hat{b}}{^{\alpha }} \tensor{\hat{b}}{^{\rho }} \tensor{\hat{b}}{^{\sigma }}\right)  \right.\\
		+& \left(\tensor{\Pi}{^{\alpha }^{\sigma }}+3 \tensor{\hat{b}}{^{\alpha }} \tensor{\hat{b}}{^{\sigma }} \right)\left(  D_{35}\left(u_{1}^{\beta } u_{2}^{\nu } (\tensor{a}{_2}\cdot \tensor{u}{_1})+a_{2}^{\nu } u_{2\perp}^{\beta}\right)  \right. \\
		&+\left.\left(D_{36}a_{1}^{\beta } \tensor{a}{_2_{\sigma }} u_{1}^{\nu } + D_{37}u_{1}^{\beta } u_{2}^{\nu }\left(\tensor{a}{_2_{\sigma }}  (\tensor{a}{_1}\cdot \tensor{u}{_2}) + \gamma \tensor{a}{_1_{\sigma }}  (\tensor{a}{_2}\cdot \tensor{u}{_1})\right) - D_{22} \tensor{a}{_1_{\sigma }} a_{2}^{\beta } u_{1}^{\nu }\right) \right) \\
		+& \left(\tensor{\Pi}{^{\sigma }^{\nu }}+2 \tensor{\hat{b}}{^{\sigma }} \tensor{\hat{b}}{^{\nu }}\right)   \left( -D_{27} \tensor{\hat{b}}{_{\sigma }}a_{2}^{\alpha } (\tensor{a}{_1}\cdot \tensor{u}{_2})\left( u_{1}^{\beta }-2 \gamma  u_{2}^{\beta } \right) -  D_{38}\tensor{\hat{b}}{^{\alpha }}  u_{1\perp}^{\beta}\tensor{a}{_1_{\sigma }}(\tensor{a}{_2}\cdot \tensor{u}{_1}) \right. \\
		&+ \tensor{\hat{b}}{_{\sigma }} u_{2}^{\beta }u_{1}^{\alpha } \left(D_{28} (\tensor{a}{_1}\cdot \tensor{a}{_2})-2 \gamma  D_{27}(\tensor{a}{_1}\cdot \tensor{u}{_2}) (\tensor{a}{_2}\cdot \tensor{u}{_1})\right)-D_{28} u_{1}^{\alpha } u_{2}^{\beta } \tensor{a}{_1_{\sigma }}(\hat{b}\cdot \tensor{a}{_2})\\
		&+ \left.  (\gamma ^2-1) D_{38}\tensor{a}{_1_{\sigma }} a_{2}^{\alpha } \tensor{\hat{b}}{^{\beta }} + \frac{1}{2} \gamma D_{28} a_{1}^{\alpha } \tensor{a}{_2_{\sigma }} \tensor{\hat{b}}{^{\beta }} + \frac{D_{28}}{2 \gamma }(\tensor{a}{_1}\cdot \tensor{u}{_2})\tensor{\hat{b}}{^{\alpha }} \tensor{a}{_2_{\sigma }} (\gamma  u_{1}^{\beta }+u_{2}^{\beta }) \right) \\
		-& D_{28}\left(\tensor{\Pi}{^{\sigma }^{\rho }}+2 \tensor{\hat{b}}{^{\rho }} \tensor{\hat{b}}{^{\sigma }}\right)\tensor{a}{_1_{\sigma }} \tensor{a}{_2_{\rho }} \tensor{\hat{b}}{^{\alpha }} u_{1}^{\beta } u_{2}^{\nu }  + D_{39}a_{1}^{\alpha } u_{1}^{\beta } u_{2}^{\nu } (\tensor{a}{_2}\cdot \hat{b}) - D_{28}a_{2}^{\alpha } u_{1}^{\beta } u_{2}^{\nu } (\tensor{a}{_1}\cdot \hat{b}) \\
		+& a_{1}^{\alpha } \tensor{\hat{b}}{^{\beta }}  (\tensor{a}{_2}\cdot \tensor{u}{_1}) \left(D_{16}u_{1}^{\nu } + D_{40}u_{2}^{\nu } \right) + \tensor{\hat{b}}{^{\alpha }} u_{1}^{\beta } u_{2}^{\nu }\left(D_{28}  (\tensor{a}{_1}\cdot \tensor{a}{_2}) + D_{41}  (\tensor{a}{_1}\cdot \tensor{u}{_2}) (\tensor{a}{_2}\cdot \tensor{u}{_1}) \right) \\
		+&  a_{2}^{\alpha } \tensor{\hat{b}}{^{\beta }} (\tensor{a}{_1}\cdot \tensor{u}{_2})\left(D_{42}u_{1}^{\nu } + D_{41}u_{2}^{\nu } \right)   + D_{37} a_{2}^{\alpha } u_{1}^{\beta } u_{2}^{\nu } (\tensor{a}{_1}\cdot \tensor{u}{_2}) + D_{43} a_{1}^{\alpha } u_{1}^{\beta } u_{2}^{\nu } (\tensor{a}{_2}\cdot \tensor{u}{_1}) \\
		+& \left. \gamma D_{40}a_{1}^{\alpha } a_{2}^{\beta } \tensor{\hat{b}}{^{\nu }} - D_{36}a_{1}^{\alpha } a_{2}^{\beta } u_{1}^{\nu }  \right]\,,
	\end{aligned}
\end{equation}
\begin{equation}
	\begin{aligned}
		\Delta^{(2)}_{a^2} S_{1}^{\mu} = &\frac{\nu G^2  M^3}{|b|^2} \tensor{\epsilon}{^{\mu }_{\nu }_{\alpha }_{\beta }}\left[  D_{26} \tensor{a}{_1_{\rho }} \tensor{a}{_1_{\sigma }} u_1^{\beta} u_2^{\nu}\left(\tensor{T}{^{\alpha }^{\rho }^{\sigma }}+4 \tensor{\hat{b}}{^{\alpha }} \tensor{\hat{b}}{^{\rho }} \tensor{\hat{b}}{^{\sigma }}\right)  \right.\\
		+& a_{1}^{\nu } \tensor{a}{_1_{\sigma }} \left(\frac{2}{3}D_{5} u_1^{\beta } + D_{1}u_{2\perp}^{\beta }\right) \left(\tensor{\Pi}{^{\alpha }^{\sigma }}+3 \tensor{\hat{b}}{^{\alpha }} \tensor{\hat{b}}{^{\sigma }}\right) \\
		+&\left(\tensor{\Pi}{^{\nu }^{\sigma }}+2 \tensor{\hat{b}}{^{\sigma }} \tensor{\hat{b}}{^{\nu }}\right)\left( D_{28}   u_{1}^{\alpha }u_{2}^{\beta } \left(a_1^2\tensor{\hat{b}}{_{\sigma }} - \tensor{a}{_1_{\sigma }} (\hat{b}\cdot \tensor{a}{_1})\right) \right.\\
		&+2 \gamma  D_{27} \tensor{\hat{b}}{^{\alpha }}\tensor{a}{_1_{\sigma }} (\tensor{a}{_1}\cdot \tensor{u}{_2}) \left(u_{2\perp}^{\beta }+ \left(\gamma ^2-1\right)u_{1}^{\beta }\right) -D_{27}(\tensor{a}{_1}\cdot \tensor{u}{_2})  a_1^{\alpha} \tensor{\hat{b}}{_{\sigma }} (u_{1}^{\beta }-2 \gamma  u_{2}^{\beta })  \\
		&\left.+D_{28}  (\tensor{a}{_1}\cdot \tensor{u}{_2})\tensor{a}{_1_{\sigma }} \tensor{\hat{b}}{^{\alpha }} u_{1}^{\beta } + D_{29}\tensor{\hat{b}}{^{\beta }}a_1^{\alpha} \tensor{a}{_1_{\sigma }}  \right) + D_{28} \tensor{\hat{b}}{^{\alpha }} u_{1}^{\nu } u_{2}^{\beta } \tensor{a}{_1_{\rho }} \tensor{a}{_1_{\sigma }}\left(\tensor{\Pi}{^{\sigma }^{\rho }}+2 \tensor{\hat{b}}{^{\rho }} \tensor{\hat{b}}{^{\sigma }}\right)\\
		+ &  \tensor{\hat{b}}{^{\nu }} u_{1}^{\alpha } u_{2}^{\beta } \left(D_{30} a_{1}^{2} + D_{31} (\tensor{a}{_1}\cdot \tensor{u}{_2})^2 \right) + D_{32}a_1^{\nu} u_{1}^{\alpha } u_{2}^{\beta } (\tensor{a}{_1}\cdot \hat{b})   \\
		+ & \left. a_1^{\nu} \tensor{\hat{b}}{^{\alpha }} (\tensor{a}{_1}\cdot \tensor{u}{_2}) \left(D_{33} u_{1}^{\beta} + D_{34} u_{2}^{\beta} \right) - \frac{2}{3}D_{10} a_1^{\alpha} u_1^{\beta} u_2^{\nu} (\tensor{a}{_1}\cdot \tensor{u}{_2})  \right]\,,
	\end{aligned}
\end{equation}
with the remaining $D_i$'s also collected in Appendix \ref{data}.

\subsection{Canonical variables}\label{canonical}

In order to apply the B2B dictionary we must also understand the map to canonical variables, in particular for the orbital angular momentum. This will be useful also to compare our results with the derivations in \cite{zvispin}, obtained directly in terms of canonical spins. Here we follow closely the analysis put forward in \cite{justin1,justin2} (see also \cite{nrgrso}), which we recommend for further details, while warning the reader to pay attention to the different conventions.\vskip 4pt

The canonical (or Newton-Wigner) spin constraints may be written with the aid of a background time-like four-vector, $U^\mu$, such that  the SSC becomes, in contrast to \eqref{ssc3},
\beq
{\cal S}_{\rm can}^{\mu\nu} (U_\nu + u_\nu )=0\,,
\eeq
for the (initial) spin and velocities. One can then search a transformation 
\beq
 {\cal S}_{\rm can}^{\mu\nu} =  {\cal S}^{\mu\nu} + m \, u^{[\mu} \delta x^{\nu]}\,,
\eeq
between covariant and canonical variables, with $ x^\mu_{\rm can} = x^\mu + \delta x^\mu$ (obeying $\delta x\cdot u =0$). We proceed as follows. Firstly, we split the velocity as
\beq
u^\mu = \hat E \, U^\mu + u_\perp^\mu\,,
\eeq
with $\hat E = E/m \equiv u\cdot U $, the body's (reduced) energy in the $U$-frame. Hence, introducing the canonical spin vector as
\beq
a^\mu_{\rm can} \equiv \frac{1}{2m} \epsilon^{\mu}_{\nu\alpha\beta} U^\nu {\cal S}^{\alpha\beta}_{\rm can}\,,
\eeq
we find 
\beq
a^\mu_{\rm can} = a^\mu + \frac{u_\perp\cdot a}{\hat E}  \left( U^\mu + \frac{u_\perp^\mu}{\hat E+1}\right)\,,\label{acan0}
\eeq
for the relationship to the covariant spin four-vector, and
\beq
\delta x^\mu = -\frac{1}{\hat E + 1} {\cal S}_{\rm can}^{\mu\alpha}\, u_{\perp \alpha}\,.  
\eeq
We now move to the two-body problem and the CoM frame, and choose the background four-vector as $U^\nu=\delta^\nu_0$. Hence, using the SSC for the covariant spin, we can re-write \eqref{acan0}
\beq
\beal
a^0_{A, \rm can} &= 0\,, \quad\quad
\ba_{A, \rm can} =  \ba_A - \frac{\bu_{A,\perp}\cdot \ba_A}{\hat E_A(\hat E_A+1)}  \bu_{A,\perp}\label{acan1}\,,
\eeal
\eeq
for each particle. For the sake of comparison, it is also convenient to invert the relationship,  
\beq
\beal
a^0_A &= \frac{\bu_{A,\perp}\cdot \ba_A}{\hat E_A} =  \bu_{A,\perp}\cdot \ba_{A, \rm can}\,,\\
\ba_A &=  \ba_{A, \rm can} + \frac{\bu_{A,\perp}\cdot \ba_{A, \rm can}}{(\hat E_A+1)}\,  \bu_{A,\perp}\,.\label{acan2}
\eeal
\eeq
where the velocity is given by  $\bu_{A,\perp} = (-1)^{A+1} \,\bp/m_A$ in the CoM frame.\vskip 4pt  Finally, using that $u_\perp\cdot U=0$, we also find 
\beq
\delta \bx_A =\frac{\bu_{A,\perp}\times \ba_{\rm can}}{(\hat E_A+1)} \to \bb = \bb_{\rm can}-\frac{\bp \times \bSi_{\rm can}}{(\hat E_A+1)}  \,,\label{shiftb}
\eeq
for the change of impact parameter between covariant and canonical coordinates, where we introduced the three-vector
\beq
\bSi \equiv  \sum_A  \frac{\ba_{A , \rm can}}{m_A (\hat E_A +1)}\,.
\eeq
From \eqref{acan1}-\eqref{acan2} it follows that the spin variables remain invariant for aligned spins, for which $\bp\cdot \ba =0$, and moreover do not evolve with time. In addition, the shift in \eqref{shiftb} yields the relation between covariant and canonical orbital angular momentum in \eqref{eqLb}, which is needed for the B2B map, we implement momentarily.  These transformations also allow us to compare the results reported in this paper and those in \cite{zvispin}. After applying \eqref{acan2}-\eqref{shiftb} to our results  (see also Eq.~(2.17) in~\cite{zvispin}) we find full agreement for the NLO spin-orbit and spin$_1$-spin$_2$ momentum impulse and spin kick.\footnote{Notice that the condition for the spin tensor/vector in Eq.~(2.18) of \cite{zvispin} has an overall minus sign,  however, they also use the (opposite) convention $\epsilon_{0123}=-1$.} 

\subsection{Aligned-spin scattering angle}

For the case of spins aligned with the orbital angular momentum, the motion remains in the plane,~and the following applies 
\begin{align}
  \epsilon_{\mu\nu\alpha\sigma} \hat{b}^{\mu} u_1^{\nu} u_2^{\alpha} a_A^{\sigma} = \sqrt{\gamma^2-1}\, a_A\,,
 \end{align} 
with the sign of $a_A \,(= \ba_A\cdot \bL)$ determined by the direction of the spin w.r.t. the orbital angular momentum. The scattering angle then follows from the total change of momentum in the CoM, see \eqref{eq:apm}. The result, to ${\cal O}(G^2)$ and quadratic order in the spins, reads
\bea
\label{chiaa2}
&&{\Delta_{(a,a^2)}\chi \over \Gamma}  = -{GM\over |b|}
\Bigg(
{4 \gamma \over \sqrt{\gamma^2 {-} 1}} {a_{+} \over |b|}
- {2\gamma^2 - 1 \over 2(\gamma^2 {-} 1)}\,
{(\kappa_{+} +2)\,a_{+}^2  +  (\kappa_{+}  - 2)\, a_{-}^2  +  2\kappa_{-}\, a_{-} a_{+}
\over |b|^2}
\Bigg)
\\
&&-\pi \left({GM\over |b|}\right)^2
\Bigg(
{\gamma\,(5 \gamma^2-3) \over 4(\gamma^2-1)^{3/2}}\, {7 a_{+} + \delta\, a_{-} \over |b|}
- {3 \over 256 (\gamma^2-1)^2}\,
{\lambda_{++}\,a_{+}^2 + \lambda_{--}\,a_{-}^2 + 2\lambda_{+-}\,a_{+}a_{-} \over |b|^2}
\Bigg)
\nonumber
\eea
including finite-size effects, with
\begin{align}\label{}
\begin{aligned}\label{}
\lambda_{++} &= 
830 \gamma^4-876 \gamma^2+110 
+ (35 \gamma^4-54 \gamma^2+19)\,\delta\,\kappa_{-} 
+ (215 \gamma^4-222 \gamma^2+39)\, \kappa_{+}, 
\\
\lambda_{--} &= -450 \gamma^4+468 \gamma^2 - 82
+ (35 \gamma^4-54 \gamma^2+19)\, \delta\,\kappa_{-} 
+ (215 \gamma^4-222 \gamma^2+39)\, \kappa_{+}, 
\\ 
\lambda_{+-} &= (215 \gamma^4-222 \gamma^2+39)\, \kappa_{-} 
+ (\gamma^2-1)   \big(70\gamma^2 +10 +(35 \gamma^2-19)\,\delta\,\kappa_{+}\big)\,.\label{lambda}
\end{aligned}
\end{align}
It is straightforward to show that, for Kerr black holes, we have
\begin{equation}
\begin{aligned}
	\frac{\Delta^{(1)}_{\rm Kerr}\,\chi}{\Gamma} &= -\frac{GM  }{|b|} \left(\frac{4 \gamma  }{ \sqrt{\gamma^2-1}} \frac{a_+}{|b|} - \frac{2\left(2 \gamma^2-1\right)}{ \left(\gamma^2-1\right)} \frac{a_+^2}{|b|^2}\right)+ {\cal O}(a^3),
\end{aligned}
\end{equation}
\bea
\frac{\Delta^{(2)}_{\rm Kerr}\, \chi}{\Gamma} &=& -\pi \frac{G^2 M^2 }{|b|^2} \left(\frac{  \gamma  (5 \gamma^2 {-} 3) }{4 (\gamma^2 {-} 1)^{3/2}} \left(\delta  \frac{a_-}{|b|}+7\frac{a_+}{|b|}\right)- \frac{3  }{64 (\gamma^2 {-} 1)^2} \left[14 \delta \left(5 \gamma^4-6 \gamma^2+1\right)   \frac{a_+ a_-}{|b|^2} \right.\right.\nn\\
&&-\left.\left(5 \gamma^4-6 \gamma^2+1\right) \frac{a_-^2}{|b|^2}+\left(315 \gamma^4-330 \gamma^2+47\right) \frac{a_+^2}{|b|^2}\right]\Bigg)+ {\cal O}(a^3)\,,
\eea
such that our result is consistent with \cite{justin1} and the conjecture~in~\cite{justin2}, respectively.\vskip 4pt

 In order to apply the B2B map, we must re-write the expression in \eqref{chiaa2} as an expansion in $\ell$ and $\tilde a_\pm$, to read off the relevant PM coefficients of the scattering angle. Using the decomposition in \eqref{eq:angAB}, we find the  following  values
\begingroup
\allowdisplaybreaks
\begin{align}
\label{eq1}
\chi^{(2)}_+ &=  {(2 \gamma^2-1) \over \sqrt{\gamma^2-1}}\, {\gamma-1 \over \Gamma +1} 
- {2 \gamma  \sqrt{\gamma^2-1} \over \Gamma},
\\[0.3 em]
\chi^{(2)}_- &= - {(2 \gamma^2-1) \over \sqrt{\gamma^2-1}  } {(\gamma-1)\,\delta \over \Gamma(\Gamma +1)},
\\[0.3 em]
\chi^{(3)}_+ &= \frac{\pi}{4} \bigg(\frac{3(\gamma-1)(5\gamma^2-1)}{\Gamma(\Gamma+1)}- \frac{7\gamma(5\gamma^2-3)}{2\Gamma^2} \bigg),
\\[0.3 em]
\chi^{(3)}_- &= -\frac{\pi \delta}{4} \bigg(\frac{3(\gamma-1)(5\gamma^2-1)}{\Gamma^2(\Gamma+1)}+ \frac{(5\gamma^2-3)\gamma}{2\Gamma^2}\bigg),
\\[0.3 em]
\chi^{(3)}_{++} &= \frac{1}{8\Gamma^2} \Bigg(\frac{8(\gamma {-} 1)^2(2\gamma^2 {-} 1)\,\Gamma^2}{\sqrt{\gamma^2 {-} 1}(\Gamma {+} 1)^2} - \frac{32\Gamma\,\gamma\,(\gamma {-} 1)\sqrt{\gamma^2 {-} 1}}{\Gamma+1}
+ 2\sqrt{\gamma^2 {-} 1}\,(2\gamma^2 {-} 1) (\kappa_{+} + 2)\Bigg),
\nonumber\\[0.3 em]
\chi^{(3)}_{--} &= \frac{1}{8\Gamma^2} \bigg(\frac{8(\gamma-1)^2(2\gamma^2-1)\delta^2}{\sqrt{\gamma^2-1}(\Gamma+1)^2}+ 2\sqrt{\gamma^2-1}(2\gamma^2-1) (\kappa_{+} - 2) \bigg),
\\[0.3 em]
\chi^{(3)}_{- +} &=\chi^{(3)}_{+-} = \frac{1}{\Gamma^2}\bigg(
{(2 \gamma^2 {-} 1) (\gamma {-} 1)^2\, \delta \over \sqrt{\gamma^2 {-} 1} (\Gamma +1)^2}
+\frac{(2 \gamma {+}1) (\gamma {-} 1)^2\, \delta }{\sqrt{\gamma^2 {-} 1} (\Gamma +1)}
+ {1 \over 4}\sqrt{\gamma^2 {-} 1}\, (2\gamma^2 {-} 1) \kappa_{-}
\bigg),
\\[0.3 em]
\chi^{(4)}_{++} &= \frac{3\pi}{512\Gamma^3} \bigg(\frac{192\Gamma^2(\gamma-1)^2(5\gamma^2-1)}{(\Gamma+1)^2} -\frac{448\Gamma\, \gamma(\gamma-1)(5\gamma^2-3)}{(\Gamma+1)}+ 830\gamma^4 -876\gamma^2+110
\nonumber \\ 
&\qquad\qquad\quad
+\delta\, \kappa_{-} (35\gamma^4-54\gamma^2+19)+ \kappa_{+} (215\gamma^4-222\gamma^2+39)\bigg),
\\[0.3 em]
\chi^{(4)}_{--} &= \frac{3\pi}{512\Gamma^3} \bigg(\frac{192\delta^2(\gamma-1)^2(5\gamma^2-1)}{(\Gamma+1)^2} 
+ \frac{64 \delta^2\, \gamma(\gamma-1)(5\gamma^2-3)}{(\Gamma+1)}- 450\gamma^4 +468\gamma^2-82
\nonumber\\ 
&\qquad\qquad\quad
+\delta\, \kappa_{-} (35\gamma^4-54\gamma^2+19)+ \kappa_{+} (215\gamma^4-222\gamma^2+39) \bigg),
\\[0.3 em]
\chi^{(4)}_{- +} &=\chi^{(4)}_{+-} = \frac{3\pi}{512\Gamma^3}
\bigg(
\frac{192 (5 \gamma^2-1) (\gamma -1)^2\,\delta }{(\Gamma +1)^2}+\frac{64 (5 \gamma^4+10 \gamma^3-24 \gamma^2+6 \gamma +3)\,\delta }{\Gamma +1}
\label{eq:fin}\\
&\quad
-2 (\gamma {-} 1) (45 \gamma^3 {-} 35 \gamma^2 {-} 53 \gamma {-} 5)\,\delta 
+ (215 \gamma^4 {-} 222 \gamma^2 {+} 39) \kappa_{-} + (\gamma^2 {-} 1) (35 \gamma^2 {-} 19)\,\delta \,\kappa_{+}
\bigg).
\nonumber
\end{align}
\endgroup

\section{Bound states to 2PM: linear and bilinear (aligned) spin effects} \label{2pmo}

Once the scattering angle is computed  the radial action for the bound problem follows, via analytic continuation. Crucially, the relationship between $b$ and $\ell$ in \eqref{eqLb} also introduces spin effects. This forces us to keep not only spin-dependent terms but also the spin-independent corrections, computed in \cite{pmeft}. From the radial action it is straightforward to derive the gravitational observables through differentiation. In what follows we illustrate the procedure to 2PM order. We also discuss how to incorporate the extra terms needed to complete the NLO contributions to the binding energy linear and bilinear in spin to 3PN order. Finally, we provide the coefficients in the CoM momentum to ${\cal O}(G^2)$, and all orders in velocity.

\subsection{Radial action} 

From the terms in \eqref{eq1}-\eqref{eq:fin},  only those which are even in the total angular momentum survive the B2B map, yielding for the (bound) radial action to 2PM order:
\beq
 i^{\rm 2PM}_r(\cE,\ell,\tilde a_\pm) =- \ell + \frac{2\gamma^2-1}{\sqrt{1-\gamma^2}} + \frac{3}{4\ell}\frac{5\gamma^2-1}{\Gamma} + \frac{1}{\pi} \sum_{A=\pm} \chi^{(3)}_{A}(\gamma)  \frac{\tilde a_A}{\ell^2}  
+\frac{2}{3\pi} \sum_{\{A,B\}=\pm} \chi^{(4)}_{AB}(\gamma)  \frac{\tilde a_A\tilde a_B}{\ell^3}\label{ir2pm}\,,
\eeq
after adding the results in \cite{pmeft} for the spin-independent terms. The reader will immediately notice that the analytic continuation to negative binding energies ($\gamma <1)$ follows smoothly. While the expression in \eqref{ir2pm} allows us to compute observables, incorporating an infinite series of velocity corrections at a given order in $1/\ell$, the fact that we truncate the radial action in the PM expansion prevents us from having direct access to the information needed to consistently obtain the PN effects associated with our PM results, for example the binding energy. This, however, is easily remediated by using the $f_n$'s in \eqref{eq:Pr2} and the representation in \eqref{eq:irs2}, yielding a consistent PN-truncation in higher orders terms. As in the non-spinning case, we found  in \S\ref{secim} that the coefficients of the CoM impetus can be obtained directly from the scattering angle, albeit in a more intricate fashion. We display their full expressions shortly. Once these are known, only the static limit is needed to add the terms required to complete the knowledge of the dynamics to NLO in the PN expansion. Using \eqref{eq:irs6}, we have
\beq
i_r^{\rm 2PM/3PN} = i_r^{\rm 2PM} +  \Delta i_r^{\rm 3PN}\label{ir3pn} \,,
\eeq
with 
\bea
 \Delta i_r^{\rm 3PN}&=& \left( \frac{1}{2\pi} \sum_{A=\pm} \chi^{(5)}_{A}(\gamma)  \frac{\tilde a_A}{\ell^4}  
+\frac{2}{5\pi} \sum_{\{A,B\}=\pm} \chi^{(6)}_{AB}(\gamma)  \frac{\tilde a_A\tilde a_B}{\ell^5}\right)_{\gamma \to 1}\\
 &=& \left(\frac{3\hat p_\infty^6}{16\ell^4} \sum_{A=\pm}\tilde a_A  ( 2 f^0_1f_2^0 f_3^A + (f^0_1)^2 f_4^A )  + \frac{3\hat p_\infty^6}{16\ell^5}  \sum_{\{A,B\}=\pm} \tilde a_A \tilde a_B \left( f_4^{AB} (f_1^0)^2+2 f_3^{AB}f_1^0 f_2^0 \right) \right.\nn \\
&&  \quad + \sum_{\{A,B\}=\pm} \frac{15\hat p_\infty^8}{64\ell^5} (f_1^0)^2 f_3^A f_3^B\tilde a_A \tilde a_B \Bigg)_{\gamma \to 1}\nn
\eea
where we only keep the leading PN corrections. As discussed in \cite{paper1,paper2,pmeft,3pmeft,tidaleft}, the remaining terms in \eqref{eq:irs6} have fewer factors of $f_i$'s, thus scaling with additional powers of $\hat p_\infty^2 \sim {\cal E}$ relative to the ones displayed, and therefore contributing at higher PN order. (This is a consequence of the fact that the impetus in the CoM has well-defined static limit, so that $f_i \sim 1/\hat p_\infty^2$.)

\subsection{Observables}

\subsubsection{Periastron Advance}
The periastron advance follows from the radial action via
\beq
\frac{\Delta \Phi}{2\pi} = -\frac{\partial}{\partial \ell} (i_r+\ell)\,.
\eeq
However, this is obviously equivalent to the condition in \eqref{eq:chiphiS}, which we used to build the radial action. Using \eqref{chiaa2}, translated to orbital angular momentum space, we find to ${\cal O}(G^2)$,
\begin{align}
{\Delta \Phi \over 2 \pi} =\,& {3 (5 \gamma^2 {-} 1) \over 4 \Gamma} {1\over \ell^2} 
+ \left[ \frac{6}{\Gamma+1} (5 \gamma^2 {-} 1) (\gamma {-}1) (\Gamma\tilde{a}_+-\delta  \tilde{a}_-) 
-\gamma (5 \gamma^2 {-} 3) (\delta  \tilde{a}_-+7 \tilde{a}_+) \right] {1 \over 4 \Gamma^2\ell^3}
\nonumber\\
&+ 
\bigg[ - {64 \gamma (5 \gamma^2 {-} 3) (\gamma {-}1) \over \Gamma+1}  (\delta  \tilde{a}_-+7 \tilde{a}_+) (\Gamma  \tilde{a}_+-\delta  \tilde{a}_-) 
\label{phil2}
\\
&
+ {192(5 \gamma^2 {-} 1) (\gamma {-}1)^2 \over (\Gamma+1)^2}  (\Gamma\tilde{a}_{+} -\delta\,\tilde{a}_{-})^2 
+ \lambda_{--}\,\tilde{a}_{-}^2 
+ 2\lambda_{+-}\, \tilde{a}_{+} \tilde{a}_{-} 
+ \lambda_{++}\, \tilde{a}_+^2  \bigg] {3 \over 256 \Gamma^3\ell^4}
+\cdots\,\,,
\nonumber
\end{align}
where the $\lambda_{AB}$ coefficients are given in \eqref{lambda}. In order to compare with the PN literature, we expand it in powers of $\epsilon \equiv -2{\cal E}$, yielding
\begin{align}
	\frac{\Delta \Phi(\ell, a, \epsilon)}{2 \pi} &= \left[3+\frac{3(2 \nu-5)}{4} \epsilon+\frac{3\left(5-5 \nu+4 \nu^{2}\right)}{16} \epsilon^{2}\right] \frac{1}{\ell^{2}}  \\
	&+\Bigg[-\frac{7 \tilde{a}_{+}+\delta \tilde{a}_{-}}{2}-\frac{(\nu-6) \delta \tilde{a}_{-}+(7 \nu-18) \tilde{a}_{+}}{2}\epsilon\nn  \\
	&\quad\quad -\left.\frac{3\left(\left(15-14 \nu+2 \nu^{2}\right) \delta \tilde{a}_{-}+\left(25-38 \nu+14 \nu^{2}\right) \tilde{a}_{+}\right)}{16} \epsilon^{2} \right] \frac{1}{\ell^{3}} \nn\\
	&+\Bigg[ \frac{3}{8} \Big(\tilde{a}_-^2 \left(\kappa_{+}-2\right)+2 \tilde{a}_+ \tilde{a}_- \kappa_{-}+\tilde{a}_+^2 \left(\kappa_{+}+2\right)\Big)\nn \\
	&\quad\quad -\left.\frac{3 }{16 }\epsilon\Big(\tilde{a}_-^2 \left(\delta  \kappa_{-}+\kappa_{+} (13-3 \nu )-2 \nu -25\right)+2 \tilde{a}_+ \tilde{a}_-  \left(\kappa_{-} (13-3 \nu )+\delta  \left(\kappa_{+}+11 \right) \right)\right.\nn\\
	&\quad\quad +\tilde{a}_+^2  \left(\delta  \kappa_{-}+\kappa_{+} (13-3 \nu )-6 \nu +35\right) \Big)  + \cdots \Bigg]\frac{1}{ \ell^{4}} + \cdots. \nn
	\label{eq:Delta_Phi}
\end{align}
The comparison with the PN result given in Eq. (33) of \cite{tessmer} thus gives perfect agreement,  including finite-size effects.\footnote{The error in \cite{tessmer}, first pointed out in \cite{paper2}, turns out to be a mere factor of $2$, when looking at their result with $C_Q \neq 1$. The correct result is given by replacing $(7-\frac{33\nu}{4}+3\nu^2) \to (7-\frac{33\nu}{2}+3\nu^2)$ in their Eq. (33).} Yet, the expression in \eqref{phil2} contains all orders in $\epsilon$, at ${\cal O}(a/\ell^3)$ and ${\cal O}(a^2/\ell^4)$, extending the results in \cite{paper2} at quadratic order~in~spins.

\subsubsection{Binding Energy}

The binding energy for circular orbits can be computed in different ways. One option is to get the value of the orbital angular momentum as a function of the energy, $\ell_c(\cE_c,a_\pm)$, from the condition $i_r(\ell_c(\cE_c,a_\pm),\cE_c,a_\pm) = 0$. (Alternatively we can use the condition  $r_+ = r_-$ for the roots of \eqref{eq:Pr2} in a circular orbits~\cite{paper1}.) From the orbital angular momentum we obtain the orbital frequency, $\Omega_c$, via the first-law~\cite{letiec}
\beq
x \equiv (GM \Omega_c)^{2/3} =\left( \frac{d \ell_c}{d{\cal E}_c} \right)^{-2/3}\,,
\eeq
 from which we obtain the relationship $\cE_c(x,a_\pm)$ for circular orbits. Using the expression in \eqref{ir3pn}, which includes a few terms at higher orders in $1/\ell$  needed to account for all the contributions to NLO in the PN expansion, we find (recall~$\epsilon = -2\cE$)
\begin{align}\label{}
\epsilon_c \,=\,  x &- \frac{x^2}{12} (\nu +9) 
+ x^{5/2}\bigg(
\frac{1}{3} (\delta  \tilde{a}_{-} + 7\tilde{a}_{+}) 
+\frac{x}{18} \Big[(99-61 \nu ) \tilde{a}_{+} - (\nu -45) \delta\,\tilde{a}_{-}\Big]
\bigg)
\\
&
+ {1 \over 6}\,x^{3}\Big[
- (\kappa_{+} + 2) \tilde{a}_{+}^2 - (\kappa_{+} - 2) \tilde{a}_{-}^2
- 2\kappa_{-}\,\tilde{a}_{-} \tilde{a}_{+} 
\Big]
\nonumber\\
&
+ {5 \over 72} x^{4} \Big[
\big(6  (\nu -5) \kappa_{-} - 4(3 \kappa_{+} +5) \delta\, \big)\tilde{a}_{-} \tilde{a}_{+}
+ \big(32-6 \delta  \kappa_{-}+10 \nu +3 (\nu -5) \kappa_{+}\big) \tilde{a}_{-}^2
\nonumber\\
&\qquad\qquad
+ \big(20-6 \delta  \kappa _{-}+6 \nu +3 (\nu -5) \kappa_{+}\big) \tilde{a}_{+}^2
\Big]
+ \cdots\,,
\nonumber
\end{align}
to 3PN order and quadratic in spins. After translating between various conventions, this agrees with the known value in the literature  \cite{Faye1,prl,Porto:2007px,nrgrso,nrgrss,nrgrs2,jan1,jan2,blanchet,bohe}.
 
\subsection{Center-of-Mass Momentum} 
The PM coefficients of the impetus in \eqref{eq:Pr2}, written using in the decomposition in \eqref{eq:Pr2n}, follow by inverting the relations in~\S\ref{secim}. We~obtain
\begin{align}
 f_3^{A} &= \frac{2 \chi^{(2)}_A }{\hat p_\infty^3}\,, \quad\quad
 f_3^{AB} = \frac{\chi^{(3)}_{AB} }{\hat p_\infty^3}\\ 
 f_4^{A} &= \frac{4}{\hat p_\infty^4}\left(\frac{\chi^{(3)}_A}{\pi}- \chi^{(1)}_0 \chi^{(2)}_A \right)\,,\quad
 f_4^{AB} =  \frac{2}{\hat p_\infty^4}\left(4 \frac{\chi^{(4)}_{AB}}{3\pi} - \chi^{(1)}_0 \chi^{(3)}_{AB}-\frac{3}{2}  \chi^{(2)}_A \chi^{(2)}_B\right)\,,
\end{align}
where we used $f_1^0=\frac{2}{\hat p_\infty} \chi^{(1)}_0 $ \cite{paper1}. Then, from the data collected in \eqref{eq1}-\eqref{eq:fin}, we  find  
\begingroup
\allowdisplaybreaks
\begin{align}\label{}
f_{3}^{(+)} =\,& -\frac{4 \gamma  \Gamma ^2}{\gamma ^2-1} + \frac{2 (\gamma -1) \left(2 \gamma ^2-1\right) \Gamma ^3}{\left(\gamma ^2-1\right)^2
   (\Gamma +1)},
\\[0.3 em]
f_{3}^{(-)} =\,& -\frac{2 \left(2 \gamma ^2-1\right) \Gamma ^2 \delta }{(\gamma -1) (\gamma +1)^2 (\Gamma+1)}, \\[0.3 em]
f_{3}^{(++)} =\,& \frac{\left(2 \gamma ^2-1\right) \Gamma ^3}{(\gamma +1)^2 (\Gamma +1)^2} -\frac{4 \gamma  \Gamma ^2}{(\gamma +1) (\Gamma +1)} + \frac{\left(2 \gamma ^2-1\right) \Gamma  \left(\kappa _++2\right)}{4 (\gamma^2 -1)},\\[0.3 em]
f_{3}^{(--)} =\,& \frac{\left(2 \gamma ^2-1\right) \Gamma  \delta ^2}{(\gamma +1)^2 (\Gamma +1)^2} + \frac{\left(2 \gamma ^2-1\right) \Gamma  \left(\kappa _+-2\right)}{4 \left(\gamma
   ^2-1\right)}
\\[0.3 em]
f_{3}^{(+-)} =\,& f_{3}^{(-+)} = \frac{2 \gamma  \Gamma  \delta }{(\gamma +1) (\Gamma +1)} -\frac{\left(2 \gamma ^2-1\right) \Gamma ^2 \delta }{(\gamma +1)^2 (\Gamma +1)^2} + \frac{\left(2 \gamma ^2-1\right) \Gamma  \kappa _-}{4 (\gamma^2 -1)},\\[0.3 em]
f_{4}^{(+)} =\,& \frac{8 \gamma  \left(2 \gamma ^2-1\right) \Gamma ^3}{(\gamma^2 -1)^2} -\frac{4 \left(2 \gamma ^2 -1\right)^2 \Gamma ^4}{(\gamma -1)^2 (\gamma +1)^3 (\Gamma +1)} -\frac{7 \gamma  \left(5 \gamma ^2-3\right) \Gamma ^2}{2 \left(\gamma ^2-1\right)^2}
\\
&
+\frac{3 \left(5 \gamma ^2-1\right) \Gamma ^3}{(\gamma -1) (\gamma +1)^2 (\Gamma +1)},
\nonumber\\[0.3 em]
f_{4}^{(-)} =\,&  \frac{4 \left(2 \gamma ^2 -1\right)^2 \Gamma ^3 \delta }{(\gamma -1)^2 (\gamma +1)^3 (\Gamma +1)} -\frac{\gamma  \left(5 \gamma ^2-3\right) \Gamma ^2 \delta }{2 \left(\gamma^2-1\right)^2} -\frac{3 \left(5 \gamma ^2-1\right) \Gamma ^2 \delta }{(\gamma -1) (\gamma +1)^2 (\Gamma +1)},
\\[0.3 em]
f_{4}^{(++)} =\,&   -\frac{5 \left(2 \gamma ^2 -1\right)^2 \Gamma ^4}{(\gamma -1) (\gamma +1)^3 (\Gamma +1)^2}+\frac{20 \gamma  \left(2 \gamma ^2-1\right) \Gamma ^3}{(\gamma -1)(\gamma +1)^2 (\Gamma +1)}+\frac{3 \left(5 \gamma ^2-1\right) \Gamma ^3}{(\gamma +1)^2 (\Gamma +1)^2}
\nonumber\\
&
-\frac{\left(2 \gamma ^2 -1\right)^2 \Gamma ^2 \left(\kappa
   _++2\right)}{2 \left(\gamma ^2-1\right)^2}-\frac{12 \gamma ^2 \Gamma ^2}{\gamma ^2-1}-\frac{7 \gamma  \left(5 \gamma ^2-3\right) \Gamma ^2}{(\gamma -1) (\gamma
   +1)^2 (\Gamma +1)}
\\
&
+\frac{\left(35 \gamma ^2-19\right) \Gamma  \delta  \kappa _-}{64 \left(\gamma ^2-1\right)}+\frac{\left(215 \gamma ^4-222 \gamma ^2+39\right)\Gamma  \kappa _+}{64 \left(\gamma ^2-1\right)^2}+\frac{\left(415 \gamma ^4-438 \gamma ^2+55\right) \Gamma }{32 \left(\gamma ^2-1\right)^2},
\nonumber\\[0.3 em]
f_{4}^{(--)} =\,&  -\frac{5 \left(2 \gamma ^2 - 1\right)^2 \Gamma ^2 \delta ^2}{(\gamma -1) (\gamma +1)^3 (\Gamma +1)^2}-\frac{\left(2 \gamma ^2 - 1\right)^2 \Gamma ^2 \left(\kappa_+-2\right)}{2 \left(\gamma ^2-1\right)^2}+\frac{\gamma  \left(5 \gamma ^2-3\right) \Gamma  \delta ^2}{(\gamma -1) (\gamma +1)^2 (\Gamma +1)}
\nonumber\\
&
+\frac{3 \left(5\gamma ^2-1\right) \Gamma  \delta ^2}{(\gamma +1)^2 (\Gamma +1)^2}+\frac{\left(35 \gamma ^2-19\right) \Gamma  \delta  \kappa _-}{64 \left(\gamma^2-1\right)}+\frac{\left(215 \gamma ^4-222 \gamma ^2+39\right) \Gamma  \kappa _+}{64 \left(\gamma ^2-1\right)^2}
\\
&
-\frac{\left(225 \gamma ^4-234 \gamma^2+41\right) \Gamma }{32 \left(\gamma ^2-1\right)^2},
\nonumber\\[0.3 em]
f_{4}^{(+-)} =\,& f_{4}^{(-+)} =  \frac{5 \left(2 \gamma ^2 - 1\right)^2 \Gamma ^3 \delta }{(\gamma -1) (\gamma +1)^3 (\Gamma +1)^2}-\frac{10 \gamma  \left(2 \gamma ^2-1\right) \Gamma ^2 \delta}{(\gamma -1) (\gamma +1)^2 (\Gamma +1)}-\frac{3 \left(5 \gamma ^2-1\right) \Gamma ^2 \delta }{(\gamma +1)^2 (\Gamma +1)^2}
\nonumber\\
&
-\frac{\left(2 \gamma ^2 - 1\right)^2\Gamma ^2 \kappa _-}{2 \left(\gamma ^2-1\right)^2}+\frac{\left(35 \gamma ^2-19\right) \Gamma  \delta  \kappa _+}{64 \left(\gamma ^2-1\right)}-\frac{\gamma  \left(5 \gamma ^2-3\right) (\Gamma -7) \Gamma  \delta }{2 (\gamma -1) (\gamma +1)^2 (\Gamma +1)}\\
&
+\frac{5 \Gamma  \left(7 \gamma ^2 +1 \right)\delta}{32 \left(\gamma ^2-1\right)}+\frac{\left(215 \gamma ^4-222 \gamma ^2+39\right) \Gamma  \kappa _-}{64 \left(\gamma ^2-1\right)^2}.
\nonumber
\end{align}
\endgroup
The expansion of the CoM momentum can then be analytically continued to negative binding energies ($\gamma < 1$), yielding a local effective description of the dynamics in a quasi-isotropic gauge, that may be used to construct the Hamiltonian. (Alternatively, one can derive it from the radial action using Delaunay variables, e.g.~\cite{bini2}.) However, as emphasized in \cite{paper1,paper2} and clearly shown in this paper, from the point of view of the scattering angle, radial action and B2B map (as well as the amplitude through the impetus formula \cite{paper1}) the $f_n$'s are the most natural variables. Therefore, they are preferable to the --- much more cumbersome --- PM~coefficients of the Hamiltonian, which we refrain from displaying here.

\section{Discussion \& Outlook}\label{disc}
Building on the EFT approach developed in \cite{nrgr,nrgrs,nrgrss,nrgrs2,nrgrso}, in this paper we have extended the PM framework introduced in \cite{pmeft} to incorporate spin effects. We then used the formalism to compute the NLO momentum and spin impulses with generic initial conditions and to quadratic order in the spins, including for the first time finite-size effects beyond leading~order. Afterwards we considered aligned-spin configurations and derived the scattering angle, which we used to construct the bound radial action via the B2B correspondence. The latter allows us to compute all the gravitational observables for elliptic-like orbits. As a notable example, we obtained the periastron advance to ${\cal O}(G^2)$ and all orders in velocity. We also computed the linear and bilinear in spin contributions to the binding energy for circular orbits to 3PN order. In~addition, we derived the CoM momentum (or impetus) in a quasi-isotropic gauge, from which one can readily obtain the EoM (or the Hamiltonian) to 2PM order, if so desired. Our results are in perfect agreement with the known literature, notably spin-orbit and spin$_1$-spin$_2$ effects to NLO in the PM expansion in \cite{zvispin}, while the spin$_{1(2)}$-spin$_{1(2)}$ contributions for generic compact bodies are computed here for the first time.  We also find agreement with the conjectured value for the 2PM aligned-spin scattering angle of Kerr black holes \cite{justin2}. \vskip 4pt In order to perform the comparison with the findings using scattering amplitudes in~\cite{zvispin}, we have translated our covariant results into canonical variables. The latter have the advantage of furnishing a canonical algebra involving only a spin three-vector, yet the Lorentz covariance of the results is hidden in somewhat cumbersome vectorial expressions. In contrast, in the former the results are not only manifestly covariant, by construction, but also remarkably compact when written in terms of four-dimensional vectors, as displayed here. In~fact, due to the conservation of the SSC ($S_\mu p^\mu =0$) upon evolution, both the spin and momentum {\it rotate} in spacetime in the same fashion \cite{justin1}. This implies the simple structure
\beq
\Delta p^\mu = m \,\delta\Lambda^\mu_\alpha\, u^\alpha\,,\,\,  \Delta S^\mu =m \,\delta\Lambda^\mu_\alpha \, a^\alpha\,,
\eeq 
with $\delta\Lambda_{\mu\nu} = - \delta\Lambda_{\nu\mu}$, must hold for both impulses. As shown in \cite{justin1}, the form of the $\delta\Lambda_{\mu\nu}$ matrix can be easily found at 1PM order in terms of a four-vector, $Z_\mu$, and the velocity $u_\mu$, obeying the condition $Z\cdot u = 0$. This representation is the basis for the $``b \to b+ia"$ shift which lies at the heart of the (complex) transformation introduced in \cite{nj} connecting Schwarzschild and Kerr solutions, (see also~\cite{Arkani-Hamed:2019ymq}).\footnote{This transformation (applied to perturbations of the background) may also play a key role in understanding the vanishing of (static) tidal response for rotating black holes \cite{Chia:2020dye,LeTiec:2020bos,dis3,Charalambous:2021mea} following the pattern observed in Schwarzschild \cite{Binnington:2009bb,Damour:2009vw,Hui:2020xxx} (\scriptsize \url{http://www.youtube.com/watch?v=v9fvAohXD8g&t=45m15s}).} At NLO, however, the construction of the $\delta\Lambda_{\mu\nu}$ matrix is less straightforward, mainly due to the new directions in which a non-zero impulse appears. For the case of non-spinning bodies, the task is relatively simple and the concurrent rotation can be easily written down incorporating the impulse in the $u^\mu$ direction. However, when spin effects are included, the form of the transformation turns out to be much more involved, begging instead for a more convenient basis to decompose the impulses. Without spin, such basis exists, by combining the $b^\mu$ and $u^\mu$ impulses into a space-like vector, and it is directly connected to the eikonal representation. It would be interesting to perform the same manipulations for the case of spinning bodies, in particular in light of the remarkable algebraic structure involving the eikonal phase recently discovered in \cite{zvispin}.\vskip 4pt

Another interesting area for further study is the possibility, for the special case of Kerr black holes, to promote the worldline effective theory into a worldsheet, as advocated in \cite{spinsheet}. The motivation is also built on the $``x+ia"$ shift \cite{nj}, which suggests extending the worldline action into one more dimension. This is not entirely surprising, after all the covariant SSC implies a non-commutative (Dirac) algebra for the position, $\{\bx^i,\bx^j\}_{\rm DB} \simeq {\epsilon}^{ijk}\ba^k/m$, e.g.~\cite{hanson,nrgrs}, hinting at the {\it extendedness} of spinning particles. The worldsheet idea is also rooted on the fact that, not only the quadrupole \cite{nrgrss}, but all of the worldline (Wilson) coefficients obey $C_{ES^{2n}} = 1$ (and similarly with the magnetic terms) when matching to the multipole moments of a Kerr black hole \cite{spinsheet}. This observation allows one to resume all the derivatively coupled higher-derivative terms in the effective action, which then exponentiate into a translation operation, $e^{i a \cdot \partial}$, that is directly linked to a complex coordinate shift. However, because of the equivalence principle, finite-size effects start with two derivatives (with the $1$ and $a\partial/2$ terms already appropriated by the mass and spin) yielding the structure $(e^x-1)/x = \int_0^1 d\lambda e^{\lambda x}$, which naturally allows one to introduce a two-dimensional integral for the effective action. Furthermore, it turns out to be natural to introduce a spinor-helicity representation~\cite{spinsheet} (see \cite{Berezin:1976eg} for other possible routes). Hence, armed with an action incorporating all of the (self-induced) finite-size effects at once, we could then set up the EFT formalism described in this paper, uplifted to a worldsheet Routhian, to compute the momentum and spin impulses for Kerr black holes without having to introduce the curvature terms representing finite-size effects. This possibility is currently under investigation.\vskip 4pt

Yet one more aspect of the framework which deserves more attention is the generalization of the B2B correspondence to the case of non-aligned spins. As shown in \cite{paper2}, the planar B2B map can be extended to spinning bodies by performing an analytic continuation in the total (canonical) angular momentum. However, when $\dot \bL = -\dot \bS \neq 0$, the precession of the plane complicates matters. Moreover, only canonical variables may be associated through the B2B dictionary, which requires also transforming the spin variables when we allow for non-planar dynamics. In principle, we could consider the periastron shift in the instantaneous plane or, more likely, orbital averages over a period. For instance, because the orbital elements are related via analytic continuation --- with aligned or zero spins --- the B2B map yields a relationship between the total radiated energy in a scattering process and the integrated power over an orbit, via analytic continuation (see footnote \ref{foot}). Hence, we may expect a similar situation once we include the precession of the orbital plane, thus retaining a link between the total impulse in momentum and an {\it averaged} periastron advance. There is as well the total spin kick to be considered, and the associated change in orbital angular momentum. Provided a relationship between the orbital elements still holds, we may expect an analogous connection to the integrated change over an orbit for the bound case. Another interesting venue is to explore the modification of the first-law to the case of spinning bodies, e.g.~\cite{letiecs}, which would allow us to compute the precession frequency once the non-planar B2B map is obtained. We will return to these issues in future work. Finally, since the master integrals to N$^2$LO are known \cite{3pmeft}, the formalism is ready to march forward in the PM expansion. The~computation of spin effects to ${\cal O}(G^3)$ is currently underway.\vskip 8pt
 
{\it Note added:} While the results of this project were prepared for submission we learned of the concurrent work of \cite{andres2}, which also computed finite-size contributions to the NLO impulses via the methods discussed in \cite{zvispin}. After transforming between the different variables (as discussed in \S\ref{canonical}) their results agree with ours. We thank the authors of \cite{andres2} for confirming the perfect match with us.

\vskip 8pt
{\bf Acknowledgements}. We thank Gregor K\"alin, and Justin Vines for helpful discussions. We also thank Andres Luna and Chia-Hsien Shen for useful exchanges on the  results~in~\cite{zvispin}. This work is supported by the ERC Consolidator Grant ``Precision Gravity: From the LHC~to LISA,"  provided by the European Research Council (ERC) under the European Union's H2020 research and innovation programme, grant No.\,817791. Our work is also supported by the DFG under Germany's Excellence Strategy `Quantum Universe' (No.\,390833306). We~acknowledge extensive use of the \texttt{xAct} packages (\url{www.xact.es}). 
\appendix
\section{Supplemental Material}\label{data}

\subsection{Trajectories}

An important element in the computation of the NLO impulses is iteration of the 1PM EoM into the tree-level Routhian. For completeness, we provide here the trajectories for the position and spin to 1PM order. To simplify the notation we always refer to the dynamics of particle 1 and do not include the mirror images. The contribution from the latter, as well as the corresponding deflection for the second particle, can be derived as explained in \cite{pmeft}.

\subsubsection{Velocity \& Position}

 The velocity correction at linear order in spin is given by,
\begin{equation}
	\begin{aligned}
	\delta^{(1)}_{S_{1}} v^{\mu}_{1}(\tau_{1}) = \frac{-i m_{2}}{8M^2_{\text{Pl}}m_{1}}\int_{k} & \frac{  \hat{\delta}\left(k \cdot u_{2}\right)}{k^{2}\left(k \cdot u_{1}-i \epsilon\right)} e^{i k \cdot b} e^{i\left(k \cdot u_{1}-i \epsilon\right) \tau_{1}}\\
		&\times\tensor{k}{^{\alpha }} \left(\tensor{k}{^{\nu }} \tensor{u}{_1_{\alpha }} \left(\tensor{\mathcal{S}}{_1^{\mu }_{\nu }}+2 \tensor{u}{_2^{\beta }} \tensor{u}{_2^{\mu }} \tensor{\mathcal{S}}{_1_{\nu }_{\beta }}\right)-2 \gamma  \tensor{k}{^{\mu }} \tensor{u}{_2^{\nu }} \tensor{\mathcal{S}}{_1_{\alpha }_{\nu }}\right),  	
		\end{aligned}
\end{equation}
and one time integration yields the position correction
\begin{equation}
	\begin{aligned}
	\delta^{(1)}_{S_{1}} x^{\mu}_{1}(\tau_{1}) = \frac{- m_{2}}{8M^2_{\text{Pl}}m_{1}}\int_{k} & \frac{  \hat{\delta}\left(k \cdot u_{2}\right)}{k^{2}\left(k \cdot u_{1}-i \epsilon\right)^2} e^{i k \cdot b} e^{i\left(k \cdot u_{1}-i \epsilon\right) \tau_{1}}\\
		&\times\tensor{k}{^{\alpha }} \left(\tensor{k}{^{\nu }} \tensor{u}{_1_{\alpha }} \left(\tensor{\mathcal{S}}{_1^{\mu }_{\nu }}+2 \tensor{u}{_2^{\beta }} \tensor{u}{_2^{\mu }} \tensor{\mathcal{S}}{_1_{\nu }_{\beta }}\right)-2 \gamma  \tensor{k}{^{\mu }} \tensor{u}{_2^{\nu }} \tensor{\mathcal{S}}{_1_{\alpha }_{\nu }}\right).  	
		\end{aligned}
\end{equation}

At quadratic order in the spin we have several contributions. For the term proportional to the SSC in \eqref{routhian} we find
\begin{equation}
	\begin{aligned}
	\delta^{(1)}_{RS_1S_1} v^{\mu}_{1}(\tau_{1}) = \frac{-m_{2}}{8M^2_{\text{Pl}}m_{1}^2}\int_{k} & \frac{  \hat{\delta}\left(k \cdot u_{2}\right)}{k^{2}} e^{i k \cdot b} e^{i\left(k \cdot u_{1}-i \epsilon\right) \tau_{1}}\left[\tensor{k}{^{\nu }}(k\cdot \tensor{u}{_1}) \tensor{\mathcal{S}}{_1^{\mu }^{\beta }} \tensor{\mathcal{S}}{_1_{\nu }_{\beta }} \right.\\
		& \left.+2 \gamma  \tensor{k}{^{\nu }}\tensor{\mathcal{S}}{_1^{\mu }_{\nu }} (\tensor{k}{^{\alpha }} \tensor{u}{_2^{\beta }} \tensor{\mathcal{S}}{_1_{\alpha }_{\beta }})-2 \tensor{u}{_2^{\nu }} (k\cdot \tensor{u}{_1}) \tensor{\mathcal{S}}{_1^{\mu }_{\nu }} (\tensor{k}{^{\alpha }} \tensor{u}{_2^{\beta }} \tensor{\mathcal{S}}{_1_{\alpha }_{\beta }}) \right],
		\end{aligned}
\end{equation}
and
\begin{equation}
	\begin{aligned}
	\delta^{(1)}_{RS_1S_1} x^{\mu}_{1}(\tau_{1}) = \frac{i m_{2}}{8M^2_{\text{Pl}}m_{1}^2}\int_{k} & \frac{  \hat{\delta}\left(k \cdot u_{2}\right)}{k^{2}\left(k \cdot u_{1}-i \epsilon\right)} e^{i k \cdot b} e^{i\left(k \cdot u_{1}-i \epsilon\right) \tau_{1}}\left[\tensor{k}{^{\nu }}(k\cdot \tensor{u}{_1}) \tensor{\mathcal{S}}{_1^{\mu }^{\beta }} \tensor{\mathcal{S}}{_1_{\nu }_{\beta }} \right.\\
		& \left.+2 \gamma  \tensor{k}{^{\nu }}\tensor{\mathcal{S}}{_1^{\mu }_{\nu }} (\tensor{k}{^{\alpha }} \tensor{u}{_2^{\beta }} \tensor{\mathcal{S}}{_1_{\alpha }_{\beta }})-2 \tensor{u}{_2^{\nu }} (k\cdot \tensor{u}{_1}) \tensor{\mathcal{S}}{_1^{\mu }_{\nu }} (\tensor{k}{^{\alpha }} \tensor{u}{_2^{\beta }} \tensor{\mathcal{S}}{_1_{\alpha }_{\beta }}) \right],
		\end{aligned}
\end{equation}
whereas from the finite-size effects we arrive at
\begin{align}
	\delta^{(1)}_{ES_1^2} v^{\mu}_{1}(\tau_{1}) =\,& \frac{m_{2}}{16M^2_{\text{Pl}}m_{1}^2}\int_{k}\frac{  \hat{\delta}\left(k \cdot u_{2}\right)\tensor{k}{^{\alpha }} \tensor{k}{^{\nu }} }{k^{2}\left(k \cdot u_{1}-i \epsilon\right)} e^{i k \cdot b} e^{i\left(k \cdot u_{1}-i \epsilon\right) \tau_{1}}\\
	&\times\left[\tensor{k}{^{\mu }} \left(2 \gamma ^2-1\right) \tensor{\mathcal{S}}{_1_{\alpha }^{\beta }} \tensor{\mathcal{S}}{_1_{\nu }_{\beta }} +\tensor{k}{^{\mu }}\tensor{u}{_1_{\alpha }} \tensor{u}{_1_{\nu }} (\tensor{\mathcal{S}}{_1_{\beta }_{\rho }} \tensor{\mathcal{S}}{_1^{\beta }^{\rho }}-2 \tensor{u}{_2^{\beta }} \tensor{u}{_2^{\rho }} \tensor{\mathcal{S}}{_1_{\beta }^{\gamma }} \tensor{\mathcal{S}}{_1_{\rho }_{\gamma }}) \right. \nonumber\\
		&\quad \left. +2 \tensor{k}{^{\beta }} \tensor{u}{_1_{\alpha }} (\tensor{\mathcal{S}}{_1_{\beta }_{\rho }} \tensor{\mathcal{S}}{_1_{\nu }^{\rho }} (\tensor{u}{_1^{\mu }}-2 \gamma  \tensor{u}{_2^{\mu }})+\tensor{u}{_1_{\nu }} (2 \tensor{u}{_2^{\mu }} \tensor{u}{_2^{\rho }} \tensor{\mathcal{S}}{_1_{\beta }^{\gamma }} \tensor{\mathcal{S}}{_1_{\rho }_{\gamma }}-\tensor{\mathcal{S}}{_1_{\beta }_{\rho }} \tensor{\mathcal{S}}{_1^{\mu }^{\rho }}))\right]
\nonumber
\end{align}
so that 
\begin{align}
	\delta^{(1)}_{ES_1^2} x^{\mu}_{1}(\tau_{1}) =\,& \frac{-im_{2}}{16M^2_{\text{Pl}}m_{1}^2}\int_{k}\frac{  \hat{\delta}\left(k \cdot u_{2}\right)\tensor{k}{^{\alpha }} \tensor{k}{^{\nu }} }{k^{2}\left(k \cdot u_{1}-i \epsilon\right)^2} e^{i k \cdot b} e^{i\left(k \cdot u_{1}-i \epsilon\right) \tau_{1}}\\
	&\times\left[\tensor{k}{^{\mu }} \left(2 \gamma ^2-1\right) \tensor{\mathcal{S}}{_1_{\alpha }^{\beta }} \tensor{\mathcal{S}}{_1_{\nu }_{\beta }} +\tensor{k}{^{\mu }}\tensor{u}{_1_{\alpha }} \tensor{u}{_1_{\nu }} (\tensor{\mathcal{S}}{_1_{\beta }_{\rho }} \tensor{\mathcal{S}}{_1^{\beta }^{\rho }}-2 \tensor{u}{_2^{\beta }} \tensor{u}{_2^{\rho }} \tensor{\mathcal{S}}{_1_{\beta }^{\gamma }} \tensor{\mathcal{S}}{_1_{\rho }_{\gamma }}) \right. 
\nonumber\\
		&\quad \left. +2 \tensor{k}{^{\beta }} \tensor{u}{_1_{\alpha }} (\tensor{\mathcal{S}}{_1_{\beta }_{\rho }} \tensor{\mathcal{S}}{_1_{\nu }^{\rho }} (\tensor{u}{_1^{\mu }}-2 \gamma  \tensor{u}{_2^{\mu }})+\tensor{u}{_1_{\nu }} (2 \tensor{u}{_2^{\mu }} \tensor{u}{_2^{\rho }} \tensor{\mathcal{S}}{_1_{\beta }^{\gamma }} \tensor{\mathcal{S}}{_1_{\rho }_{\gamma }}-\tensor{\mathcal{S}}{_1_{\beta }_{\rho }} \tensor{\mathcal{S}}{_1^{\mu }^{\rho }}))\right].
\nonumber
\end{align}

Finally, we also have spin$_{1}$-spin$_{2}$ corrections to the trajectories given by
\begin{align}
	\delta^{(1)}_{S_{1}S_{2}} v^{\mu}_{1}(\tau_{1}) =\,& \frac{1}{8M^2_{\text{Pl}}m_{1}}\int_{k}\frac{  \hat{\delta}\left(k \cdot u_{2}\right) \tensor{k}{^{\alpha }} \tensor{k}{^{\nu }} }{k^{2}\left(k \cdot u_{1}-i \epsilon\right)} e^{i k \cdot b} e^{i\left(k \cdot u_{1}-i \epsilon\right) \tau_{1}}\\
		&\times \left[  \tensor{k}{^{\mu }} \tensor{\mathcal{S}}{_2_{\nu }_{\beta }} \left(\gamma  \tensor{\mathcal{S}}{_1_{\alpha }^{\beta }}+\tensor{u}{_1^{\beta }} \tensor{u}{_2^{\rho }} \tensor{\mathcal{S}}{_1_{\alpha }_{\rho }}\right)+\tensor{k}{^{\beta }} \tensor{u}{_1_{\alpha }} \left(\tensor{u}{_2^{\rho }} \tensor{\mathcal{S}}{_1_{\nu }_{\rho }} \tensor{\mathcal{S}}{_2^{\mu }_{\beta }}-\tensor{u}{_2^{\mu }} \tensor{\mathcal{S}}{_1_{\nu }^{\rho }} \tensor{\mathcal{S}}{_2_{\beta }_{\rho }}\right)\right],
\nonumber
\end{align}
and
\begin{align}
	\delta^{(1)}_{S_{1}S_{2}} x^{\mu}_{1}(\tau_{1}) =\,& \frac{-i}{8M^2_{\text{Pl}}m_{1}}\int_{k}\frac{  \hat{\delta}\left(k \cdot u_{2}\right) \tensor{k}{^{\alpha }} \tensor{k}{^{\nu }} }{k^{2}\left(k \cdot u_{1}-i \epsilon\right)^2} e^{i k \cdot b} e^{i\left(k \cdot u_{1}-i \epsilon\right) \tau_{1}}\\
		&\times \left[  \tensor{k}{^{\mu }} \tensor{\mathcal{S}}{_2_{\nu }_{\beta }} \left(\gamma  \tensor{\mathcal{S}}{_1_{\alpha }^{\beta }}+\tensor{u}{_1^{\beta }} \tensor{u}{_2^{\rho }} \tensor{\mathcal{S}}{_1_{\alpha }_{\rho }}\right)+\tensor{k}{^{\beta }} \tensor{u}{_1_{\alpha }} \left(\tensor{u}{_2^{\rho }} \tensor{\mathcal{S}}{_1_{\nu }_{\rho }} \tensor{\mathcal{S}}{_2^{\mu }_{\beta }}-\tensor{u}{_2^{\mu }} \tensor{\mathcal{S}}{_1_{\nu }^{\rho }} \tensor{\mathcal{S}}{_2_{\beta }_{\rho }}\right)\right].
\nonumber
\end{align}

\subsubsection{Spin}
Integrating the spin equation in \eqref{eom} we find (with the notation $ k^{[\alpha} q^{\beta]} = k^{\alpha} q^{\beta} - k^{\beta} q^{\alpha}$)
\begin{equation}
	\begin{aligned}
		\delta^{(1)}_{S_{1}} S^{\alpha\beta}_{1}(\tau_{1}) = -\frac{m_{2}}{8M^2_{\text{Pl}}} \int_{k} & \frac{  \hat{\delta}\left(k \cdot u_{2}\right)  }{k^{2}\left(k \cdot u_{1}-i \epsilon\right)} e^{i k \cdot b} e^{i\left(k \cdot u_{1}-i \epsilon\right) \tau_{1}} \\
		&\times\left(2 \gamma  \tensor{u}{_2^{\rho }}\tensor{k}{^{[\alpha }} \tensor{\mathcal{S}}{_1^{\beta] }_{\rho }}  -\tensor{k}{^{\rho }}\tensor{\mathcal{S}}{_1^{[\alpha }_{\rho }} \left(\tensor{u}{_1} -2 \gamma  \tensor{u}{_2}\right)^{\beta] }\right),
	\end{aligned}
\end{equation}
for the linear term, whereas at quadratic order
\begin{equation}
	\begin{aligned}
		\delta^{(1)}_{S_{1}S_{2}} S^{\alpha\beta}_{1}(\tau_{1}) = -\frac{i}{8M^2_{\text{Pl}}} \int_{k} & \frac{  \hat{\delta}\left(k \cdot u_{2}\right) \tensor{k}{^{\rho }}}{k^{2}\left(k \cdot u_{1}-i \epsilon\right)} e^{i k \cdot b} e^{i\left(k \cdot u_{1}-i \epsilon\right) \tau_{1}}  \\
		\times &\left(  \tensor{k}{^{\sigma }}\tensor{u}{_1^{\nu }} \tensor{u}{_2^{[\alpha }}\tensor{\mathcal{S}}{_1^{\beta] }_{\rho }} \tensor{\mathcal{S}}{_2_{\sigma }_{\nu }}  +\gamma  \tensor{k}{^{\sigma }}\tensor{\mathcal{S}}{_1^{[\alpha }_{\rho }}\tensor{\mathcal{S}}{_2^{\beta] }_{\sigma }} \right.\\
		& \left.-\tensor{\mathcal{S}}{_2_{\rho }_{\sigma }}  \left(\gamma  \tensor{k}{^{[\alpha }}\tensor{\mathcal{S}}{_1^{\beta] }^{\sigma }}+\tensor{u}{_1^{\sigma }} \tensor{u}{_2^{\nu }} \tensor{k}{^{[\alpha }}\tensor{\mathcal{S}}{_1^{\beta] }_{\nu }}\right) \right),
	\end{aligned}
\end{equation}
\begin{align}
		\delta^{(1)}_{ES_1^2} S^{\alpha\beta}_{1}(\tau_{1}) = \frac{-im_{2}}{8M^2_{\text{Pl}}m_{1}} \tensor{\mathcal{S}}{_1_{\rho }_{\sigma }}\int_{k} & \frac{  \hat{\delta}\left(k \cdot u_{2}\right)}{k^{2}\left(k \cdot u_{1}-i \epsilon\right)} e^{i k \cdot b} e^{i\left(k \cdot u_{1}-i \epsilon\right) \tau_{1}} \\
		 \times&\left((k\cdot \tensor{u}{_1})^2 2  \tensor{u}{_2^{\sigma }}  \tensor{u}{_2^{[\alpha }} \tensor{\mathcal{S}}{_1^{\beta] }^{\rho }} +\tensor{k}{^{\rho }}   (k\cdot \tensor{u}{_1}) \tensor{\mathcal{S}}{_1^{[\alpha }^{\sigma }}(\tensor{u}{_1}-2 \gamma  \tensor{u}{_2})^{\beta] } \right.
\nonumber\\
		&\left.-\tensor{k}{^{[\alpha }} \tensor{\mathcal{S}}{_1^{\beta] }^{\sigma }}  \left(\left(2 \gamma ^2-1\right) \tensor{k}{^{\rho }}-2 \gamma  \tensor{u}{_2^{\rho }} (k\cdot \tensor{u}{_1})\right)\right)
\nonumber
\end{align}
\begin{equation}
	\begin{aligned}
		\delta^{(1)}_{RS_1S_1} S^{\alpha\beta}_{1}(\tau_{1}) = \frac{im_{2}}{8M^2_{\text{Pl}}m_{1}} \int_{k} & \frac{  \hat{\delta}\left(k \cdot u_{2}\right) \tensor{k}{^{\rho }}}{k^{2}\left(k \cdot u_{1}-i \epsilon\right)} e^{i k \cdot b} e^{i\left(k \cdot u_{1}-i \epsilon\right) \tau_{1}} \left(2 \gamma  \tensor{k}{^{\sigma }} \tensor{u}{_2^{\nu }} \tensor{\mathcal{S}}{_1_{\sigma }_{\nu }} \tensor{u}{_1^{[\alpha }} \tensor{\mathcal{S}}{_1^{\beta] }_{\rho }}\right.\\
		& \left.   +(k\cdot \tensor{u}{_1}) \left(\tensor{u}{_1^{[\alpha }} \tensor{\mathcal{S}}{_1^{\beta] }^{\sigma }} \tensor{\mathcal{S}}{_1_{\rho }_{\sigma }} - 2 \tensor{u}{_2^{\nu }} \tensor{u}{_2^{\sigma }} \tensor{\mathcal{S}}{_1_{\rho }_{\nu }} \tensor{u}{_1^{[\alpha }} \tensor{\mathcal{S}}{_1^{\beta] }_{\sigma }}\right)\right).
	\end{aligned}
\end{equation}

\subsection{Scattering Data}

The value of the $D_i$ coefficients for the total momentum and spin impulses are given by: 
\begingroup
\allowdisplaybreaks
\begin{align}
		D_{1} =& \frac{\pi  \gamma  \left(5 \gamma ^2-3\right) (\delta +7)}{8 \left(\gamma ^2-1\right)^{3/2}} \\
		D_{2} =& \frac{1}{\gamma ^2-1} \left[\left(8 \gamma ^3+4 \gamma ^2-4 \gamma -1 \right) + \left(8 \gamma ^3-4 \gamma ^2-4 \gamma +1\right) \delta \right] \\
		D_{3} =& \frac{\gamma }{\left(\gamma ^2-1\right)^2}\left[ (2 \gamma +1) \left(8 \gamma ^2+2 \gamma -5\right) + (2 \gamma -1) \left(8 \gamma ^2-2 \gamma -5\right) \delta\right] \\
		D_{4} =& \frac{1}{\left(\gamma ^2-1\right)^2}\left[  \left(-8 \gamma ^4-16 \gamma ^3+8 \gamma +1\right) + \left(8 \gamma ^4-16 \gamma ^3+8 \gamma -1\right) \delta\right] \\
		D_{5} =& \frac{3 \pi }{128 \sqrt{\gamma ^2-1}} \left(35 \gamma ^2 (\delta +1)+9 \delta +1\right) \\
		&+\frac{3 \pi C_{ES}^{(1)}}{128 \left(\gamma ^2-1\right)^{3/2}} \left(5 \gamma ^4 (7 \delta +55)-58 \gamma ^2 (\delta +5)+23 \delta +47\right)  \\
		D_{6} =& -\frac{2 \gamma }{\gamma ^2-1}\left[ (2 \gamma +1) +(2 \gamma -1) \delta \right] \\
		&-\frac{\left(2 \gamma ^2-1\right)  }{\left(\gamma ^2-1\right)^2} C_{ES}^{(1)}\left[\left(2 \gamma ^2+\gamma -1\right) + \left(2 \gamma ^2-\gamma -1\right) \delta \right] \nonumber\\
		D_{7} =& \frac{ \left(2 \gamma ^2 - 1\right)^2   }{\left(\gamma ^2-1\right)^2}C_{ES}^{(1)}\left[(\gamma +1) + (\gamma -1)\delta\right], \quad D_{8} = D_{7} (\delta \leftrightarrow -\delta) \\ 
		D_{9} =& \frac{3 \pi }{128 \sqrt{\gamma ^2-1}} \left(5 \gamma ^2 (7 \delta +31)+\delta -15\right) \\
		&+ \frac{3 \pi  C_{ES}^{(1)} }{128 \left(\gamma ^2-1\right)^{3/2}}\left(5 \gamma ^4 (7 \delta +31)-2 \gamma ^2 (25 \delta +77)+15 \delta +31\right) \nonumber\\
		D_{10} =& \frac{3 \pi }{64 \left(\gamma ^2-1\right)^{3/2}} \left(5 \gamma ^2 (7 \delta +19)-3 \delta +1\right) \\
		&+\frac{3 \pi  C_{ES}^{(1)} }{64 \left(\gamma ^2-1\right)^{5/2}}\left(5 \gamma ^4 (7 \delta +19)-2 \gamma ^2 (23 \delta +43)+11 \delta +23\right) \nonumber\\ 
		D_{11} =& \frac{4  \gamma ^2}{\left(\gamma ^2-1\right)^2}\left[(3 \gamma +2)+(3 \gamma -2)\delta  \right] \\
		&+ \frac{\left(2 \gamma ^2-1\right)}{\left(\gamma ^2-1\right)^3} C_{ES}^{(1)}\left[  (\gamma +1) \left(6 \gamma ^2-2 \gamma -1\right) + (\gamma -1) \left(6 \gamma ^2+2 \gamma -1\right) \delta\right] \nonumber\\
		D_{12} =& \frac{4 \gamma }{\left(\gamma ^2-1\right)^2}\left[ \left(-\gamma ^2-3 \gamma -1\right) +  \left(\gamma ^2-3 \gamma +1\right) \delta \right] \\
		&+\frac{\left(2 \gamma ^2-1\right) }{\left(\gamma ^2-1\right)^3} C_{ES}^{(1)}\left[(-\gamma -1) \left(2 \gamma ^2+4 \gamma -3\right) + (\gamma -1) \left(2 \gamma ^2-4 \gamma -3\right) \delta \right]
\nonumber\\
		D_{13} =& \frac{4  \gamma ^2}{\gamma ^2-1}\left[(\gamma +1) + (\gamma -1)\delta \right] + D_{7} , \quad D_{14} = D_{13} (\delta \leftrightarrow -\delta) \\ 		
		D_{15} =& \frac{3 \pi  \left(20 \gamma ^4-21 \gamma ^2+3\right)}{4 \left(\gamma ^2-1\right)^{3/2}} \\
		D_{16} =& \frac{1}{\left(\gamma ^2-1\right)^2} \left[(\gamma +1) \left(8 \gamma ^3-4 \gamma ^2-4 \gamma +1\right) -(\gamma -1) \left(8 \gamma ^3+4 \gamma ^2-4 \gamma -1\right) \delta \right] \\
		D_{17} =&  2\frac{ \left(2 \gamma ^2 - 1\right)^2   }{\left(\gamma ^2-1\right)^2}\left[(\gamma +1) + (\gamma -1)\delta\right] \\
		D_{18} =& \frac{3 \pi  \gamma ^3 \left(4-5 \gamma ^2\right)}{\left(\gamma ^2-1\right)^{5/2}} \\
		D_{19}  =& -\frac{2}{\left(\gamma ^2-1\right)^3} \left[(\gamma +1) \left(8 \gamma ^5+8 \gamma ^4-4 \gamma ^3-8 \gamma ^2-2 \gamma +1\right) \right. \\
	 &\left. +(\gamma -1) \left(8 \gamma ^5-8 \gamma ^4-4 \gamma ^3+8 \gamma ^2-2 \gamma -1\right) \delta \right] 
	          \nonumber\\
		D_{20}  =& \frac{2}{\left(\gamma ^2-1\right)^2} \left[(\gamma +1) \left(8 \gamma ^4-8 \gamma ^2+1\right) + (\gamma -1) \left(8 \gamma ^4-8 \gamma ^2+1\right) \delta    \right] \\
		D_{21} =& -\frac{\left(8 \gamma ^4-8 \gamma ^2+1\right) (\delta -1)}{2 \left(\gamma ^2-1\right)} \\
		D_{22} =& \frac{3 \pi  \left(5 \gamma ^2-1\right)}{4 \sqrt{\gamma ^2-1}} \\
		D_{23} =& 4\left(1-2 \gamma ^2\right)^2 D_{25} \\
		D_{24} =& \frac{\left(2 \gamma ^2-1\right) \left(2 \gamma ^2-4 \gamma +1\right)}{(\gamma -1)^2 (\gamma +1)} -\frac{\left(2 \gamma ^2-1\right) \left(2 \gamma ^2+4 \gamma +1\right) \delta }{(\gamma -1) (\gamma +1)^2} \\
		D_{25} =& \frac{\delta -1}{2 \left(\gamma ^2-1\right)^2} \\
		D_{26} =& -\frac{2 \gamma  \left(2 \gamma ^2-1\right) (\delta -1) (C_{ES^2}^{(1)}+1)}{\gamma ^2-1} =\frac{ (C_{ES^2}^{(1)}+1)}{2} D_{28}\\
		D_{27} =& -\frac{4 \gamma  (\delta -1)}{\gamma ^2-1}\\
		D_{28} =& (2\gamma^2 -1) D_{27} \\
		D_{29} =& -\frac{2 \left(1-2 \gamma ^2\right)^2 (\delta -1) C_{ES^2}^{(1)}}{\gamma ^2-1}\\
		D_{30} =& -\frac{8 \gamma  \left(2 \gamma ^2-1\right) (C_{ES^2}^{(1)}-3) \nu  }{\left(\gamma ^2-1\right) (\delta +1)} \\
		D_{31} =& \frac{8 \gamma  \left(2 \gamma ^2-1\right) (C_{ES^2}^{(1)}-1) \nu }{\left(\gamma ^2-1\right)^2 (\delta +1)}
\\
		D_{32} =& -\frac{8 \gamma  }{\left(\gamma ^2-1\right) (\delta +1)}\left(8 \gamma ^2 \nu +\gamma  (\delta -2 \nu +1)-3 \nu\right) \nonumber\\
		& +\frac{2 \left(2 \gamma ^2-1\right) C_{ES^2}^{(1)} }{\left(\gamma ^2-1\right)^2 (\delta +1)} \left(8 \gamma ^3 \nu +\gamma ^2 (2 \delta -4 \nu +2)-6 \gamma  \nu -\delta +2 \nu -1\right) \\
		D_{33} =& -\frac{4 \gamma  }{\left(\gamma ^2-1\right)^2 (\delta +1)}\left(8 \gamma ^4 \nu +\gamma ^3 (12 \nu -6 (\delta +1))-22 \gamma ^2 \nu +\gamma  (2 \delta -4 \nu +2)+7 \nu \right)  \nonumber\\
		&+ \frac{2 \left(2 \gamma ^2-1\right) C_{ES^2}^{(1)} }{\left(\gamma ^2-1\right)^2 (\delta +1)}\left(8 \gamma ^3 \nu +\gamma ^2 (-2 \delta +4 \nu -2)-10 \gamma  \nu +\delta -2 \nu +1\right)\\
		D_{34} =& \frac{4 }{\left(\gamma ^2-1\right)^2 (\delta +1)}\left(-8 \gamma ^4 \nu +\gamma ^3 (-8 \delta +16 \nu -8)+\gamma  (4 \delta -8 \nu +4)+\nu \right) \\
		D_{35} =& \frac{\pi  \gamma  \left(5 \gamma ^2-3\right) (7 - \delta )}{8 \left(\gamma ^2-1\right)^{3/2}} = \frac{7 - \delta }{\delta +7}  D_{1} \\
		D_{36} =& \frac{\pi  \gamma ^2 \left(3-5 \gamma ^2\right)}{4 \left(\gamma ^2-1\right)^{3/2}} = -\frac{2 \gamma  }{\delta +7} D_{1}\\
		D_{37} =& -\frac{D_{22}}{\gamma^2 -1} \\
		D_{38} =& \frac{2 \left(2 \gamma ^2+1\right) (\delta -1)}{\gamma ^2-1} \\
		D_{39} =& \frac{\gamma  (1-\delta )+\delta +1}{\left(\gamma ^2-1\right)^2}\\
		D_{40} =& \frac{  \left(8 \gamma ^3-4 \gamma ^2-4 \gamma +1\right) \delta }{\gamma ^2-1}+\frac{  \left(-8 \gamma ^3-4 \gamma ^2+4 \gamma +1\right)}{\gamma ^2-1} \\
		D_{41} =& -\frac{\gamma (2 \gamma +1) \left(8 \gamma ^2+2 \gamma -5\right)}{\left(\gamma ^2-1\right)^2} + \frac{\gamma (2 \gamma -1) \left(8 \gamma ^2-2 \gamma -5\right) \delta}{\left(\gamma ^2-1\right)^2}\\
		D_{42} =& \frac{\gamma  \left(-8 \gamma ^4+12 \gamma ^3-4 \gamma ^2-5 \gamma +4\right) \delta }{\left(\gamma ^2-1\right)^2}+\frac{\gamma  \left(8 \gamma ^4+12 \gamma ^3+4 \gamma ^2-5 \gamma -4\right)}{\left(\gamma ^2-1\right)^2} \\
		D_{43} =& \frac{\pi  \gamma  \left(5 \gamma ^4+2 \gamma ^2-3\right)}{4 \left(\gamma ^2-1\right)^{5/2}} = \frac{\gamma ^2+1}{\gamma(1 -\gamma ^2)} D_{36}
\end{align}
\endgroup

\section{One-loop Integration}\label{bubble}

At 2PM order, all expressions for momentum and spin impulses from Feynman diagrams contain one-loop tensor integrals of the form:
\begin{align}\label{}
I^{\mu_1\cdots \mu_m}_{(a_1,a_2,a_3)} =  
\int_k  {\hat \delta(k\cdot u_j)\, k^{\mu_1} \cdots k^{\mu_m} \over [k^2]^{a_1}\, [(k-q)^2]^{a_2}\, (\pm k\cdot u_\slashed{j} - i\epsilon)^{a_3} },
\end{align}
with $q\cdot u_1 = q\cdot u_2=0$ and we use the convention $\{\slashed{1} = 2,\, \slashed{2} =1\}$, introduced in \cite{3pmeft}. The linear propagators appear due to the iterations where we input the trajectories shown in Appendix~\ref{data} in the tree-level Routhian/action. In~non-spinning cases \cite{pmeft,3pmeft,tidaleft}, all integrals have at most one Lorentz index in the numerator. The situation changes when spin is included, and we find various tensor integrals of rank $m\in\{0,1,2,3\}$. 
Following the standard method first proposed by Passarino and Veltman~\cite{Passarino:1978jh},we reduce all the one-loop tensor integral to a linear combination of scalar integrals. The idea is simple, Lorentz covariance implies that the tensor structure in the final results can be constructed in terms of only the external data $\{q^\mu, u_1^\mu, u_2^\mu\}$ and the metric tensor $g^{\mu\nu}$. Let us consider for example the integral, which we encountered in the non-spinning case~\cite{pmeft},
\begin{align}\label{app-int-exp-tensor-1}
I^{\mu}_{(a_1,a_2,a_3)}  =  
\int_k \, {\hat \delta(k\cdot u_2)\, k^{\mu} \over [k^2]^{a_1}\, [(k-q)^2]^{a_2}\, (\pm k\cdot u_1 - i\epsilon)^{a_3} }\,,
\end{align}
with $a_3>0$. Hence, the tensor decomposition yields
\beq
I^{\mu}_{(a_1,a_2,a_3)}= q^\mu I_q\, + u_1^\mu\, I_{u_1} + u_2^\mu\, I_{u_2}\,,
\eeq
with the scalar integrals $I_q$, $I_{u_1}$ and $I_{u_2}$ often denoted as `form factors' in the literature. We can now solve the form factors by performing Lorentz contractions on both sides, with $q$, $u_1$ and $u_2$. In particular, by contracting with $u_2$ we immediately find $I_{u_2} = -\gamma\,I_{u_1}$ (reflecting the fact that the integral in \eqref{app-int-exp-tensor-1} must be perpendicular to $u_2$). As a result, we must compute only the scalar integrals
\begin{align}\label{}
I_q  &=  
{1 \over q^2} \int_k {\hat \delta(k\cdot u_2)\, k\cdot q \over [k^2]^{a_1}\, [(k-q)^2]^{a_2}\, (\pm k\cdot u_1 - i\epsilon)^{a_3}},
\\[0.3 em]
I_{u_1}  &=  -{1 \over \gamma^2 - 1}\,
\int_k {\hat \delta(k\cdot u_2) \over [k^2]^{a_1}\, [(k-q)^2]^{a_2}\, (\pm k\cdot u_1 - i\epsilon)^{a_3 - 1} }\,.
\end{align}
At the end of the day, going to the rest frame of particle 2 to resolve the delta function, all of these integrals belong to the following family:\footnote{Alternatively, as shown in \cite{pmeft}, we can also perform the $k^0$ integral in the rest frame of particle 1 and pick up the (conservative) pole from the linear propagator only.} 
\begin{align}\label{2pm-int-family}
\int_k \, {1 \over (\vecbf{k}^{2})^{a_1} [(\vecbf{k} - \vecbf{q})^2]^{a_2} (\pm \vecbf{k} \!\cdot\! \vecbf{u} - i\epsilon)^{a_{3}}},
\end{align}
with $\vecbf{q} \cdot \vecbf{u} = 0$\ and $\vecbf{u}^2 = \gamma^2 {-} 1$, in $D=3-2\epsilon$ dimensions. While analytical expressions for any $\{a_1, a_2, a_3\}$ are known, e.g.~\cite{Smirnov}, it is often convenient to use integration-by-parts (IBP) relations to reduce the integrals in \eqref{2pm-int-family} to a combination of the following masters
\begin{align}\label{master-1L-1}
\int {d^Dk \over \pi^{D/2}}\, {1 \over \vecbf{k}^{2} (\vecbf{k} - \vecbf{q})^2}
\,&=\, {1 \over (\vecbf{q}^2)^{2 - D/2}}\, {\Gamma(2 - {D}/{2})\, \Gamma^2({D}/{2} - 1) \over \Gamma (D - 2)},
\\[0.3 em]
\label{master-1L-2}
\int {d^Dk \over \pi^{D/2}}\, {1 \over \vecbf{k}^{2} (\vecbf{k} - \vecbf{q})^2 (\pm \vecbf{k} \!\cdot\! \vecbf{u} - i \epsilon)}
\,&=i \, { \sqrt{\pi} \over (\vecbf{q}^2)^{(5-D)/{2}}\, \sqrt{\gamma^2-1}}\,
{\Gamma((5 {-} D)/{2})\,\Gamma^2((D{-}3)/{2}) \over \sqrt{\gamma^2 {-} 1}\, \Gamma(D-3)}\,,
\end{align}
with the factor of $\pi^{D/2}$ introduced to comply with the present literature. Similar considerations apply to higher-ranked tensor decompositions.\vskip 4pt

Finally, we must perform the Fourier transform to impact parameter space, which can be written in terms of derivatives w.r.t. to $b^\mu$,
\begin{align}\label{app-frr-tensor}
\int_q {e^{iq\cdot b}\,\hat\delta(q\cdot u_1) \hat\delta(q\cdot u_2)\,  q^{\mu_1} \cdots q^{\mu_m} \over (-q^2)^n}
= \big({-}i \partial_b^{\mu_1}\big)\cdots \big({-}i \partial_b^{\mu_m}\big)
 \int_q{e^{iq\cdot b}\,\hat\delta(q\cdot u_1) \hat\delta(q\cdot u_2) \over (-q^2)^n}\,.
\end{align}
We first notice that the results must lie in the plane orthogonal to both $u_1$ and $u_2$. We can then construct a projected metric \cite{justin1,donalvines}
\begin{align}\label{app-frr-pi}
{\partial \over \partial b_\mu} b^\nu \,=\, \Pi^{\mu\nu} \,= \,\eta^{\mu\nu}
+ {u_{1}^{\mu}(u_1^{\nu} - \gamma u_2^{\nu}) + u_{2}^{\mu}(u_2^{\nu} - \gamma u_1^{\nu}) \over \gamma^2 - 1}\,,
\end{align}
which we can use to reduce into scalar integrals. Using \eqref{app-frr-tensor}, together with \eqref{app-frr-pi}, and the master integral
\begin{align}\label{app-frr-scalar}
 \int_q\,{e^{iq\cdot b}\,\hat\delta(q\cdot u_1)\hat\delta(q\cdot u_2) \over (-q^2)^n}
 \,&=\, {4^{-n} \pi^{(2-D)/2} \over \sqrt{\gamma^2 - 1}\, |b|^{D-2 - 2n}}\,{\Gamma( \tfrac{D-2}{2} {-} n)  \over  \Gamma(n)}\,,
\end{align}
it is straightforward to generate the Fourier integrals of any rank, e.g.
\begin{align}\label{app-frr-rank2}
\int_q & {e^{iq\cdot b}\,\hat \delta(q\cdot u_1)\hat \delta(q\cdot u_2)\,  q^{\mu} q^{\nu} \over (-q^2)^n}
\nonumber\\
&= - {2^{1 - 2n}\pi^{(2-D)/2}  \over \sqrt{\gamma^2-1}\,|b|^{D + 2 - 2n}}\,
   {\Gamma(D/2 {-} n) \over \Gamma (n)}\,
   \Big(   (D {-} 2n)\,b^\mu b^\nu + |b|^2\, \Pi^{\mu\nu}  \Big).
\end{align}

\bibliographystyle{JHEP}

\bibliography{ref3PM}

\providecommand{\href}[2]{#2}\begingroup\raggedright\begin{thebibliography}{100}

\bibitem{LIGO}
{\scshape LIGO Scientific, Virgo} collaboration, \emph{{Open data from the
  first and second observing runs of Advanced LIGO and Advanced Virgo}},
  \href{https://arxiv.org/abs/1912.11716}{{\ttfamily 1912.11716}}.

\bibitem{buosathya}
A.~Buonanno and B.~Sathyaprakash, \emph{{Sources of Gravitational Waves: Theory
  and Observations}},  \href{https://arxiv.org/abs/1410.7832}{{\ttfamily
  1410.7832}}.

\bibitem{tune}
R.~A. Porto, \emph{{The Tune of Love and the Nature(ness) of Spacetime}},
  \href{https://doi.org/10.1002/prop.201600064}{\emph{Fortsch. Phys.}
  {\bfseries 64} (2016) 723}
  [\href{https://arxiv.org/abs/1606.08895}{{\ttfamily 1606.08895}}].

\bibitem{music}
R.~A. Porto, \emph{{The Music of the Spheres: The Dawn of Gravitational Wave
  Science}},  \href{https://arxiv.org/abs/1703.06440}{{\ttfamily 1703.06440}}.

\bibitem{salvo}
S.~Vitale, R.~Lynch, J.~Veitch, V.~Raymond and R.~Sturani, \emph{{Measuring the
  spin of black holes in binary systems using gravitational waves}},
  \href{https://doi.org/10.1103/PhysRevLett.112.251101}{\emph{Phys. Rev. Lett.}
  {\bfseries 112} (2014) 251101}
  [\href{https://arxiv.org/abs/1403.0129}{{\ttfamily 1403.0129}}].

\bibitem{Zackay:2019tzo}
B.~Zackay, T.~Venumadhav, L.~Dai, J.~Roulet and M.~Zaldarriaga, \emph{{Highly
  spinning and aligned binary black hole merger in the Advanced LIGO first
  observing run}},
  \href{https://doi.org/10.1103/PhysRevD.100.023007}{\emph{Phys. Rev.}
  {\bfseries D100} (2019) 023007}
  [\href{https://arxiv.org/abs/1902.10331}{{\ttfamily 1902.10331}}].

\bibitem{axiverse}
A.~Arvanitaki and S.~Dubovsky, \emph{{Exploring the String Axiverse with
  Precision Black Hole Physics}},
  \href{https://doi.org/10.1103/PhysRevD.83.044026}{\emph{Phys. Rev. D}
  {\bfseries 83} (2011) 044026}
  [\href{https://arxiv.org/abs/1004.3558}{{\ttfamily 1004.3558}}].

\bibitem{qcd1}
A.~Arvanitaki, M.~Baryakhtar, S.~Dimopoulos, S.~Dubovsky and R.~Lasenby,
  \emph{{Black Hole Mergers and the QCD Axion at Advanced LIGO}},
  \href{https://doi.org/10.1103/PhysRevD.95.043001}{\emph{Phys. Rev. D}
  {\bfseries 95} (2017) 043001}
  [\href{https://arxiv.org/abs/1604.03958}{{\ttfamily 1604.03958}}].

\bibitem{cardoso}
R.~Brito, S.~Ghosh, E.~Barausse, E.~Berti, V.~Cardoso, I.~Dvorkin et~al.,
  \emph{{Stochastic and resolvable gravitational waves from ultralight
  bosons}}, \href{https://doi.org/10.1103/PhysRevLett.119.131101}{\emph{Phys.
  Rev. Lett.} {\bfseries 119} (2017) 131101}
  [\href{https://arxiv.org/abs/1706.05097}{{\ttfamily 1706.05097}}].

\bibitem{salvo2}
K.~K. Ng, M.~Isi, C.-J. Haster and S.~Vitale, \emph{{Multiband
  gravitational-wave searches for ultralight bosons}},
  \href{https://doi.org/10.1103/PhysRevD.102.083020}{\emph{Phys. Rev. D}
  {\bfseries 102} (2020) 083020}
  [\href{https://arxiv.org/abs/2007.12793}{{\ttfamily 2007.12793}}].

\bibitem{gcollider1}
D.~Baumann, H.~S. Chia and R.~A. Porto, \emph{{Probing Ultralight Bosons with
  Binary Black Holes}},
  \href{https://doi.org/10.1103/PhysRevD.99.044001}{\emph{Phys. Rev. D}
  {\bfseries 99} (2019) 044001}
  [\href{https://arxiv.org/abs/1804.03208}{{\ttfamily 1804.03208}}].

\bibitem{gcollider2}
D.~Baumann, H.~S. Chia, R.~A. Porto and J.~Stout, \emph{{Gravitational Collider
  Physics}}, \href{https://doi.org/10.1103/PhysRevD.101.083019}{\emph{Phys.
  Rev. D} {\bfseries 101} (2020) 083019}
  [\href{https://arxiv.org/abs/1912.04932}{{\ttfamily 1912.04932}}].

\bibitem{Kerr}
R.~P. Kerr, \emph{{Rotating black holes and the Kerr metric}},
  \href{https://doi.org/10.1063/1.3012288}{\emph{AIP Conf. Proc.} {\bfseries
  1059} (2008) 9}.

\bibitem{nrgr}
W.~D. Goldberger and I.~Z. Rothstein, \emph{{An Effective field theory of
  gravity for extended objects}},
  \href{https://doi.org/10.1103/PhysRevD.73.104029}{\emph{Phys. Rev.}
  {\bfseries D73} (2006) 104029}
  [\href{https://arxiv.org/abs/hep-th/0409156}{{\ttfamily hep-th/0409156}}].

\bibitem{nrgrs}
R.~A. Porto, \emph{{Post-Newtonian Corrections to the Motion of Spinning Bodies
  in NRGR}}, \href{https://doi.org/10.1103/PhysRevD.73.104031}{\emph{Phys. Rev.
  D} {\bfseries 73} (2006) 104031}
  [\href{https://arxiv.org/abs/gr-qc/0511061}{{\ttfamily gr-qc/0511061}}].

\bibitem{walterLH}
W.~D. Goldberger, \emph{{Les Houches lectures on effective field theories and
  gravitational radiation}},  in \emph{Les Houches Summer School - Session 86},
  1, 2007, \href{https://arxiv.org/abs/hep-ph/0701129}{{\ttfamily
  hep-ph/0701129}}.

\bibitem{Foffa:2013qca}
S.~Foffa and R.~Sturani, \emph{{Effective Field Theory Methods to Model Compact
  Binaries}}, \href{https://doi.org/10.1088/0264-9381/31/4/043001}{\emph{Class.
  Quant. Grav.} {\bfseries 31} (2014) 043001}
  [\href{https://arxiv.org/abs/1309.3474}{{\ttfamily 1309.3474}}].

\bibitem{review}
R.~A. Porto, \emph{{The effective field theorist's approach to gravitational
  dynamics}}, \href{https://doi.org/10.1016/j.physrep.2016.04.003}{\emph{Phys.
  Rept.} {\bfseries 633} (2016) 1}
  [\href{https://arxiv.org/abs/1601.04914}{{\ttfamily 1601.04914}}].

\bibitem{Hinder:2018fsy}
I.~Hinder, S.~Ossokine, H.~P. Pfeiffer and A.~Buonanno, \emph{{Gravitational
  waveforms for high spin and high mass-ratio binary black holes: A synergistic
  use of numerical-relativity codes}},
  \href{https://doi.org/10.1103/PhysRevD.99.061501}{\emph{Phys. Rev. D}
  {\bfseries 99} (2019) 061501}
  [\href{https://arxiv.org/abs/1810.10585}{{\ttfamily 1810.10585}}].

\bibitem{Faye1}
G.~Faye, L.~Blanchet and A.~Buonanno, \emph{{Higher-order spin effects in the
  dynamics of compact binaries. I. Equations of motion}},
  \href{https://doi.org/10.1103/PhysRevD.74.104033}{\emph{Phys. Rev. D}
  {\bfseries 74} (2006) 104033}
  [\href{https://arxiv.org/abs/gr-qc/0605139}{{\ttfamily gr-qc/0605139}}].

\bibitem{prl}
R.~A. Porto and I.~Rothstein, \emph{{The Hyperfine Einstein-Infeld-Hoffmann
  Potential}}, \href{https://doi.org/10.1103/PhysRevLett.97.021101}{\emph{Phys.
  Rev. Lett.} {\bfseries 97} (2006) 021101}
  [\href{https://arxiv.org/abs/gr-qc/0604099}{{\ttfamily gr-qc/0604099}}].

\bibitem{Porto:2007px}
R.~A. Porto, \emph{{New results at 3PN via an effective field theory of
  gravity}},  in \emph{{11th Marcel Grossmann Meeting on General Relativity}},
  pp.~2493--2496, 1, 2007,
  \href{https://arxiv.org/abs/gr-qc/0701106}{{\ttfamily gr-qc/0701106}}.

\bibitem{nrgrss}
R.~A. Porto and I.~Z. Rothstein, \emph{{Spin(1)Spin(2) Effects in the Motion of
  Inspiralling Compact Binaries at Third Order in the Post-Newtonian
  Expansion}},
  \href{https://doi.org/10.1103/PhysRevD.78.044012}{\emph{Phys.Rev.} {\bfseries
  D78} (2008) 044012} [\href{https://arxiv.org/abs/0802.0720}{{\ttfamily
  0802.0720}}].

\bibitem{nrgrs2}
R.~A. Porto and I.~Z. Rothstein, \emph{{Next to Leading Order Spin(1)Spin(1)
  Effects in the Motion of Inspiralling Compact Binaries}},
  \href{https://doi.org/10.1103/PhysRevD.78.044013}{\emph{Phys.Rev.} {\bfseries
  D78} (2008) 044013} [\href{https://arxiv.org/abs/0804.0260}{{\ttfamily
  0804.0260}}].

\bibitem{jan1}
J.~Steinhoff, S.~Hergt and G.~Schaefer, \emph{{On the next-to-leading order
  gravitational spin(1)-spin(2) dynamics}},
  \href{https://doi.org/10.1103/PhysRevD.77.081501}{\emph{Phys. Rev. D}
  {\bfseries 77} (2008) 081501}
  [\href{https://arxiv.org/abs/0712.1716}{{\ttfamily 0712.1716}}].

\bibitem{jan2}
J.~Steinhoff, S.~Hergt and G.~Schaefer, \emph{{Spin-squared Hamiltonian of
  next-to-leading order gravitational interaction}},
  \href{https://doi.org/10.1103/PhysRevD.78.101503}{\emph{Phys. Rev. D}
  {\bfseries 78} (2008) 101503}
  [\href{https://arxiv.org/abs/0809.2200}{{\ttfamily 0809.2200}}].

\bibitem{nrgrso}
R.~A. Porto, \emph{{Next-to-Leading Order Spin-Orbit Effects in the Motion of
  Inspiralling Compact Binaries}},
  \href{https://doi.org/10.1088/0264-9381/27/20/205001}{\emph{Class. Quant.
  Grav.} {\bfseries 27} (2010) 205001}
  [\href{https://arxiv.org/abs/1005.5730}{{\ttfamily 1005.5730}}].

\bibitem{Levi:2015uxa}
M.~Levi and J.~Steinhoff, \emph{{Next-to-next-to-leading order gravitational
  spin-orbit coupling via the effective field theory for spinning objects in
  the post-Newtonian scheme}},
  \href{https://doi.org/10.1088/1475-7516/2016/01/011}{\emph{JCAP} {\bfseries
  01} (2016) 011} [\href{https://arxiv.org/abs/1506.05056}{{\ttfamily
  1506.05056}}].

\bibitem{Levi:2016ofk}
M.~Levi and J.~Steinhoff, \emph{{Complete conservative dynamics for
  inspiralling compact binaries with spins at fourth post-Newtonian order}},
  \href{https://arxiv.org/abs/1607.04252}{{\ttfamily 1607.04252}}.

\bibitem{Levi:2020uwu}
M.~Levi, A.~J. Mcleod and M.~Von~Hippel, \emph{{NNNLO gravitational
  quadratic-in-spin interactions at the quartic order in G}},
  \href{https://arxiv.org/abs/2003.07890}{{\ttfamily 2003.07890}}.

\bibitem{Levi:2020kvb}
M.~Levi, A.~J. Mcleod and M.~Von~Hippel, \emph{{N$^3$LO gravitational
  spin-orbit coupling at order $G^4$}},
  \href{https://arxiv.org/abs/2003.02827}{{\ttfamily 2003.02827}}.

\bibitem{Faye2}
L.~Blanchet, A.~Buonanno and G.~Faye, \emph{{Higher-order spin effects in the
  dynamics of compact binaries. II. Radiation field}},
  \href{https://doi.org/10.1103/PhysRevD.81.089901}{\emph{Phys. Rev. D}
  {\bfseries 74} (2006) 104034}
  [\href{https://arxiv.org/abs/gr-qc/0605140}{{\ttfamily gr-qc/0605140}}].

\bibitem{rads1}
R.~A. Porto, A.~Ross and I.~Z. Rothstein, \emph{{Spin induced multipole moments
  for the gravitational wave flux from binary inspirals to third Post-Newtonian
  order}}, \href{https://doi.org/10.1088/1475-7516/2011/03/009}{\emph{JCAP}
  {\bfseries 1103} (2011) 009}
  [\href{https://arxiv.org/abs/1007.1312}{{\ttfamily 1007.1312}}].

\bibitem{amps}
R.~A. Porto, A.~Ross and I.~Z. Rothstein, \emph{{Spin induced multipole moments
  for the gravitational wave amplitude from binary inspirals to 2.5
  Post-Newtonian order}},
  \href{https://doi.org/10.1088/1475-7516/2012/09/028}{\emph{JCAP} {\bfseries
  1209} (2012) 028} [\href{https://arxiv.org/abs/1203.2962}{{\ttfamily
  1203.2962}}].

\bibitem{bohe}
A.~Boh\'e, G.~Faye, S.~Marsat and E.~K. Porter, \emph{{Quadratic-in-spin
  effects in the orbital dynamics and gravitational-wave energy flux of compact
  binaries at the 3PN order}},
  \href{https://doi.org/10.1088/0264-9381/32/19/195010}{\emph{Class. Quant.
  Grav.} {\bfseries 32} (2015) 195010}
  [\href{https://arxiv.org/abs/1501.01529}{{\ttfamily 1501.01529}}].

\bibitem{natalia1}
N.~T. Maia, C.~R. Galley, A.~K. Leibovich and R.~A. Porto, \emph{{Radiation
  reaction for spinning bodies in effective field theory I: Spin-orbit
  effects}}, \href{https://doi.org/10.1103/PhysRevD.96.084064}{\emph{Phys.
  Rev.} {\bfseries D96} (2017) 084064}
  [\href{https://arxiv.org/abs/1705.07934}{{\ttfamily 1705.07934}}].

\bibitem{natalia2}
N.~T. Maia, C.~R. Galley, A.~K. Leibovich and R.~A. Porto, \emph{{Radiation
  reaction for spinning bodies in effective field theory II: Spin-spin
  effects}}, \href{https://doi.org/10.1103/PhysRevD.96.084065}{\emph{Phys.
  Rev.} {\bfseries D96} (2017) 084065}
  [\href{https://arxiv.org/abs/1705.07938}{{\ttfamily 1705.07938}}].

\bibitem{zixin}
Z.~Yang and A.~K. Leibovich, \emph{{Analytic Solutions to Compact Binary
  Inspirals With Leading Order Spin-Orbit Contribution Using The Dynamical
  Renormalization Group}},
  \href{https://doi.org/10.1103/PhysRevD.100.084021}{\emph{Phys. Rev. D}
  {\bfseries 100} (2019) 084021}
  [\href{https://arxiv.org/abs/1908.05688}{{\ttfamily 1908.05688}}].

\bibitem{Pardo}
B.~A. Pardo and N.~T. Maia, \emph{{Next-to-leading order spin-orbit effects in
  the equations of motion, energy loss and phase evolution of binaries of
  compact bodies in the effective field theory approach}},
  \href{https://doi.org/10.1103/PhysRevD.102.124020}{\emph{Phys. Rev. D}
  {\bfseries 102} (2020) 124020}
  [\href{https://arxiv.org/abs/2009.05628}{{\ttfamily 2009.05628}}].

\bibitem{andirad}
W.~D. Goldberger and A.~Ross, \emph{{Gravitational radiative corrections from
  effective field theory}},
  \href{https://doi.org/10.1103/PhysRevD.81.124015}{\emph{Phys. Rev.}
  {\bfseries D81} (2010) 124015}
  [\href{https://arxiv.org/abs/0912.4254}{{\ttfamily 0912.4254}}].

\bibitem{radnrgr}
A.~K. Leibovich, N.~T. Maia, I.~Z. Rothstein and Z.~Yang, \emph{{Second
  post-Newtonian order radiative dynamics of inspiralling compact binaries in
  the Effective Field Theory approach}},
  \href{https://doi.org/10.1103/PhysRevD.101.084058}{\emph{Phys. Rev. D}
  {\bfseries 101} (2020) 084058}
  [\href{https://arxiv.org/abs/1912.12546}{{\ttfamily 1912.12546}}].

\bibitem{dis1}
W.~Goldberger and I.~Rothstein, \emph{{Dissipative Effects in the Worldline
  Approach to Black Hole Dynamics}},
  \href{https://doi.org/10.1103/PhysRevD.73.104030}{\emph{Phys. Rev. D}
  {\bfseries 73} (2006) 104030}
  [\href{https://arxiv.org/abs/hep-th/0511133}{{\ttfamily hep-th/0511133}}].

\bibitem{dis2}
R.~A. Porto, \emph{{Absorption Effects due to Spin in the Worldline Approach to
  Black Hole Dynamics}},
  \href{https://doi.org/10.1103/PhysRevD.77.064026}{\emph{Phys. Rev. D}
  {\bfseries 77} (2008) 064026}
  [\href{https://arxiv.org/abs/0710.5150}{{\ttfamily 0710.5150}}].

\bibitem{dis3}
W.~D. Goldberger, J.~Li and I.~Z. Rothstein, \emph{{Non-conservative effects on
  Spinning Black Holes from World-Line Effective Field Theory}},
  \href{https://arxiv.org/abs/2012.14869}{{\ttfamily 2012.14869}}.

\bibitem{Blanchet:2003gy}
L.~Blanchet, T.~Damour and G.~Esposito-Farese, \emph{{Dimensional
  regularization of the third postNewtonian dynamics of point particles in
  harmonic coordinates}},
  \href{https://doi.org/10.1103/PhysRevD.69.124007}{\emph{Phys. Rev. D}
  {\bfseries 69} (2004) 124007}
  [\href{https://arxiv.org/abs/gr-qc/0311052}{{\ttfamily gr-qc/0311052}}].

\bibitem{nrgr3pn}
S.~Foffa and R.~Sturani, \emph{{Effective field theory calculation of
  conservative binary dynamics at third post-Newtonian order}},
  \href{https://doi.org/10.1103/PhysRevD.84.044031}{\emph{Phys. Rev. D}
  {\bfseries 84} (2011) 044031}
  [\href{https://arxiv.org/abs/1104.1122}{{\ttfamily 1104.1122}}].

\bibitem{Foffa:2012rn}
S.~Foffa and R.~Sturani, \emph{{Dynamics of the gravitational two-body problem
  at fourth post-Newtonian order and at quadratic order in the Newton
  constant}}, \href{https://doi.org/10.1103/PhysRevD.87.064011}{\emph{Phys.
  Rev. D} {\bfseries 87} (2013) 064011}
  [\href{https://arxiv.org/abs/1206.7087}{{\ttfamily 1206.7087}}].

\bibitem{tail}
C.~Galley, A.~Leibovich, R.~A. Porto and A.~Ross, \emph{{Tail Effect in
  Gravitational Radiation Reaction: Time Nonlocality and Renormalization Group
  Evolution}}, \href{https://doi.org/10.1103/PhysRevD.93.124010}{\emph{Phys.
  Rev. D} {\bfseries 93} (2016) 124010}
  [\href{https://arxiv.org/abs/1511.07379}{{\ttfamily 1511.07379}}].

\bibitem{nrgrG5}
S.~Foffa, P.~Mastrolia, R.~Sturani and C.~Sturm, \emph{{Effective field theory
  approach to the gravitational two-body dynamics, at fourth post-Newtonian
  order and quintic in the Newton constant}},
  \href{https://doi.org/10.1103/PhysRevD.95.104009}{\emph{Phys. Rev. D}
  {\bfseries 95} (2017) 104009}
  [\href{https://arxiv.org/abs/1612.00482}{{\ttfamily 1612.00482}}].

\bibitem{apparent}
R.~A. Porto and I.~Rothstein, \emph{{Apparent Ambiguities in the Post-Newtonian
  Expansion for Binary Systems}},
  \href{https://doi.org/10.1103/PhysRevD.96.024062}{\emph{Phys. Rev. D}
  {\bfseries 96} (2017) 024062}
  [\href{https://arxiv.org/abs/1703.06433}{{\ttfamily 1703.06433}}].

\bibitem{Damour:2014jta}
T.~Damour, P.~Jaranowski and G.~Sch{\"a}fer, \emph{{Nonlocal-In-Time Action for
  the Fourth Post-Newtonian Conservative Dynamics of Two-Body Systems}},
  \href{https://doi.org/10.1103/PhysRevD.89.064058}{\emph{Phys. Rev. D}
  {\bfseries 89} (2014) 064058}
  [\href{https://arxiv.org/abs/1401.4548}{{\ttfamily 1401.4548}}].

\bibitem{Marchand:2017pir}
T.~Marchand, L.~Bernard, L.~Blanchet and G.~Faye, \emph{{Ambiguity-Free
  Completion of the Equations of Motion of Compact Binary Systems at the Fourth
  Post-Newtonian Order}},
  \href{https://doi.org/10.1103/PhysRevD.97.044023}{\emph{Phys. Rev. D}
  {\bfseries 97} (2018) 044023}
  [\href{https://arxiv.org/abs/1707.09289}{{\ttfamily 1707.09289}}].

\bibitem{nrgr4pn1}
S.~Foffa and R.~Sturani, \emph{{Conservative dynamics of binary systems to
  fourth Post-Newtonian order in the EFT approach I: Regularized Lagrangian}},
  \href{https://doi.org/10.1103/PhysRevD.100.024047}{\emph{Phys. Rev. D}
  {\bfseries 100} (2019) 024047}
  [\href{https://arxiv.org/abs/1903.05113}{{\ttfamily 1903.05113}}].

\bibitem{nrgr4pn2}
S.~Foffa, R.~A. Porto, I.~Rothstein and R.~Sturani, \emph{{Conservative
  dynamics of binary systems to fourth Post-Newtonian order in the EFT approach
  II: Renormalized Lagrangian}},
  \href{https://doi.org/10.1103/PhysRevD.100.024048}{\emph{Phys. Rev.}
  {\bfseries D100} (2019) 024048}
  [\href{https://arxiv.org/abs/1903.05118}{{\ttfamily 1903.05118}}].

\bibitem{5pn1}
S.~Foffa, P.~Mastrolia, R.~Sturani, C.~Sturm and W.~J. Torres~Bobadilla,
  \emph{{Static two-body potential at fifth post-Newtonian order}},
  \href{https://doi.org/10.1103/PhysRevLett.122.241605}{\emph{Phys. Rev. Lett.}
  {\bfseries 122} (2019) 241605}
  [\href{https://arxiv.org/abs/1902.10571}{{\ttfamily 1902.10571}}].

\bibitem{5pn2}
J.~Bl{\"u}mlein, A.~Maier and P.~Marquard, \emph{{Five-Loop Static Contribution
  to the Gravitational Interaction Potential of Two Point Masses}},
  {\emph{Phys. Lett. B} {\bfseries 800} (2020) 135100}
  [\href{https://arxiv.org/abs/1902.11180}{{\ttfamily 1902.11180}}].

\bibitem{tail2}
S.~Foffa and R.~Sturani, \emph{{Hereditary terms at next-to-leading order in
  two-body gravitational dynamics}},
  \href{https://doi.org/10.1103/PhysRevD.101.064033}{\emph{Phys. Rev. D}
  {\bfseries 101} (2020) 064033}
  [\href{https://arxiv.org/abs/1907.02869}{{\ttfamily 1907.02869}}].

\bibitem{Blumlein:2020pyo}
J.~Bl\"umlein, A.~Maier, P.~Marquard and G.~Sch\"afer, \emph{{The fifth-order
  post-Newtonian Hamiltonian dynamics of two-body systems from an effective
  field theory approach: potential contributions}},
  \href{https://arxiv.org/abs/2010.13672}{{\ttfamily 2010.13672}}.

\bibitem{blum}
J.~Bl{\"u}mlein, A.~Maier, P.~Marquard and G.~Sch{\"a}fer, \emph{{Testing
  binary dynamics in gravity at the sixth post-Newtonian level}},
  \href{https://doi.org/10.1016/j.physletb.2020.135496}{\emph{Phys. Lett. B}
  {\bfseries 807} (2020) 135496}
  [\href{https://arxiv.org/abs/2003.07145}{{\ttfamily 2003.07145}}].

\bibitem{blum2}
J.~Bl\"umlein, A.~Maier, P.~Marquard and G.~Sch\"afer, \emph{{The 6th
  Post-Newtonian Potential Terms at $O(G_N^4)$}},
  \href{https://arxiv.org/abs/2101.08630}{{\ttfamily 2101.08630}}.

\bibitem{bini2}
D.~Bini, T.~Damour and A.~Geralico, \emph{{Sixth post-Newtonian local-in-time
  dynamics of binary systems}},
  \href{https://arxiv.org/abs/2004.05407}{{\ttfamily 2004.05407}}.

\bibitem{tail3}
L.~Blanchet, S.~Foffa, F.~Larrouturou and R.~Sturani, \emph{{Logarithmic tail
  contributions to the energy function of circular compact binaries}},
  \href{https://doi.org/10.1103/PhysRevD.101.084045}{\emph{Phys. Rev. D}
  {\bfseries 101} (2020) 084045}
  [\href{https://arxiv.org/abs/1912.12359}{{\ttfamily 1912.12359}}].

\bibitem{paper1}
G.~K{\"a}lin and R.~A. Porto, \emph{{From Boundary Data to Bound States}},
  \href{https://doi.org/10.1007/JHEP01(2020)072}{\emph{JHEP} {\bfseries 01}
  (2020) 072} [\href{https://arxiv.org/abs/1910.03008}{{\ttfamily
  1910.03008}}].

\bibitem{paper2}
G.~K{\"a}lin and R.~A. Porto, \emph{{From boundary data to bound states. Part
  II. Scattering angle to dynamical invariants (with twist)}},
  \href{https://doi.org/10.1007/JHEP02(2020)120}{\emph{JHEP} {\bfseries 02}
  (2020) 120} [\href{https://arxiv.org/abs/1911.09130}{{\ttfamily
  1911.09130}}].

\bibitem{pmeft}
G.~K\"alin and R.~A. Porto, \emph{{Post-Minkowskian Effective Field Theory for
  Conservative Binary Dynamics}},
  \href{https://doi.org/10.1007/JHEP11(2020)106}{\emph{JHEP} {\bfseries 11}
  (2020) 106} [\href{https://arxiv.org/abs/2006.01184}{{\ttfamily
  2006.01184}}].

\bibitem{3pmeft}
G.~K\"alin, Z.~Liu and R.~A. Porto, \emph{{Conservative Dynamics of Binary
  Systems to Third Post-Minkowskian Order from the Effective Field Theory
  Approach}}, \href{https://doi.org/10.1103/PhysRevLett.125.261103}{\emph{Phys.
  Rev. Lett.} {\bfseries 125} (2020) 261103}
  [\href{https://arxiv.org/abs/2007.04977}{{\ttfamily 2007.04977}}].

\bibitem{tidaleft}
G.~K\"alin, Z.~Liu and R.~A. Porto, \emph{{Conservative Tidal Effects in
  Compact Binary Systems to Next-to-Leading Post-Minkowskian Order}},
  \href{https://doi.org/10.1103/PhysRevD.102.124025}{\emph{Phys. Rev. D}
  {\bfseries 102} (2020) 124025}
  [\href{https://arxiv.org/abs/2008.06047}{{\ttfamily 2008.06047}}].

\bibitem{blanchet}
L.~Blanchet, \emph{Gravitational {Radiation} from {Post}-{Newtonian} {Sources}
  and {Inspiralling} {Compact} {Binaries}},
  \href{https://doi.org/10.12942/lrr-2014-2}{\emph{Living Reviews in
  Relativity} {\bfseries 17} (2014) 2}.

\bibitem{Mathisson}
M.~Mathisson, \emph{{Neue mechanik materieller systemes}}, {\emph{Acta Phys.
  Polon.} {\bfseries 6} (1937) 163}.

\bibitem{Papa}
A.~Papapetrou, \emph{{Spinning test particles in general relativity. 1.}},
  \href{https://doi.org/10.1098/rspa.1951.0200}{\emph{Proc. Roy. Soc. Lond. A}
  {\bfseries 209} (1951) 248}.

\bibitem{Dixon}
W.~Dixon, \emph{{Dynamics of extended bodies in general relativity. I. Momentum
  and angular momentum}},
  \href{https://doi.org/10.1098/rspa.1970.0020}{\emph{Proc. Roy. Soc. Lond. A}
  {\bfseries 314} (1970) 499}.

\bibitem{hanson}
A.~J. Hanson and T.~Regge, \emph{{The Relativistic Spherical Top}},
  \href{https://doi.org/10.1016/0003-4916(74)90046-3}{\emph{Annals Phys.}
  {\bfseries 87} (1974) 498}.

\bibitem{yee}
K.~Yee and M.~Bander, \emph{{Equations of motion for spinning particles in
  external electromagnetic and gravitational fields}},
  \href{https://doi.org/10.1103/PhysRevD.48.2797}{\emph{Phys. Rev. D}
  {\bfseries 48} (1993) 2797}
  [\href{https://arxiv.org/abs/hep-th/9302117}{{\ttfamily hep-th/9302117}}].

\bibitem{elvang}
H.~Elvang and Y.-t. Huang, \emph{{Scattering Amplitudes in Gauge Theory and
  Gravity}}. Cambridge University Press, 4, 2015.

\bibitem{reviewdc}
Z.~Bern, J.~J. Carrasco, M.~Chiodaroli, H.~Johansson and R.~Roiban, \emph{{The
  Duality Between Color and Kinematics and its Applications}},
  \href{https://arxiv.org/abs/1909.01358}{{\ttfamily 1909.01358}}.

\bibitem{Henn:2014qga}
J.~M. Henn, \emph{{Lectures on differential equations for Feynman integrals}},
  \href{https://doi.org/10.1088/1751-8113/48/15/153001}{\emph{J. Phys. A}
  {\bfseries 48} (2015) 153001}
  [\href{https://arxiv.org/abs/1412.2296}{{\ttfamily 1412.2296}}].

\bibitem{ira1}
D.~Neill and I.~Z. Rothstein, \emph{{Classical Space-Times from the S Matrix}},
  \href{https://doi.org/10.1016/j.nuclphysb.2013.09.007}{\emph{Nucl. Phys.}
  {\bfseries B877} (2013) 177}
  [\href{https://arxiv.org/abs/1304.7263}{{\ttfamily 1304.7263}}].

\bibitem{cheung}
C.~Cheung, I.~Z. Rothstein and M.~P. Solon, \emph{{From Scattering Amplitudes
  to Classical Potentials in the Post-Minkowskian Expansion}},
  \href{https://doi.org/10.1103/PhysRevLett.121.251101}{\emph{Phys. Rev. Lett.}
  {\bfseries 121} (2018) 251101}
  [\href{https://arxiv.org/abs/1808.02489}{{\ttfamily 1808.02489}}].

\bibitem{zvi1}
Z.~Bern, C.~Cheung, R.~Roiban, C.-H. Shen, M.~P. Solon and M.~Zeng,
  \emph{{Scattering Amplitudes and the Conservative Hamiltonian for Binary
  Systems at Third Post-Minkowskian Order}},
  \href{https://doi.org/10.1103/PhysRevLett.122.201603}{\emph{Phys. Rev. Lett.}
  {\bfseries 122} (2019) 201603}
  [\href{https://arxiv.org/abs/1901.04424}{{\ttfamily 1901.04424}}].

\bibitem{zvi2}
Z.~Bern, C.~Cheung, R.~Roiban, C.-H. Shen, M.~P. Solon and M.~Zeng,
  \emph{{Black Hole Binary Dynamics from the Double Copy and Effective
  Theory}}, \href{https://doi.org/10.1007/JHEP10(2019)206}{\emph{JHEP}
  {\bfseries 10} (2019) 206}
  [\href{https://arxiv.org/abs/1908.01493}{{\ttfamily 1908.01493}}].

\bibitem{donal}
D.~A. Kosower, B.~Maybee and D.~O'Connell, \emph{{Amplitudes, Observables, and
  Classical Scattering}},
  \href{https://doi.org/10.1007/JHEP02(2019)137}{\emph{JHEP} {\bfseries 02}
  (2019) 137} [\href{https://arxiv.org/abs/1811.10950}{{\ttfamily
  1811.10950}}].

\bibitem{donalvines}
B.~Maybee, D.~O'Connell and J.~Vines, \emph{{Observables and amplitudes for
  spinning particles and black holes}},
  \href{https://doi.org/10.1007/JHEP12(2019)156}{\emph{JHEP} {\bfseries 12}
  (2019) 156} [\href{https://arxiv.org/abs/1906.09260}{{\ttfamily
  1906.09260}}].

\bibitem{withchad}
C.~Galley and R.~A. Porto, \emph{{Gravitational Self-Force in the
  Ultra-Relativistic Limit: the ``Large-$N$" Expansion}},
  \href{https://doi.org/10.1007/JHEP11(2013)096}{\emph{JHEP} {\bfseries 11}
  (2013) 096} [\href{https://arxiv.org/abs/1302.4486}{{\ttfamily 1302.4486}}].

\bibitem{Holstein:2008sx}
B.~R. Holstein and A.~Ross, \emph{{Spin Effects in Long Range Gravitational
  Scattering}},  \href{https://arxiv.org/abs/0802.0716}{{\ttfamily 0802.0716}}.

\bibitem{Bjerrum-Bohr:2013bxa}
N.~Bjerrum-Bohr, J.~F. Donoghue and P.~Vanhove, \emph{{On-shell Techniques and
  Universal Results in Quantum Gravity}},
  \href{https://doi.org/10.1007/JHEP02(2014)111}{\emph{JHEP} {\bfseries 02}
  (2014) 111} [\href{https://arxiv.org/abs/1309.0804}{{\ttfamily 1309.0804}}].

\bibitem{Vaidya:2014kza}
V.~Vaidya, \emph{{Gravitational spin Hamiltonians from the S matrix}},
  \href{https://doi.org/10.1103/PhysRevD.91.024017}{\emph{Phys. Rev.}
  {\bfseries D91} (2015) 024017}
  [\href{https://arxiv.org/abs/1410.5348}{{\ttfamily 1410.5348}}].

\bibitem{Guevara:2017csg}
A.~Guevara, \emph{{Holomorphic Classical Limit for Spin Effects in
  Gravitational and Electromagnetic Scattering}},
  \href{https://doi.org/10.1007/JHEP04(2019)033}{\emph{JHEP} {\bfseries 04}
  (2019) 033} [\href{https://arxiv.org/abs/1706.02314}{{\ttfamily
  1706.02314}}].

\bibitem{Chung:2018kqs}
M.-Z. Chung, Y.-T. Huang, J.-W. Kim and S.~Lee, \emph{{The simplest massive
  S-matrix: from minimal coupling to Black Holes}},
  \href{https://doi.org/10.1007/JHEP04(2019)156}{\emph{JHEP} {\bfseries 04}
  (2019) 156} [\href{https://arxiv.org/abs/1812.08752}{{\ttfamily
  1812.08752}}].

\bibitem{Guevara:2018wpp}
A.~Guevara, A.~Ochirov and J.~Vines, \emph{{Scattering of Spinning Black Holes
  from Exponentiated Soft Factors}},
  \href{https://doi.org/10.1007/JHEP09(2019)056}{\emph{JHEP} {\bfseries 09}
  (2019) 056} [\href{https://arxiv.org/abs/1812.06895}{{\ttfamily
  1812.06895}}].

\bibitem{Guevara:2019fsj}
A.~Guevara, A.~Ochirov and J.~Vines, \emph{{Black-hole scattering with general
  spin directions from minimal-coupling amplitudes}},
  \href{https://doi.org/10.1103/PhysRevD.100.104024}{\emph{Phys. Rev. D}
  {\bfseries 100} (2019) 104024}
  [\href{https://arxiv.org/abs/1906.10071}{{\ttfamily 1906.10071}}].

\bibitem{bohr}
N.~E.~J. Bjerrum-Bohr, P.~H. Damgaard, G.~Festuccia, L.~Plante and P.~Vanhove,
  \emph{{General Relativity from Scattering Amplitudes}},
  \href{https://doi.org/10.1103/PhysRevLett.121.171601}{\emph{Phys. Rev. Lett.}
  {\bfseries 121} (2018) 171601}
  [\href{https://arxiv.org/abs/1806.04920}{{\ttfamily 1806.04920}}].

\bibitem{cristof1}
A.~Cristofoli, N.~E.~J. Bjerrum-Bohr, P.~H. Damgaard and P.~Vanhove, \emph{{On
  Post-Minkowskian Hamiltonians in General Relativity}},
  \href{https://arxiv.org/abs/1906.01579}{{\ttfamily 1906.01579}}.

\bibitem{simon}
S.~Caron-Huot and Z.~Zahraee, \emph{{Integrability of Black Hole Orbits in
  Maximal Supergravity}},
  \href{https://doi.org/10.1007/JHEP07(2019)179}{\emph{JHEP} {\bfseries 07}
  (2019) 179} [\href{https://arxiv.org/abs/1810.04694}{{\ttfamily
  1810.04694}}].

\bibitem{Arkani-Hamed:2019ymq}
N.~Arkani-Hamed, Y.-t. Huang and D.~O'Connell, \emph{{Kerr black holes as
  elementary particles}},
  \href{https://doi.org/10.1007/JHEP01(2020)046}{\emph{JHEP} {\bfseries 01}
  (2020) 046} [\href{https://arxiv.org/abs/1906.10100}{{\ttfamily
  1906.10100}}].

\bibitem{Bjerrum-Bohr:2019kec}
N.~E.~J. Bjerrum-Bohr, A.~Cristofoli and P.~H. Damgaard,
  \emph{{Post-Minkowskian Scattering Angle in Einstein Gravity}},
  \href{https://arxiv.org/abs/1910.09366}{{\ttfamily 1910.09366}}.

\bibitem{Chung:2019duq}
M.-Z. Chung, Y.-T. Huang and J.-W. Kim, \emph{{From quantized spins to rotating
  black holes}},  \href{https://arxiv.org/abs/1908.08463}{{\ttfamily
  1908.08463}}.

\bibitem{Bautista:2019tdr}
Y.~F. Bautista and A.~Guevara, \emph{{From Scattering Amplitudes to Classical
  Physics: Universality, Double Copy and Soft Theorems}},
  \href{https://arxiv.org/abs/1903.12419}{{\ttfamily 1903.12419}}.

\bibitem{Bautista:2019evw}
Y.~F. Bautista and A.~Guevara, \emph{{On the Double Copy for Spinning Matter}},
   \href{https://arxiv.org/abs/1908.11349}{{\ttfamily 1908.11349}}.

\bibitem{KoemansCollado:2019ggb}
A.~Koemans~Collado, P.~Di~Vecchia and R.~Russo, \emph{{Revisiting the second
  post-Minkowskian eikonal and the dynamics of binary black holes}},
  \href{https://doi.org/10.1103/PhysRevD.100.066028}{\emph{Phys. Rev.}
  {\bfseries D100} (2019) 066028}
  [\href{https://arxiv.org/abs/1904.02667}{{\ttfamily 1904.02667}}].

\bibitem{Johansson:2019dnu}
H.~Johansson and A.~Ochirov, \emph{{Double copy for massive quantum particles
  with spin}}, \href{https://doi.org/10.1007/JHEP09(2019)040}{\emph{JHEP}
  {\bfseries 09} (2019) 040}
  [\href{https://arxiv.org/abs/1906.12292}{{\ttfamily 1906.12292}}].

\bibitem{Aoude:2020onz}
R.~Aoude, K.~Haddad and A.~Helset, \emph{{On-shell heavy particle effective
  theories}},  \href{https://arxiv.org/abs/2001.09164}{{\ttfamily 2001.09164}}.

\bibitem{Cristofoli:2020uzm}
A.~Cristofoli, P.~H. Damgaard, P.~Di~Vecchia and C.~Heissenberg,
  \emph{{Second-order Post-Minkowskian scattering in arbitrary dimensions}},
  \href{https://arxiv.org/abs/2003.10274}{{\ttfamily 2003.10274}}.

\bibitem{Chung:2020rrz}
M.-Z. Chung, Y.-t. Huang, J.-W. Kim and S.~Lee, \emph{{Complete Hamiltonian for
  spinning binary systems at first post-Minkowskian order}},
  \href{https://arxiv.org/abs/2003.06600}{{\ttfamily 2003.06600}}.

\bibitem{zvispin}
Z.~Bern, A.~Luna, R.~Roiban, C.-H. Shen and M.~Zeng, \emph{{Spinning Black Hole
  Binary Dynamics, Scattering Amplitudes and Effective Field Theory}},
  \href{https://arxiv.org/abs/2005.03071}{{\ttfamily 2005.03071}}.

\bibitem{Bern:2020gjj}
Z.~Bern, H.~Ita, J.~Parra-Martinez and M.~S. Ruf, \emph{{Universality in the
  classical limit of massless gravitational scattering}},
  \href{https://arxiv.org/abs/2002.02459}{{\ttfamily 2002.02459}}.

\bibitem{DiVecchia:2019myk}
P.~Di~Vecchia, A.~Luna, S.~G. Naculich, R.~Russo, G.~Veneziano and C.~D. White,
  \emph{{A tale of two exponentiations in ${\cal N}=8$ supergravity}},
  \href{https://doi.org/10.1016/j.physletb.2019.134927}{\emph{Phys. Lett. B}
  {\bfseries 798} (2019) 134927}
  [\href{https://arxiv.org/abs/1908.05603}{{\ttfamily 1908.05603}}].

\bibitem{Antonelli:2019ytb}
A.~Antonelli, A.~Buonanno, J.~Steinhoff, M.~van~de Meent and J.~Vines,
  \emph{{Energetics of two-body Hamiltonians in post-Minkowskian gravity}},
  \href{https://doi.org/10.1103/PhysRevD.99.104004}{\emph{Phys. Rev.}
  {\bfseries D99} (2019) 104004}
  [\href{https://arxiv.org/abs/1901.07102}{{\ttfamily 1901.07102}}].

\bibitem{Brandhuber:2019qpg}
A.~Brandhuber and G.~Travaglini, \emph{{On higher-derivative effects on the
  gravitational potential and particle bending}},
  \href{https://doi.org/10.1007/JHEP01(2020)010}{\emph{JHEP} {\bfseries 01}
  (2020) 010} [\href{https://arxiv.org/abs/1905.05657}{{\ttfamily
  1905.05657}}].

\bibitem{Cheung:2020gyp}
C.~Cheung and M.~P. Solon, \emph{{Classical gravitational scattering at $
  \mathcal{O} $(G$^{3}$) from Feynman diagrams}},
  \href{https://doi.org/10.1007/JHEP06(2020)144}{\emph{JHEP} {\bfseries 06}
  (2020) 144} [\href{https://arxiv.org/abs/2003.08351}{{\ttfamily
  2003.08351}}].

\bibitem{Parra}
J.~Parra-Martinez, M.~S. Ruf and M.~Zeng, \emph{{Extremal black hole scattering
  at $O(G^3)$: graviton dominance, eikonal exponentiation, and differential
  equations}},  \href{https://arxiv.org/abs/2005.04236}{{\ttfamily
  2005.04236}}.

\bibitem{soloncheung}
C.~Cheung and M.~P. Solon, \emph{{Tidal Effects in the Post-Minkowskian
  Expansion}},  \href{https://arxiv.org/abs/2006.06665}{{\ttfamily
  2006.06665}}.

\bibitem{AccettulliHuber:2020oou}
M.~Accettulli~Huber, A.~Brandhuber, S.~De~Angelis and G.~Travaglini,
  \emph{{Eikonal phase matrix, deflection angle and time delay in effective
  field theories of gravity}},
  \href{https://doi.org/10.1103/PhysRevD.102.046014}{\emph{Phys. Rev. D}
  {\bfseries 102} (2020) 046014}
  [\href{https://arxiv.org/abs/2006.02375}{{\ttfamily 2006.02375}}].

\bibitem{Bern:2020uwk}
Z.~Bern, J.~Parra-Martinez, R.~Roiban, E.~Sawyer and C.-H. Shen, \emph{{Leading
  Nonlinear Tidal Effects and Scattering Amplitudes}},
  \href{https://arxiv.org/abs/2010.08559}{{\ttfamily 2010.08559}}.

\bibitem{Cheung:2020gbf}
C.~Cheung, N.~Shah and M.~P. Solon, \emph{{Mining the Geodesic Equation for
  Scattering Data}},  \href{https://arxiv.org/abs/2010.08568}{{\ttfamily
  2010.08568}}.

\bibitem{spinsheet}
A.~Guevara, B.~Maybee, A.~Ochirov, D.~O'Connell and J.~Vines, \emph{{A
  worldsheet for Kerr}},  \href{https://arxiv.org/abs/2012.11570}{{\ttfamily
  2012.11570}}.

\bibitem{4pmzvi}
Z.~Bern, J.~Parra-Martinez, R.~Roiban, M.~S. Ruf, C.-H. Shen, M.~P. Solon
  et~al., \emph{{Scattering Amplitudes and Conservative Binary Dynamics at
  ${\cal O}(G^4)$}},  \href{https://arxiv.org/abs/2101.07254}{{\ttfamily
  2101.07254}}.

\bibitem{justin1}
J.~Vines, \emph{{Scattering of two spinning black holes in post-Minkowskian
  gravity, to all orders in spin, and effective-one-body mappings}},
  \href{https://doi.org/10.1088/1361-6382/aaa3a8}{\emph{Class. Quant. Grav.}
  {\bfseries 35} (2018) 084002}
  [\href{https://arxiv.org/abs/1709.06016}{{\ttfamily 1709.06016}}].

\bibitem{justin2}
J.~Vines, J.~Steinhoff and A.~Buonanno, \emph{{Spinning-black-hole scattering
  and the test-black-hole limit at second post-Minkowskian order}},
  \href{https://doi.org/10.1103/PhysRevD.99.064054}{\emph{Phys. Rev. D}
  {\bfseries 99} (2019) 064054}
  [\href{https://arxiv.org/abs/1812.00956}{{\ttfamily 1812.00956}}].

\bibitem{Parra2}
E.~Herrmann, J.~Parra-Martinez, M.~S. Ruf and M.~Zeng, \emph{{Gravitational
  Bremsstrahlung from Reverse Unitarity}},
  \href{https://arxiv.org/abs/2101.07255}{{\ttfamily 2101.07255}}.

\bibitem{Gabriele}
P.~Di~Vecchia, C.~Heissenberg, R.~Russo and G.~Veneziano, \emph{{Radiation
  Reaction from Soft Theorems}},
  \href{https://arxiv.org/abs/2101.05772}{{\ttfamily 2101.05772}}.

\bibitem{janmogul}
G.~Mogull, J.~Plefka and J.~Steinhoff, \emph{{Classical black hole scattering
  from a worldline quantum field theory}},
  \href{https://arxiv.org/abs/2010.02865}{{\ttfamily 2010.02865}}.

\bibitem{janmogul2}
G.~U. Jakobsen, G.~Mogull, J.~Plefka and J.~Steinhoff, \emph{{Classical
  Gravitational Bremsstrahlung from a Worldline Quantum Field Theory}},
  \href{https://arxiv.org/abs/2101.12688}{{\ttfamily 2101.12688}}.

\bibitem{Mougiakakos:2021ckm}
S.~Mougiakakos, M.~M. Riva and F.~Vernizzi, \emph{{Gravitational Bremsstrahlung
  in the Post-Minkowskian Effective Field Theory}},
  \href{https://arxiv.org/abs/2102.08339}{{\ttfamily 2102.08339}}.

\bibitem{iwasaki}
Y.~Iwasaki, \emph{{Quantum theory of gravitation vs. classical theory. -
  fourth-order potential}},
  \href{https://doi.org/10.1143/PTP.46.1587}{\emph{Prog. Theor. Phys.}
  {\bfseries 46} (1971) 1587}.

\bibitem{firsov}
O.~B. Firsov, \emph{Determination of the forces acting between atoms using the
  differential effective cross-section for elastic scattering}, {\emph{ZhETP}
  {\bfseries 24} (1953) 279}.

\bibitem{bini3}
D.~Bini, T.~Damour and A.~Geralico, \emph{{Sixth post-Newtonian
  nonlocal-in-time dynamics of binary systems}},
  \href{https://doi.org/10.1103/PhysRevD.102.084047}{\emph{Phys. Rev. D}
  {\bfseries 102} (2020) 084047}
  [\href{https://arxiv.org/abs/2007.11239}{{\ttfamily 2007.11239}}].

\bibitem{tessmer}
M.~Tessmer, J.~Hartung and G.~Schafer, \emph{{Aligned Spins: Orbital Elements,
  Decaying Orbits, and Last Stable Circular Orbit to high post-Newtonian
  Orders}}, \href{https://doi.org/10.1088/0264-9381/30/1/015007}{\emph{Class.
  Quant. Grav.} {\bfseries 30} (2013) 015007}
  [\href{https://arxiv.org/abs/1207.6961}{{\ttfamily 1207.6961}}].

\bibitem{letiec}
A.~Le~Tiec, L.~Blanchet and B.~F. Whiting, \emph{{The First Law of Binary Black
  Hole Mechanics in General Relativity and Post-Newtonian Theory}},
  \href{https://doi.org/10.1103/PhysRevD.85.064039}{\emph{Phys. Rev.}
  {\bfseries D85} (2012) 064039}
  [\href{https://arxiv.org/abs/1111.5378}{{\ttfamily 1111.5378}}].

\bibitem{nj}
E.~T. Newman and A.~I. Janis, \emph{{Note on the Kerr spinning particle
  metric}}, \href{https://doi.org/10.1063/1.1704350}{\emph{J. Math. Phys.}
  {\bfseries 6} (1965) 915}.

\bibitem{Chia:2020dye}
H.~S. Chia, \emph{{Probing Particle Physics with Gravitational Waves}}, Ph.D.
  thesis, Amsterdam U., 2020.
\newblock \href{https://arxiv.org/abs/2012.09167}{{\ttfamily 2012.09167}}.

\bibitem{LeTiec:2020bos}
A.~Le~Tiec, M.~Casals and E.~Franzin, \emph{{Tidal Love Numbers of Kerr Black
  Holes}},  \href{https://arxiv.org/abs/2010.15795}{{\ttfamily 2010.15795}}.

\bibitem{Charalambous:2021mea}
P.~Charalambous, S.~Dubovsky and M.~M. Ivanov, \emph{{On the Vanishing of Love
  Numbers for Kerr Black Holes}},
  \href{https://arxiv.org/abs/2102.08917}{{\ttfamily 2102.08917}}.

\bibitem{Binnington:2009bb}
T.~Binnington and E.~Poisson, \emph{{Relativistic theory of tidal Love
  numbers}}, \href{https://doi.org/10.1103/PhysRevD.80.084018}{\emph{Phys. Rev.
  D} {\bfseries 80} (2009) 084018}
  [\href{https://arxiv.org/abs/0906.1366}{{\ttfamily 0906.1366}}].

\bibitem{Damour:2009vw}
T.~Damour and A.~Nagar, \emph{{Relativistic tidal properties of neutron
  stars}}, \href{https://doi.org/10.1103/PhysRevD.80.084035}{\emph{Phys. Rev.
  D} {\bfseries 80} (2009) 084035}
  [\href{https://arxiv.org/abs/0906.0096}{{\ttfamily 0906.0096}}].

\bibitem{Hui:2020xxx}
L.~Hui, A.~Joyce, R.~Penco, L.~Santoni and A.~R. Solomon, \emph{{Static
  response and Love numbers of Schwarzschild black holes}},
  \href{https://arxiv.org/abs/2010.00593}{{\ttfamily 2010.00593}}.

\bibitem{Berezin:1976eg}
F.~A. Berezin and M.~S. Marinov, \emph{{Particle Spin Dynamics as the Grassmann
  Variant of Classical Mechanics}},
  \href{https://doi.org/10.1016/0003-4916(77)90335-9}{\emph{Annals Phys.}
  {\bfseries 104} (1977) 336}.

\bibitem{letiecs}
L.~Blanchet, A.~Buonanno and A.~Le~Tiec, \emph{{First law of mechanics for
  black hole binaries with spins}},
  \href{https://doi.org/10.1103/PhysRevD.87.024030}{\emph{Phys. Rev. D}
  {\bfseries 87} (2013) 024030}
  [\href{https://arxiv.org/abs/1211.1060}{{\ttfamily 1211.1060}}].

\bibitem{andres2}
A.~Luna and D.~Kosmopoulos, \emph{{\it to appear}}, .

\bibitem{Passarino:1978jh}
G.~Passarino and M.~J.~G. Veltman, \emph{{One Loop Corrections for $e^+ e^-$
  Annihilation Into $\mu^+ \mu^-$ in the Weinberg Model}},
  \href{https://doi.org/10.1016/0550-3213(79)90234-7}{\emph{Nucl. Phys. B}
  {\bfseries 160} (1979) 151}.

\bibitem{Smirnov}
V.~A. Smirnov, \emph{{Analytic tools for Feynman integrals}}. {Springer}, 2012,
  \href{https://doi.org/10.1007/978-3-642-34886-0}{10.1007/978-3-642-34886-0}.

\end{thebibliography}\endgroup

\end{document}